\documentclass[iop,apj]{emulateapj}

\usepackage{amsfonts}
\usepackage{amsmath}
\usepackage{times}

\newcommand{\mShock}{M_{\mbox{\tiny{Sh}}}}
\newcommand{\rShock}{R_{\mbox{\tiny{Sh}}}}
\newcommand{\rShockAvg}{\bar{R}_{\mbox{\tiny{Sh}}}}
\newcommand{\rPNS}{R_{\mbox{\tiny PNS}}}
\newcommand{\sShock}{\partial V_{\mbox{\tiny{Sh}}}}
\newcommand{\sPNS}{\partial V_{\mbox{\tiny PNS}}}
\newcommand{\vShock}{V_{\mbox{\tiny{Sh}}}}
\newcommand{\vPNS}{V_{\mbox{\tiny PNS}}}
\newcommand{\vShell}{\delta V_{i}}
\newcommand{\mSun}{M_{\odot}}
\newcommand{\edens}[1]{e_{\mbox{\tiny #1}}}
\newcommand{\eShock}[1]{E_{\mbox{\tiny #1}}}
\newcommand{\edensk}[1]{\widehat{e}_{\mbox{\tiny #1}}}
\newcommand{\kMin}{k_{\mbox{\tiny min}}}
\newcommand{\kMax}{k_{\mbox{\tiny max}}}
\newcommand{\kTurb}{k_{\mbox{\tiny tur}}}
\newcommand{\kMeanMag}{\bar{k}_{\mbox{\tiny mag}}}
\newcommand{\kDynamic}{k_{\mbox{\tiny dyn}}}
\newcommand{\lambdaCurvature}{\lambda_{\mbox{\tiny c}}}
\newcommand{\lambdaRMS}{\lambda_{\mbox{\tiny rms}}}
\newcommand{\lambdaMeanMag}{\bar{\lambda}_{\mbox{\tiny mag}}}
\newcommand{\lambdaTurb}{\lambda_{\mbox{\tiny tur}}}
\newcommand{\lambdaTaylor}{\lambda_{\mbox{\tiny T}}}
\newcommand{\lambdaDissipation}{\lambda_{\mbox{\tiny d}}}
\newcommand{\lambdaDynamic}{\lambda_{\mbox{\tiny dyn}}}
\newcommand{\EKinTurb}{E_{\mbox{\tiny kin}}^{\mbox{\tiny tur}}}
\newcommand{\uRMSTurb}{u_{\mbox{\tiny rms}}^{\mbox{\tiny tur}}}
\newcommand{\uRMS}{u_{\mbox{\tiny rms}}}
\newcommand{\tauEddy}{\tau_{\mbox{\tiny eddy}}}
\newcommand{\fMagEncl}{f_{\mbox{\tiny mag}}(k)}
\newcommand{\genasis}{GenASiS}
\newcommand{\model}[3]{\mbox{B}{#1}\mbox{L}{#2}\mbox{E}{#3}}
\newcommand{\f}[2]{\frac{#1}{#2}}
\newcommand{\pderiv}[2]{\frac{\partial #1}{\partial #2}}
\newcommand{\dderiv}[2]{\frac{d #1}{d #2}}
\newcommand{\vect}[1]{\mathbf{#1}}
\newcommand{\divergence}[1]{\mathbf{\nabla}\cdot {#1}}
\newcommand{\gradient}[1]{\mathbf{\nabla} {#1}}
\newcommand{\curl}[1]{\mathbf{\nabla}\times {#1}}
\newcommand{\timeAverage}[3]{\langle{#1}\rangle_{#2\hspace{0.05cm}\mbox{\tiny s}}^{#3\hspace{0.05cm}\mbox{\tiny s}}}
\newcommand{\volumeAverage}[2]{\langle{#1}\rangle_{#2}}
\newcommand{\rateLorentzWork}{\tau_{\vect{J}\times\vect{B}}^{-1}}
\newcommand{\timeLorentzWork}{\tau_{\vect{J}\times\vect{B}}}
\newcommand{\ratePoynting}{\tau_{\vect{P}}^{-1}}
\newcommand{\timePoynting}{\tau_{\vect{P}}}
\newcommand{\rateDissipation}{\tau_{\mbox{\tiny J}}^{-1}}
\newcommand{\timeDissipation}{\tau_{\mbox{\tiny J}}}
\newcommand{\rateCompression}{\tau_{\divergence{\vect{u}}}^{-1}}
\newcommand{\rateStretching}{\tau_{\gradient{\vect{u}}}^{-1}}
\newcommand{\rateTotal}{\tau_{\mbox{\tiny tot}}^{-1}}
\newcommand{\timeTotal}{\tau_{\mbox{\tiny tot}}}
\newcommand{\reynoldsMag}{R_{\mbox{\tiny m}}}

\defcitealias{endeve_etal_2010}{Paper I}
\defcitealias{obergaulinger_etal_2009}{Obergaulinger et al. 2009}

\shorttitle{TURBULENT MAGNETIC FIELD AMPLIFICATION FROM SASI}
\shortauthors{ENDEVE ET AL.}

\begin{document}

\slugcomment{Accepted for publication in the Astrophysical Journal}

\title{Turbulent Magnetic Field Amplification from Spiral SASI Modes:  \\
Implications for Core-Collapse Supernovae and Proto-Neutron Star Magnetization}

\author{Eirik Endeve\altaffilmark{1}, Christian Y. Cardall\altaffilmark{2,3}, Reuben D. Budiardja\altaffilmark{2,3,4}, Samuel W. Beck\altaffilmark{4}, Alborz Bejnood\altaffilmark{4}, Ross J. Toedte\altaffilmark{5}, Anthony Mezzacappa\altaffilmark{1,2,3}, and John M. Blondin\altaffilmark{6}}
\email{endevee@ornl.gov}

\altaffiltext{1}{Computer Science and Mathematics Division, Oak Ridge National Laboratory, Oak Ridge, TN 37831-6354, USA}
\altaffiltext{2}{Physics Division, Oak Ridge National Laboratory, Oak Ridge, TN 37831-6354, USA}
\altaffiltext{3}{Department of Physics and Astronomy, University of Tennessee, Knoxville, TN 37996-1200, USA}
\altaffiltext{4}{Joint Institute for Heavy Ion Research, Oak Ridge National Laboratory, Oak Ridge, TN 37831-6374, USA}
\altaffiltext{5}{National Center for Computational Sciences, Oak Ridge National Laboratory, Oak Ridge, TN 37831-6354, USA}
\altaffiltext{6}{Physics Department, North Carolina State University, Raleigh, NC 27695-8202, USA}

\begin{abstract}
We extend our investigation of magnetic field evolution in three-dimensional flows driven by the stationary accretion shock instability (SASI) with a suite of higher-resolution idealized models of the post-bounce core-collapse supernova environment.  
Our magnetohydrodynamic simulations vary in initial magnetic field strength, rotation rate, and grid resolution.  
Vigorous SASI-driven turbulence inside the shock amplifies magnetic fields exponentially; but while the amplified fields reduce the kinetic energy of small-scale flows, they do not seem to affect the global shock dynamics.  
The growth rate and final magnitude of the magnetic energy are very sensitive to grid resolution, and both are underestimated by the simulations.  
Nevertheless our simulations suggest that neutron star magnetic fields exceeding $10^{14}$~G can result from dynamics driven by the SASI, \emph{even for non-rotating progenitors}.  
\end{abstract}


\section{INTRODUCTION}

Not long after the discovery of pulsars---whose characteristic signal was linked to magnetic fields \citep{hewish_etal_1968}---the potential role of magnetic fields in the core-collapse supernova (CCSN) explosion mechanism began to be investigated  \citep[e.g.,][]{leblanc_wilson_1970,bisnovatyi-kogan_etal_1976,meier_etal_1976,symbalisty_1984}.  
In principle, a differentially rotating proto-neutron star (PNS) could both amplify magnetic fields and serve as an energy reservoir available to be tapped by those fields, giving rise to magnetically powered explosions.  
An early conclusion, however, was that both unrealistically rapid rotation \emph{and} unrealistically strong magnetic fields would be needed at the pre-collapse stage for magnetic fields to play a principal role in the explosion dynamics \citep{leblanc_wilson_1970,symbalisty_1984}.  

In more recent years interest in strong magnetic fields has returned in connection with a number of observables related to core-collapse supernovae, including asymmetries in the explosion ejecta \citep{wheeler_etal_2002}, natal neutron star kick velocities \citep{laiQian_1998}, and especially the high-energy electromagnetic activity connected to some neutron stars known as magnetars, or Anomalous X-ray Pulsars (AXPs) and Soft Gamma Repeaters (SGRs) \citep[e.g.,][]{duncanThompson_1992,thompsonDuncan_2001,hurley_etal_2005,woodsThompson_2006}.  
AXPs and SGRs are characterized by quiescent X-ray luminosities as high as $10^{35}$~erg s$^{-1}$, with sporadic outbursts releasing up to $10^{41}$~erg per event.  
Gamma-ray outbursts from SGRs are even more energetic, an extreme example being the giant flare from SGR 1806-20, which released an estimated $10^{46}$~erg over 380~s \citep{hurley_etal_2005}.  
Furthermore, AXPs and SGRs are neutron stars characterized by relatively long rotation periods ($P\gtrsim1$~s) and high spin-down rates ($\dot{P}\gtrsim10^{-12}$~ss$^{-1}$) \citep[e.g.,][]{lorimerKramer_2005}.  
As their rotational energy cannot account for the electromagnetic emission, and because of the strong magnetic torques implied by high spin-down rates, they are believed to be young neutron stars powered by dissipation of extremely strong surface magnetic fields \citep[$10^{14}$-$10^{15}$~G,][]{duncanThompson_1996,thompsonDuncan_2001}.  

On the theoretical side, the discovery of the magneto-rotational instability (MRI) by \citet{balbusHawley_1991} and its application to CCSNe \citep[initiated by][]{akiyama_etal_2003} relaxed the requirement of strong pre-collapse $B$-fields,
 renewing interest in magnetic fields as a possible key ingredient in the explosion mechanism of some supernovae \citep[i.e., those from rapidly rotating progenitor cores; e.g.,][]{wheeler_etal_2002,obergaulinger_etal_2005,moiseenko_etal_2006,burrows_etal_2007,takiwaki_etal_2009}. 
(The MRI results in exponential growth of the magnetic energy on the rotation timescale.)  
However, the rotational energy falls off quadratically with increasing rotation period, and is about $5\times10^{49}$~erg for a 20~ms period PNS---much less than the characteristic CCSN explosion energy of $\sim10^{51}~\mbox{erg}\equiv1~\mbox{Bethe (B)}$.  
Thus any magneto-rotationally driven supernovae likely would be peculiar events, since magnetic progenitor cores tend to rotate slowly at the pre-collapse stage \citep{heger_etal_2005}.  

Leaving aside the explosion mechanism, the relationship between the formation of neutron star magnetic fields and CCSNe is still an open and interesting question, particularly in the case of magnetars (AXPs and SGRs) \citep{lorimerKramer_2005}.  
\citet{thompsonDuncan_1993} argued that such strong fields must be generated during the neutrino cooling epoch after the collapse of the progenitor's iron core, and possibly before the explosion is initiated ($\sim1$~s after core collapse).  
Their model remains one of the prevailing theories for magnetar formation, and includes a convective $\alpha-\Omega$ dynamo, which operates when the rotation period is comparable to the turnover time of entropy-driven convection ($\lesssim3$~ms) near the surface of the PNS.  
The rapid turnover time may suggest that magnetars are formed in the magnetically-driven explosion of collapsed, rapidly rotating progenitors, whose remnant is spun down by MHD processes at later times.  
\citet{bonanno_etal_2003,bonanno_etal_2005} found that neutron finger instabilities \citep[e.g.,][]{bruennDineva_1996} may also result in dynamo action in PNSs with rotation periods as long as 1~s.  
In this scenario, the formation of neutron star magnetic fields may be slow (compared to the explosion time scale), and their creation is not necessarily tied to dynamics in the supernova explosion.  
The MRI may also operate near the surface of the PNS, and contribute to neutron star magnetization.  

The lack of sufficient rotational energy in magnetized pre-collapse progenitor cores, as predicted by stellar evolution models \citep{heger_etal_2005}, has sparked some recent interest in MHD processes in non-rotating CCSN environments \citep{endeve_etal_2010,guilet_etal_2011,obergaulingerJanka_2011}.  
These studies investigate field amplification mechanisms and the possible role of amplified $B$-fields on the dynamics of slowly  or non-rotating collapsed progenitors, in which rotational MHD processes are insignificant.  
In particular, \citet[][hereafter \citetalias{endeve_etal_2010}]{endeve_etal_2010} studied magnetic field amplification by the stationary accretion shock instability \citep[SASI,][]{blondin_etal_2003}.  
The SASI is central to the theory of CCSNe: recent simulations lead to the conclusion that it likely plays an important role in neutrino-powered explosions \citep{bruenn_etal_2006,buras_etal_2006,mezzacappa_etal_2007,scheck_etal_2008,marekJanka_2009,suwa_etal_2010,muller_etal_2012}, and may also explain certain observables of pulsars, including their proper motion \citep{scheck_etal_2004} and spin \citep{blondinMezzacappa_2007}.  
Thus, magnetic fields may be an important part of a supernova model if the SASI is found to be sensitive to their presence.  

In \citetalias{endeve_etal_2010} we adopted the idealized model of \citet{blondin_etal_2003} and \citet{blondinMezzacappa_2007}, and added a weak radial (split monopole) magnetic field.  
We presented results from 2D (axisymmetric) and 3D MHD simulations of the SASI, and found that SASI-driven flows may result in significant magnetic field amplification.  
Magnetic field evolution in axisymmetric simulations was found to be geometrically constrained.  
Moreover, the non-axisymmetric spiral SASI mode \citep{blondinMezzacappa_2007} dominates the post-shock flows in 3D simulations at late times.  
The nonlinear evolution of the spiral mode drives vigorous turbulence below the shock, which results in exponential amplification of $B$-fields due to ``stretching" \citep[e.g.,][]{ott_1998}, and the magnetic energy becomes concentrated in intense, intermittent magnetic flux ropes.  
We presented results from models with non-rotating and weakly rotating initial conditions, and weak ($10^{10}$~G) and stronger ($10^{12}$~G) initial magnetic fields.  
The magnetic fields were not found to reach dynamically significant levels (i.e., components of the Maxwell stress tensor did not contribute significantly to the total stress), and hence no impact of magnetic fields on local or global dynamics was demonstrated.  
However, we found that SASI-induced turbulent magnetic field amplification is very sensitive to the spatial resolution adopted in the numerical simulations.  
Most of the 3D models presented in \citetalias{endeve_etal_2010} were performed at ``medium" spatial resolution (grid cells with sides $\Delta l=1.56$~km), while one model was performed with ``high" spatial resolution ($\Delta l=1.17$~km).  
The thickness of magnetic flux ropes was found to decrease in proportion to $\Delta l$.  
We did not observe convergence of $B$-field amplification with increasing spatial resolution.  
Nevertheless, the simulations implied neutron star magnetization as a result of SASI-induced magnetic field amplification.  

This paper continues and extends the investigations initiated in \citetalias{endeve_etal_2010}.  
It improves on our previous study in several important ways, including (1) coverage of a larger parameter space, (2) higher spatial resolution (up to $1280^{3}$ zones), and (3) computation of kinetic and magnetic energy spectra.  
With the new set of simulations we investigate in some detail the nature of SASI-driven turbulence, and the growth and impact of magnetic fields during operation of the SASI.  
We investigate the saturation level of magnetic energy in our simulations, the (kinetic) energy reservoir available for magnetic field amplification, and the factors determining the magnetic energy growth rate.  
We also consider as in \citetalias{endeve_etal_2010} the impact of amplified magnetic fields on global shock dynamics, in particular any impact they may have on the SASI.  
Finally, we attempt to quantify the levels of neutron star magnetization that may be expected from SASI dynamics.  

We find that the SASI-driven turbulence shares several similarities with non-helical turbulence \citep[e.g.,][]{brandenburg_etal_1996,haugen_etal_2004}, and results in an efficient small-scale dynamo.  
Magnetic fields grow exponentially in the turbulent flows driven by the SASI as long as the ``kinematic regime'' remains valid.  
The kinematic regime ends when the magnetic energy becomes comparable (locally) to the kinetic energy of the turbulent flows---the magnetic energy source.  
From the computed energy spectra we estimate the ``turbulent" kinetic energy, $\EKinTurb$, available for magnetic field amplification, and, in our idealized model, $\EKinTurb$ constitutes about $10\%$ of the total kinetic energy below the shock ($E_{\mbox{\tiny kin}}\sim5\times10^{49}$~erg).  
The total magnetic energy saturates at about $E_{\mbox{\tiny mag}}\sim5\times10^{47}$~erg.  
The presence of amplified magnetic fields results in less kinetic energy on small spatial scales, but we find no impact of magnetic fields on global shock dynamics, which is consistent with considerations of the energetics.  
However, magnetic field evolution remains sensitive to numerical resolution, and magnetic fields are subject to significant numerical dissipation during the saturated state, and our ability to quantify fully the impact of magnetic fields in a more realistic situation is therefore limited.  
The magnetic energy growth time decreases with increasing resolution, and, based on the turnover time of the SASI-driven turbulence, is estimated to be a few milliseconds.  
We argue that the MHD processes studied in this paper may contribute significantly to strong, small-scale neutron star magnetic fields, and provide a connection between the magnetic fields of neutron stars at birth and supernova dynamics.  
The saturation energies may be sufficient to power flaring activity of AXPs, and possibly SGRs.  
Moreover, their formation does not require progenitor rotation.  

\section{SETUP OF NUMERICAL SIMULATIONS}

We employ the same numerical code and three-dimensional initial conditions we used in \citetalias{endeve_etal_2010}, which follow closely the adiabatic setup described in \citet{blondin_etal_2003} and \citet{blondinMezzacappa_2007}: a stationary accretion shock is placed at a radius $r=\rShock=200$~km, and a highly supersonic flow is nearly free-falling towards the shock for $r>\rShock$ with $\edens{kin}+\edens{grav}\approx0$.  
Between the shock and the PNS the flow settles subsonically---obeying the Bernoulli equation $\edens{kin}+\edens{int}+P+\edens{grav}=0$---and is nearly in hydrostatic equilibrium.  
Matter is allowed to flow through an inner boundary placed at $r=\rPNS=40$~km.  
The mass density and pressure just inside $\rPNS$ are determined from values just outside $\rPNS$ using power-law extrapolations: $\rho\propto r^{-3}$ and $P\propto r^{-4}$, respectively (a procedure that proved necessary in order to maintain the steady state of the unperturbed initial condition).  

\begin{figure}
  \epsscale{1.0}
  \plotone{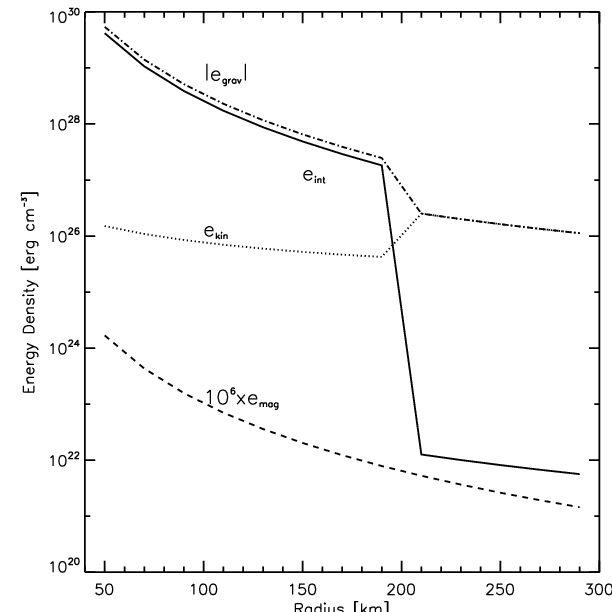}
  \caption{Plot of the initial condition for the non-rotating weak-field model with $B_{0}=10^{10}$~G ($\model{10}{0.0}{00}$): internal energy density ($\edens{int}$, solid line), magnitude of gravitational potential energy density ($|\edens{grav}|$, dash-dot line), kinetic energy density ($\edens{kin}$, dotted line), and magnetic energy density ($\edens{mag}$, dashed line) versus radial distance from the center of the PNS.  The surface of the PNS (our inner boundary) is fixed at $r=\rPNS=40$~km and the shock is initially located at $r=R_{\mbox{\tiny Sh}}=200$~km.  Inside the shock $|\edens{grav}|$, $\edens{int}$ and $\edens{mag}$ follow roughly the same power-law ($\propto r^{-4}$), while $\edens{kin}\propto r^{-1}$.  The flow is in steady state free-fall outside $\rShock$, with $\edens{kin}$ and $|\edens{grav}|$ proportional to $r^{-2.5}$, and $\edens{int}\propto r^{-2}$.  $\edens{mag}$ has been multiplied by $10^{6}$ to become visible on the plot.  (The dashed line is also identical to $\edens{mag}$ in the strong-field model ($\model{13}{0.0}{00}$), cf. Table \ref{tab:computedModels}.)  \label{fig:initialCondition}}
\end{figure}

Figure \ref{fig:initialCondition} displays the initial configuration of a spherically symmetric, non-rotating stationary accretion shock with a weak radial magnetic field ($B_{0}=1\times10^{10}$~G; the initial magnetic fields in our simulations are discussed in further detail below).  
We plot internal energy density $\edens{int}=P/(\gamma-1)$, kinetic energy density 
$\edens{kin}= \rho\vect{u}\cdot\vect{u}/2$, 
magnetic energy density 
$\edens{mag}=\vect{B}\cdot\vect{B}/(2\mu_{0})$, 
and the magnitude of the gravitational potential energy density $\edens{grav}=\rho\Phi$ versus radial distance from the center of the star.  
Here $\rho$, $\vect{u}$, $P$, $\vect{B}$, and $\Phi$ are the mass density, fluid velocity, fluid pressure, magnetic flux density (magnetic field), and gravitational potential, respectively.  
The vacuum permeability is denoted $\mu_{0}$.  
We adopt the ideal gas equation of state, with the ratio of specific heats set to $\gamma=4/3$.  
The time-independent point-mass gravitational potential is $\Phi=-GM/r$, where $G$ is Newton's constant and $M=1.2~M_{\odot}$ is the mass of the central object.  
The accretion rate ahead of the shock is $\dot{M}=0.36~M_{\odot}\mbox{ s}^{-1}$, which is held fixed during the simulations.  

Our numerical simulation code, \genasis, solves the adiabatic, non-relativistic, ideal MHD equations including gravity \citepalias[cf. Eqs. (1)-(4) in][]{endeve_etal_2010}.  
Starting from the semi-analytic initial condition, balance equations for mass density $\rho$, momentum density $\vect{S}=\rho\vect{u}$, and magneto-fluid energy density $\edens{fluid}=\edens{int}+\edens{kin}+\edens{mag}$ are evolved with a second-order HLL-type ideal MHD scheme in a manner that ensures conservation of mass and energy (i.e., volume integrals of $\rho$ and $\edens{fluid}+\edens{grav}$) to numerical precision.  
The magnetic induction equation is evolved in a divergence-free manner via the method of constrained transport \citep{evansHawley_1988}.  \citepalias[See][and the references therein for further details.  See also Appendix \ref{app:numericalDissipation} in this paper.]{endeve_etal_2010}  

Without initial perturbations the initial configuration in Figure \ref{fig:initialCondition} remains stationary.  
In order to initiate the SASI we perturb the initial condition by adding small ($\sim1\%$) random perturbations to the initial pressure profile in the region $r\in[\rPNS,\rShock]$.  
These perturbations initiate the SASI and allow us to study the evolution of magnetic fields in SASI-driven flows.  

The topology, strength and distribution of magnetic fields in core-collapse supernova progenitors are highly uncertain.  
A similar uncertainty applies to our knowledge of the angular momentum distribution in the progenitor core.  
These uncertainties then apply directly to the initial conditions of simulations aimed at studying the evolution and impact of magnetic fields in core-collapse supernovae.  

Rotation and magnetic fields in stellar interiors are intimately coupled in a complex multidimensional interplay.  
Stellar core rotation can drive the evolution of magnetic fields, while the magnetic fields can play an important role in distributing the core's angular momentum \citep[e.g.,][]{spruit_2002}.  
Three-dimensional stellar evolution models (even without magnetic fields) extending all the way to iron core collapse are currently not available.  
However, some insight into the issue of core magnetic fields (and rotation) is provided by recent stellar evolution calculations \citep[e.g.,][]{heger_etal_2005,meynet_etal_2011}.  
In particular, \citet{heger_etal_2005} included magnetic fields in their calculations, and found that magnetic torques can significantly reduce the rotation rate of the pre-collapse iron core.  
The resulting magnetic fields were dominated by a toroidal component $B_{\varphi}$ ($B_{\varphi}/B_{r}=10^{3}-10^{4}$, where $B_{r}$ is the poloidal (radial) component of the magnetic field).  
They also reported that the core rotation rate and magnetic field strength at the pre-supernova stage is an increasing function of progenitor mass.  
In the iron core of their 15~$\mSun$ model, the toroidal and poloidal magnetic fields are $B_{\varphi}\approx5\times10^{9}$~G and $B_{r}\approx8\times10^{5}$~G, respectively, while in their 35~$\mSun$ model, the toroidal and poloidal magnetic fields are $B_{\varphi}\approx1\times10^{10}$~G and $B_{r}\approx1\times10^{7}$~G, respectively.  
Accounting for the three orders of magnitude increase attained during core-collapse, the \citet[][]{heger_etal_2005} models predict the post-bounce toroidal and poloidal magnetic fields to be in the range of $10^{12}-10^{13}$~G and $10^{9}-10^{10}$~G, respectively.  
This is in the range of `common pulsars', inferred from observations of pulsar spin periods and corresponding spin-down rates \citep{lorimerKramer_2005}, but significantly lower than that of magnetars \citep[][]{duncanThompson_1992}.  

\begin{table}
  \begin{center}
    \caption{Tabular overview of computed models.  \label{tab:computedModels}}
    \begin{tabular}{cccc}
    Model & $B_{0}$ (G) & $l_{0}$ (cm$^{2}$ s$^{-1}$) & $t_{\mbox{\tiny end}}$ (ms) \\
    \tableline
    \tableline
    \model{10}{0.0}{00} & $1\times10^{10}$ & 0.0 & 1100 \\
    \model{10}{1.5}{15} & $1\times10^{10}$ & $1.5\times10^{15}$ & 878 \\
    \model{10}{4.0}{15} & $1\times10^{10}$ & $4.0\times10^{15}$ & 678 \\
    \tableline
    \model{12}{0.0}{00}\tablenotemark{a} & $1\times10^{12}$ & 0.0 & 1126 \\
    \model{12}{1.5}{15} & $1\times10^{12}$ & $1.5\times10^{15}$ & 1000 \\
    \model{12}{4.0}{15} & $1\times10^{12}$ & $4.0\times10^{15}$ & 644 \\
    \tableline
    \model{13}{0.0}{00} & $1\times10^{13}$ & 0.0 & 1100 \\
    \tableline
    \tableline
    \end{tabular}
    \tablecomments{$^{\rm a}$ Model computed with multiple grid resolutions.}
  \end{center}
\end{table}

Investigating the role of initial $B$-field topology in our simulations is beyond the scope of this study, which is restricted to an initially radial (split monopole) magnetic field configuration; $B_{r}=\mbox{sign}(\cos\vartheta)\times B_{0}(\rPNS/r)^{2}$, where $\vartheta$ is the polar angle.  
We only vary the strength of the initial magnetic field $B_{0}$ at the surface of the PNS ($r=\rPNS$).  
In particular, we vary $B_{0}$ in the range from $1\times10^{10}$~G to $1\times10^{13}$~G (cf. Table \ref{tab:computedModels}).  
The initial magnetic energy density profile for the model with $B_{0}=1\times10^{10}$~G is represented by the dashed line in Figure \ref{fig:initialCondition}, where it has been boosted by a factor of $10^{6}$ to become visible on the plot.  
(The corresponding profile for the model with $B_{0}=1\times10^{13}$~G is identical to the dashed line in Figure \ref{fig:initialCondition}.)  
Clearly, when comparing the magnetic energy density to $\edens{kin}$ and $\edens{int}$, all our models are initiated with weak magnetic fields.  

From the perspective of the \citet{heger_etal_2005} models, our initial magnetic fields are purely poloidal and stronger than their predicted poloidal fields, but comparable (in magnitude) to their predicted toroidal magnetic fields.  
Based on the expected multidimensional character of the post-bounce supernova dynamics (e.g., convection and the SASI) and the strength of the magnetic fields ($\edens{mag}$ is small relative to $\edens{kin}$ and $\edens{int}$) we do not expect the magnetic fields to retain the anisotropic ($B_{\varphi}/B_{r}\gg 1$) configuration predicted by the stellar evolution calculations of \citet{heger_etal_2005}.  
Progenitors from multidimensional stellar evolution calculations may deviate significantly from their spherically symmetric counterparts \citep[][]{arnettMeakin2011}.  
We believe the initial magnetic field configuration we have chosen has only (at best) a secondary impact on our results, and that initial insight into the MHD evolution in core-collapse supernovae can be obtained from the simulations presented here.  

For comparison, the non-rotating simulations recently presented by \citet{obergaulingerJanka_2011} with weak initial magnetic fields (models s15-B10 and s15-B11 in that study)  start with purely poloidal pre-collapse core magnetic fields of $1\times10^{10}$~G and $1\times10^{11}$~G, respectively.  
After core-collapse and shock stagnation, the strength of the magnetic field in the stable layer separating the PNS convection zone and the gain region is about $4\times10^{12}$~G and $3\times10^{13}$~G, respectively (cf. Table 2 in \citet{obergaulingerJanka_2011}).  
This layer coincides roughly with our inner boundary at $r=\rPNS$.  
Thus, the magnetic field strength in their collapsed weak-field models is comparable (initially) to that of our strongest-field model.  

Our rotating models are initiated by setting the pre-shock gas into rotation about the $z$-axis by specifying the azimuthal velocity according to $u_{\varphi}=l_{0}\sin\vartheta/r$, where $l_{0}$ is the (constant) specific angular momentum.  
We have computed rotating models where $l_{0}$ has been set to $1.5\times10^{15}$~cm$^{2}$ s$^{-1}$ and $4.0\times10^{15}$~cm$^{2}$ s$^{-1}$.  

The discretized ideal MHD equations are solved in a cubic computational domain with sides $L$ and volume $V_{L}=L^{3}$.  
Cartesian coordinates are employed.  
The computational domain is divided into $N$ zones in each spatial dimension, resulting in $N^{3}$ cubic zones with sides $\Delta l=L/N$.  
To conserve computational resources we start our simulations in a relatively small computational domain with $L=L_{\mbox{\tiny min}}=600$~km and $N=512$ (resulting in $\Delta l\approx1.17$~km).  
The time-step in our simulations (limited by the Courant-Friedrichs-Lewy condition) is about 5-10~$\mu$s, depending on the stage of the particular run.  
The runs are evolved to a physical time of about 1~s, which results in about $10^{5}$ time-steps taken per simulation.  
The MHD solver is parallelized using the Message Passing Interface (MPI), and the computational domain is subdivided into blocks containing an equal number of zones, which are distributed among MPI processes.  
During the simulations we keep the number of zones per block (MPI process) fixed to $32^{3}$.  
Once the SASI evolves into the nonlinear regime the volume encompassed by the shock grows, and the shock eventually interacts with the boundary of the computational domain.  
When this happens, we expand the computational domain by adding a layer of $32^{3}$-zones blocks (i.e., we add 64 zones in each coordinate direction) and restart the simulation from the last checkpoint written before the shock interacted with the boundary of the computational domain.  
We repeat this process, and run our simulations until the shock interacts with the boundary of the largest computational box $L=L_{\mbox{\tiny max}}=1500$~km, or the simulation time reaches $t=1100$~ms, whichever occurs first.  
Since we keep $\Delta l$ fixed during the simulations, the largest computational domain is covered by 1280 zones in each spatial dimension.  
During a run, we write simulation output for analysis and visualization every 2~ms of physical time, resulting in tens of Terabytes of data from each model.  

\section{SIMULATION RESULTS}

\begin{figure*}
  \epsscale{1.0}
  \plotone{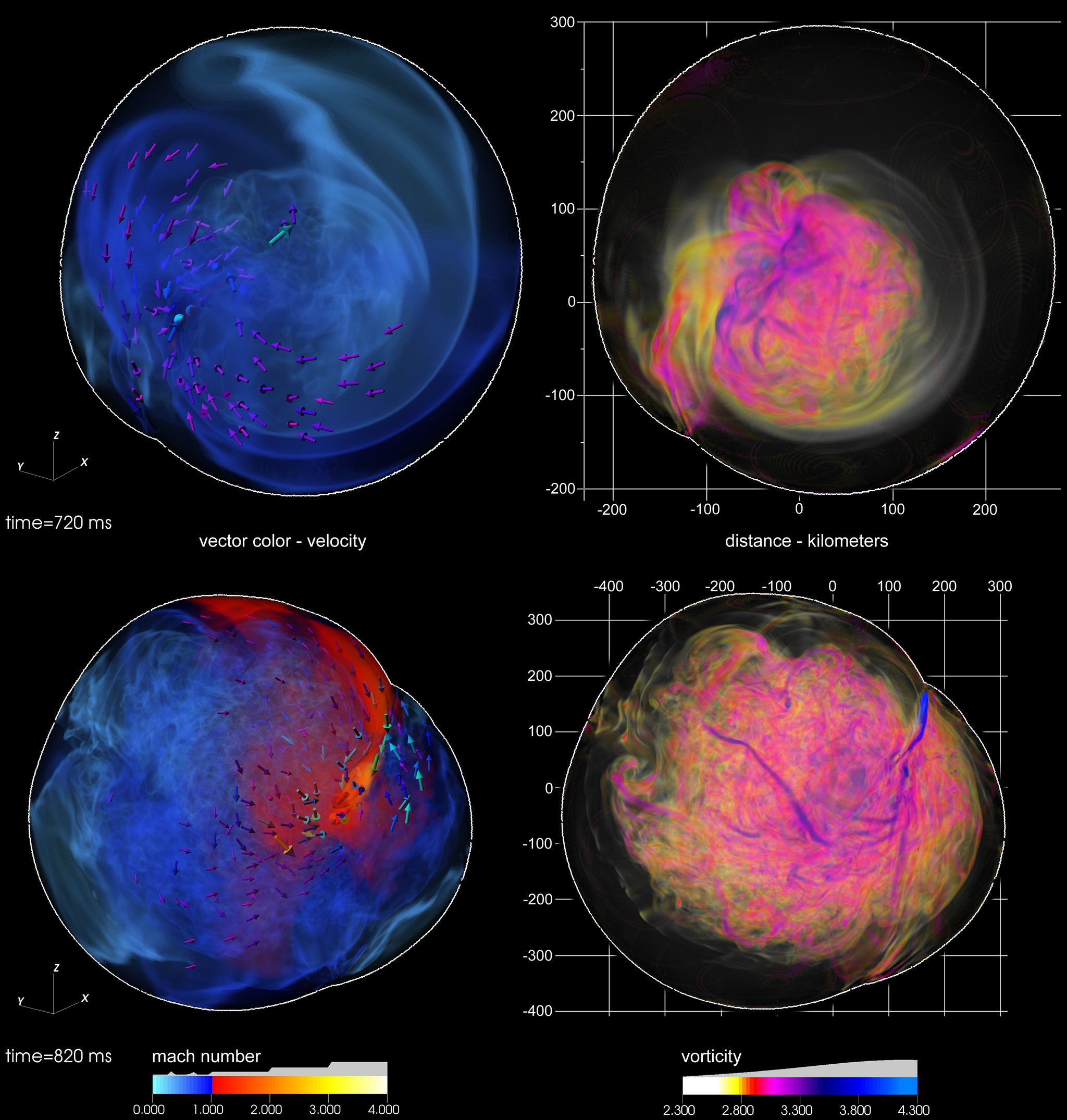}
  \caption{Nonlinear operation of the spiral SASI mode in model $\model{13}{0.0}{00}$: flow Mach number $|\vect{u}|/c_{S}$ (left panels) and magnitude of fluid vorticity $|\boldsymbol\omega|(\equiv|\curl{\vect{u}}|)$ (right panels).  The adiabatic sound speed is $c_{S}=\sqrt{\gamma P/\rho}$.  Snapshots are taken at $t=720$~ms (upper panels) and $t=820$~ms (lower panels).  To highlight the spiral mode pattern in each panel, the viewing normal is aligned with the total angular momentum in $\vShock$.  The shock surface is traced out by the white contour.  Velocity vectors where $|\vect{u}|\ge10^{4}$~km s$^{-1}$ are shown in the left panels.  \label{fig:machNumberAndVorticity}}
\end{figure*}

We focus on the magnetic field evolution during the nonlinear phase of the SASI, during which magnetic fields are amplified most effectively and the potential for back-reaction of the induced fields on the fluid flow is greatest.  
We do not apply any rigorous criterion for the onset of nonlinearity; we simply note when the accretion shock deviates noticeably from its spherically symmetric initial shape, and the post-shock velocity field has developed a significant non-radial component.  
(The upper panels of Figure \ref{fig:machNumberAndVorticity} are representative of the early nonlinear phase.)  

Our simulations vary in initial magnetic field strength and spatial resolution, and feature both non-rotating and rotating configurations (see Table \ref{tab:computedModels}).  
We focus first---and predominantly---on non-rotating models, often referring to model $\model{10}{0.0}{00}$ (non-rotating model with $B_{0}=1\times10^{10}$~G) as the ``weak-field model,'' and to model $\model{13}{0.0}{00}$ (non-rotating model with $B_{0}=1\times10^{13}$~G) as the ``strong-field model.''  
The rotating models are also briefly discussed, but we find that the turbulent magnetic field amplification is mostly unaffected by rotation.  

We initiate the SASI with random pressure perturbations in the post-shock flow in order to avoid biased excitation of particular modes (i.e. ``sloshing'' vs. ``spiral''), and find that all our simulations exhibit flows typical of the spiral mode. This is consistent with our results in \citetalias{endeve_etal_2010}, and with \citet{blondinMezzacappa_2007}, who found the spiral mode to dominate the late-time evolution independent of the initial perturbation. It is also consistent with the conclusions of \cite{fernandez_2010}, who demonstrated that the spiral modes of the SASI can be viewed as a superposition of sloshing modes out of phase, and that any superposition of sloshing modes with non-zero relative phase leads to spiral modes and angular momentum redistribution in the post-shock flow, which potentially spins up the underlying PNS \citep[see also][]{blondinShaw_2007}.  
The development of spiral SASI modes thus seems to be a general outcome (in 3D) of perturbing the (convectively stable) spherically symmetric initial condition.  
Moreover, a recent laboratory experiment---a shallow water analogue to a shock instability \citep[SWASI,][]{foglizzo_etal_2012}---found spiral modes to emerge favorably from the nonlinear phase.  

\subsection{Turbulence from Spiral SASI Modes}

Figure \ref{fig:machNumberAndVorticity} illustrates the flows that develop from the nonlinear spiral SASI mode.  
The renderings are created from the strong-field model, but the hydrodynamic developments exhibited by this model and highlighted in Figure \ref{fig:machNumberAndVorticity} are typical of all our non-rotating models.  
The flow Mach number is shown in the left panels, and in the right panels the magnitude of fluid vorticity is displayed.  

The upper panels ($t=720$~ms) depict the early development of the nonlinear spiral SASI mode.  
The shock surface is still quasi-spherical, but significant angular momentum redistribution has occurred in the post-shock flow, and the presence of strong counterrotating flows is apparent (cf. velocity vectors).  
The shock triple-point \citep[cf.][]{blondinMezzacappa_2007}, positioned to the lower left ($\sim$seven o'clock position), has just formed, and is visible as the kink in the shock surface (cf. white contour).  
The shock triple-point (a line segment extending across the shock surface) connects the pre-shock accretion flow and the two counterrotating flows in the post-shock gas.  
It moves on the shock surface in the counterclockwise direction in Figure \ref{fig:machNumberAndVorticity}.  
A layer of strongly sheared flows extends from the triple-point, downstream from the shock.  
This is clearly seen in both plots of fluid vorticity.  
This shear flow is one site of post-shock vorticity generation in our simulations.  
The post-shock flows are still subsonic for $t=720$~ms.  

In the lower panels ($t=820$~ms) the shock triple-point has completed about one and a half revolutions along the shock surface and is now positioned to the upper right ($\sim$two o'clock position).  
The shock volume has grown by more than a factor of three compared to the upper panels, and the shape of the accretion shock and the mass distribution in the shocked cavity are even more aspherical.  
The supersonic pre-shock accretion flow impinges on the shock at an oblique angle due to the aspherical shock and its off-center position.  
The significant tangential velocity component (relative to the shock surface), which is preserved across the shock, leads to supersonic post-shock flows ahead of (and directed towards) the triple-point.  
These supersonic flows, which strengthen the shear flow discussed above, are directed down towards the PNS and result in further vorticity generation as they decelerate up the density gradient or impinge on the PNS.  

\begin{figure}
  \epsscale{1.0}
  \plotone{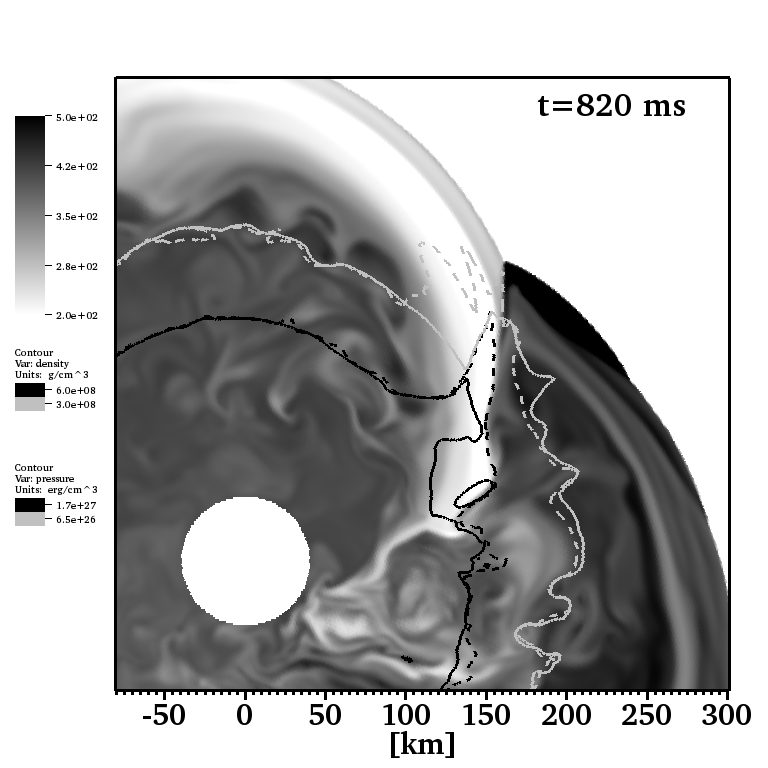}
  \caption{Slice through model \model{13}{0.0}{00} at $t=820$~ms showing the distribution of the polytropic constant $\kappa=P/\rho^{\gamma}$ around the shear layer associated with the shock triple-point.  Contours of constant density ($\rho=6\times10^{8}$~g~cm$^{-3}$ and $\rho=3\times10^{8}$~g~cm$^{-3}$; dashed black and gray, respectively) and pressure ($P=1.7\times10^{27}$~erg~cm$^{-3}$ and $P=6.5\times10^{26}$~erg~cm$^{-3}$; solid black and gray, respectively) are also plotted.   \label{fig:vorticityProduction}}
\end{figure}

The inviscid vorticity equation is obtained by taking the curl of Euler's equation \citep[e.g.,][]{landauLifshitz_1959}
\begin{equation}
  \pderiv{\boldsymbol\omega}{t}
  =\curl{(\vect{u}\times\boldsymbol\omega)}
  +\f{1}{\rho^{2}}\gradient{\rho}\times\gradient{P}, 
  \label{eq:vorticityEquation}
\end{equation}
where the first term on the right-hand side describes changes in vorticity due to fluid motions, and the second term is the baroclinic vector.  
(Magnetic fields are neglected in Eq. (\ref{eq:vorticityEquation}), but it remains valid for weak magnetic fields.)  
In particular, vorticity may be generated in regions where isosurfaces of density and pressure intersect.  
Figure \ref{fig:vorticityProduction} displays the polytropic constant $\kappa=P/\rho^{\gamma}$ (a proxy for entropy) in a slice through model $\model{13}{0.0}{00}$ at $t=820$~ms with focus on the shear flows emanating from the shock triple-point.  
Contours of constant density (dashed) and pressure (solid) are also plotted.  
The density and pressure contours are mostly parallel, but diverge strongly in the shear layer, indicating intersecting density and pressure isosurfaces and vorticity generation through the baroclinic term in Eq. (\ref{eq:vorticityEquation}).  
(Also, significant vorticity amplification occurs in the shear layer and elsewhere through the ``advection" term in Eq. (\ref{eq:vorticityEquation}).)  
Vorticity is generated, amplified, and distributed in a large fraction of the post-shock volume during operation of the SASI.  
(Movie 1 in the online material shows the generation and evolution of vorticity in the time interval from $t=720$~ms to $t=820$~ms.)  
The vorticity field exhibits strong intermittency in the late stages of SASI evolution.  
We also note the development of vorticity tube structures (vortex tubes) during the operation of the spiral mode.  
(See also Movie 2, which shows a full revolution of a vorticity still-frame at $t=820$~ms.)  
\citet{meeBrandenburg_2006} pointed out that the presence of vorticity may be helpful for turbulent magnetic field amplification.  

\begin{figure*}
  \epsscale{1.0}
  \plottwo{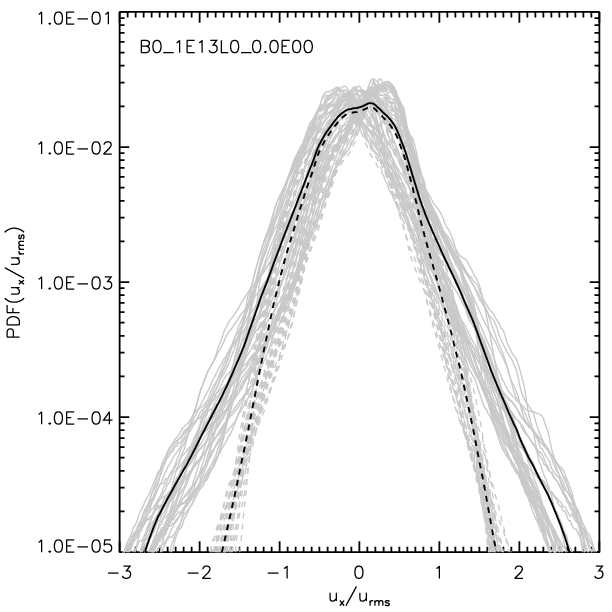}
                {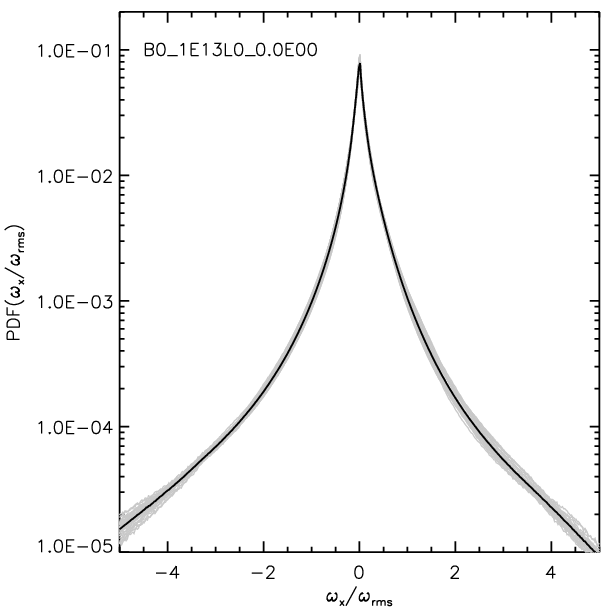}
  \caption{Normalized probability density functions (PDFs) of the $x$-component of velocity (left) and the $x$-component of vorticity (right).  The PDFs are constructed from the post-shock flows in model \model{13}{0.0}{00} during the nonlinear operation of the spiral SASI mode (from $t=810$~ms to $t=922$~ms, which corresponds to roughly one full revolution of the shock triple-point about the PNS).  The rms values of velocity and vorticity below the shock are $\uRMS=\sqrt{2E_{\mbox{\tiny kin}}/\mShock}$ and $\omega_{\mbox{\tiny rms}}=\sqrt{2\Omega/\vShock}$, respectively, and $E_{\mbox{\tiny kin}}$, $\mShock$, and $\Omega$ are the kinetic energy, mass, and enstrophy in $\vShock$, the volume bounded by the shock surface and the surface of the PNS.  We show PDFs for individual time states in gray and the average over all the time states in black.  In the left panel, the total PDFs are represented by the solid lines, while the dashed lines are the PDFs constructed from subsonic flows only ($|\vect{u}|/c_{S}<1$).  The (averaged) PDF associated with the subsonic flow fits well with the Gaussian $0.01975\times\exp{[-3.1\times(u_{x}/\uRMS)^{2}]}$.  (The PDFs constructed from the other velocity and vorticity components look similar.)  \label{fig:pdfVelocityAndVorticity}}
\end{figure*}

Strongly forced accretion-driven turbulence develops as a result of the SASI, and the post-shock flow becomes roughly divided into a supersonic (driving) component and a subsonic (volume-filling) turbulent component (cf. lower left panel in Figure \ref{fig:machNumberAndVorticity}).  
This is also reflected in the probability density function (PDF) of the velocity field below the shock.  
In the left panel in Figure \ref{fig:pdfVelocityAndVorticity} we plot normalized PDFs of the $x$-component of the velocity.  
We plot the \emph{total} PDF (solid black line), associated with the subsonic \emph{and} supersonic flow, and the PDF associated with the subsonic flow only (dashed black line).  
The supersonic flows contribute only to the tails of the distribution.  
The center of the distribution moves in response to the triple-point's movement along the shock surface (cf. gray curves), but when averaged over one revolution about the PNS, the PDF is practically centered on zero: we find $\left[\xi=u_{x}/\uRMS\right]_{\mbox{\tiny PDF}}=\int_{-\infty}^{\infty}\xi\,\mbox{PDF}(\xi)\,d\xi\approx0.019$.  
In the right panel of Figure \ref{fig:pdfVelocityAndVorticity} we plot the PDF of the $x$-component of the vorticity.  
The vorticity PDF is more peaked, has extended exponential tails, and is also centered about zero; $\left[\omega_{x}/\omega_{\mbox{\tiny rms}}\right]_{\mbox{\tiny PDF}}\approx0.002$.  
Similar vorticity distributions were found in simulations of convectively driven turbulence by \cite{brandenburg_etal_1996} and attributed to intermittency in hydrodynamic turbulence \citep[see also][]{kraichnan_1990,ishihara_etal_2009}.  
\citet{brandenburg_etal_1996} characterized the intermittency of a variable $f$ by the kurtosis of its PDF
\begin{equation}
  \mbox{Kurt}(f)=\left[f^{4}\right]_{\mbox{\tiny PDF}}/\left[f^{2}\right]_{\mbox{\tiny PDF}}^{2}, 
\end{equation}
where $\left[f^{n}\right]_{\mbox{\tiny PDF}}=\int_{-\infty}^{\infty}\,f^{n}\,\mbox{PDF}(f)\,df$ is the $n$-th moment of the PDF (assuming zero mean).  
For the $x$-component of the velocity below the shock we find $\mbox{Kurt}(u_{x}/\uRMS)\approx4.6$, and for the $x$-component of the vorticity we find $\mbox{Kurt}(\omega_{x}/\omega_{\mbox{\tiny rms}})\approx26.7$ \citep[i.e., similar to][]{brandenburg_etal_1996}.  

\subsection{Amplification of Weak Magnetic Fields from Turbulence: Elementary Concepts}

The SASI-driven hydrodynamic developments result in turbulent amplification of initially weak magnetic fields, which is the focus of this study \citepalias[see also][]{endeve_etal_2010}.  
Here we very briefly review some elementary concepts pertaining to such magnetic field amplification \citep[for details, see for example reviews by][]{ott_1998,brandenburgSubramanian_2005}.  

Stellar interiors are extremely good electrical conductors.  
In a perfectly conducting fluid the electric field is $\vect{E}=-(\vect{u}\times\vect{B})$, and Faraday's law (the induction equation), which governs the evolution of the magnetic field, becomes
\begin{equation}
  \pderiv{\vect{B}}{t}=\curl{(\vect{u}\times\vect{B})}, 
  \label{eq:inductionEquation}
\end{equation}
where the right-hand side (the induction term) describes changes to the magnetic field due to fluid motions.  
We note that, modulo the baroclinic vector, Eqs. (\ref{eq:vorticityEquation}) and (\ref{eq:inductionEquation}) have identical form, suggesting a possible analogy between $\boldsymbol\omega$ and $\vect{B}$ \citep{batchelor_1950}.  
Moreover, \citet{batchelor_1950} argued that the distribution of $\boldsymbol\omega$ and $\vect{B}$ will be similar in fully developed turbulence.  
An important difference, however, is that the vorticity equation is nonlinear, while the induction equation is linear for a specified velocity field.  
Nevertheless, similarities between vorticity and magnetic field are observed in our simulations.  

Equation (\ref{eq:inductionEquation}) can be combined with the mass conservation equation to form \citep[e.g.,][]{landauLifshitz_1960}
\begin{equation}
  \dderiv{}{t}\left(\f{\vect{B}}{\rho}\right)=\left(\f{\vect{B}}{\rho}\cdot\nabla\right)\vect{u}, 
  \label{eq:inductionEquationStretching}
\end{equation}
where $d/dt=\partial/\partial t+(\vect{u}\cdot\nabla)$, and $\vect{B}/\rho$ changes due to gradients in the velocity field.  
Equation (\ref{eq:inductionEquationStretching}) has an important physical interpretation.  
It has the exact same form as the evolution equation for an infinitesimal ``fluid line" connecting fluid elements and moving with the flow.  
Thus, two infinitely near fluid elements initially connected by a magnetic field line remain on that magnetic field line, and the value of $\vect{B}/\rho$ varies in proportion to the distance between the fluid elements \citep{landauLifshitz_1960}.  
Thus, the magnetic field is ``frozen" in a perfectly conducting fluid.  
In an approximately incompressible fluid, the magnetic field grows in direct proportion to the separation between fluid elements.  

The interpretation of Eq. (\ref{eq:inductionEquationStretching}) is equivalent to the following simple consideration of a magnetic flux tube, with strength $b$, length $l$, and cross-section $a$, which permeates (and is frozen in) a fluid element with density $\rho$: 
let the fluid element be ``stretched" by the flow to a new state (characterized by $b'$, $l'$, $a'$, and $\rho'$).  
Then, mass conservation ($\rho'l'a'=\rho la$) and magnetic flux conservation ($b'a'=ba$) gives
\begin{equation}
  \f{b'}{\rho'}=\f{b}{\rho}\times(l'/l).  
\end{equation}
In the incompressible limit, the field is amplified in direct proportion to the stretching of the tube.  
At the same time, the flux tube undergoes a decrease in the scale perpendicular to the stretching ($a'=a\times(l/l')<a$).  
The decrease in flux tube cross-section proceeds until (1) the field becomes strong enough to react back on the fluid, preventing further stretching, or (2) resistive (non-ideal) effects become important (Section \ref{sec:magneticEnergyGrowthRates}), or a combination of (1) and (2).  
Stretching is a very useful concept for turbulent $B$-field amplification.  

The frozen-in condition can result in rapid magnetic field amplification in a turbulent flow.  
Initially adjacent fluid elements separate rapidly, perhaps exponentially, in the chaotic flows that characterize turbulence \citep{ott_1998}.  
Thus, an initially weak magnetic field (i.e., $\vect{u}$ is independent of $\vect{B}$) may amplify exponentially by stretching, and the growth rate is roughly given by the inverse turnover time of turbulent eddies \citep[e.g.,][]{kulsrudAnderson_1992}.  
This is also apparent from a simplistic consideration of Eq. (\ref{eq:inductionEquationStretching}): the turbulent velocity varies significantly $\sim\mathcal{O}(\uRMSTurb$) over a turbulent eddy of size $\lambda_{\mbox{\tiny eddy}}$, hence $B^{-1}(dB/dt)\sim \uRMSTurb/\lambda_{\mbox{\tiny eddy}}$.  
Exponential amplification of weak magnetic fields is commonly seen in MHD turbulence simulations \citep[e.g.,][]{choVishniac_2000,brandenburg_2001,haugen_etal_2004}.  
Exponential growth ceases when the magnetic field becomes strong enough to cause a back-reaction on the fluid (i.e., $\vect{u}$ becomes dependent on $\vect{B}$).  

Amplification of weak magnetic fields through turbulent stretching is initially a kinematic mechanism (i.e., described by Eq. (\ref{eq:inductionEquation}) for a specified velocity field).  
As such, it differs from magnetic field amplification by the MRI in a fundamental way.  
The MRI is a dynamic mechanism, described by the full MHD system of equations, and requires the Lorentz force to be included in Euler's equation.  
(Also, the MRI requires differential rotation in the PNS to operate, while a turbulent dynamo can operate without PNS rotation.)  
However, for weak progenitor $B$-fields \citep{heger_etal_2005}, both mechanisms require high spatial resolution for simulation \citepalias[e.g.,][]{obergaulinger_etal_2009,endeve_etal_2010}, and may ultimately be computationally prohibitive to capture properly in large-scale multi-physics simulations.  

We comment here on the amount of kinetic helicity $\boldsymbol{\omega}\cdot\vect{u}$ in our simulations of SASI-driven turbulence.  
Kinetic helicity is a measure of ``handedness" (or lack of mirror symmetry) in the turbulent flows, and is an important quantity in dynamo theory \citep[e.g.,][and reference therein]{brandenburgSubramanian_2005}.  
Turbulent flows with kinetic helicity can support a so-called inverse cascade and produce large scale magnetic fields (i.e., larger than the turbulent forcing scale) in the nonlinear, saturated state \citep[e.g.,][]{meneguzzi_etal_1981,brandenburg_2001}.  
Non-helical turbulence results in mostly small-scale magnetic fields \citep[e.g.,][]{brandenburg_etal_1996,haugen_etal_2004}.  
We have constructed PDFs of the relative kinetic helicity $h_{\mbox{\tiny kin}}=(\boldsymbol\omega\cdot\vect{u})/(\omega_{\mbox{\tiny rms}}\uRMS)$ in the post-shock flow.  
The kinetic helicity distributions are similar to the vorticity distributions (strongly peaked with exponential tails), with $\left[h_{\mbox{\tiny kin}}\right]_{\mbox{\tiny PDF}}\approx-8.3\times10^{-4}$ and $\mbox{Kurt}(h_{\mbox{\tiny kin}})\approx28.4$.  
Despite the apparent handedness associated with the spiral SASI mode, the resulting turbulence is essentially non-helical.  
(This may change, however, if a rapidly---and differentially---rotating PNS is included in the model.)  
Thus, we expect that SASI-driven turbulence in our simulations results in magnetic field amplification due to a non-helical small-scale dynamo.  

\subsection{Time Evolution of Global Quantities}
\label{sec:timeGlobal}

\begin{figure*}
  \epsscale{1.0}
  \plottwo{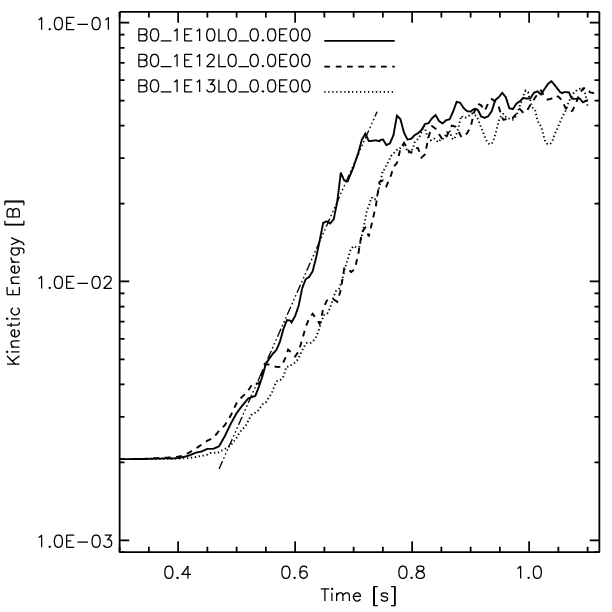}
                {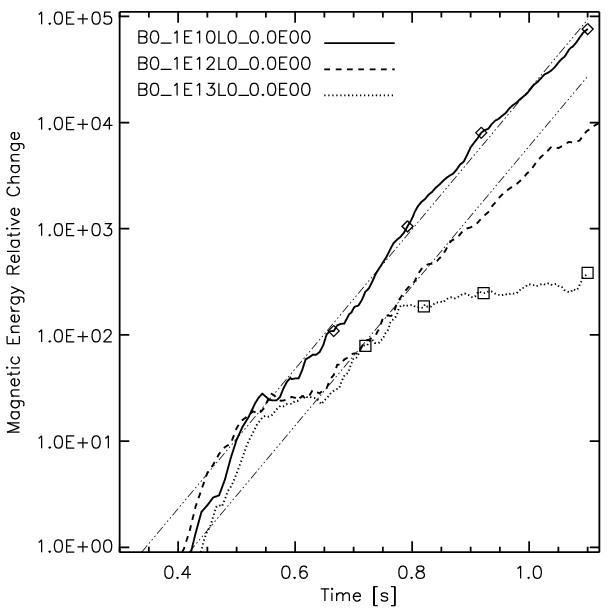}
  \plottwo{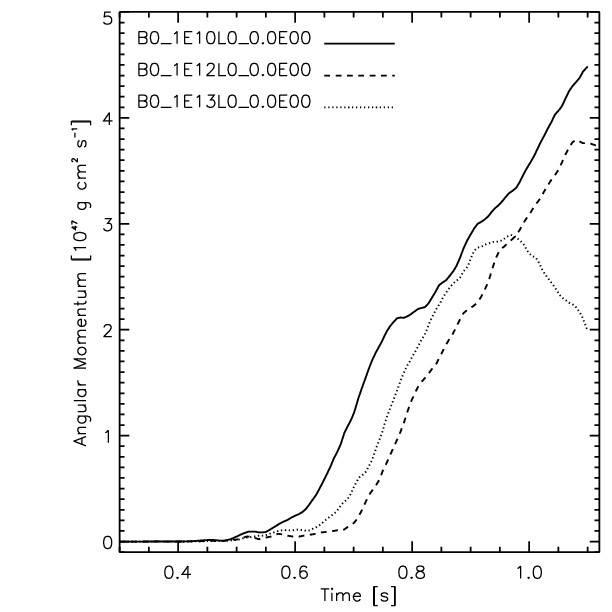}
                {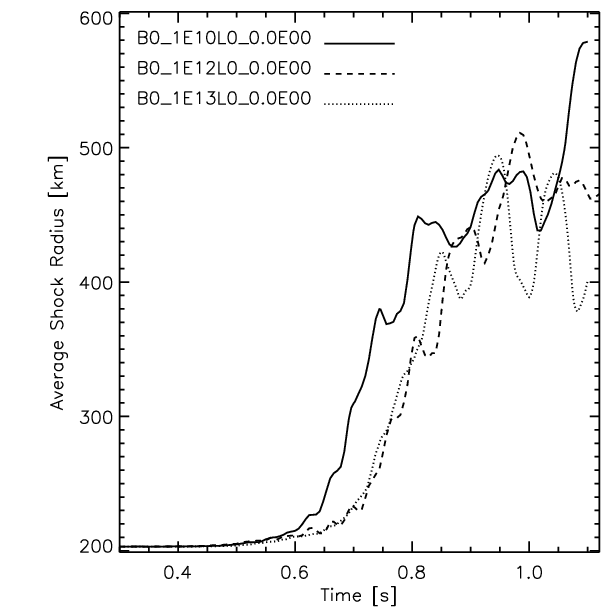}
  \caption{Time-evolution of global quantities integrated over the shock volume $\vShock$, bounded by the surface of the PNS, $\sPNS$, and the surface of the shock, $\sShock$, in non-rotating models in which the initial magnetic field has been varied from $1\times10^{10}$~G to $1\times10^{13}$~G.  Plotted are kinetic energy $E_{\mbox{\tiny kin}}$ (top left), magnetic energy change (relative to the initial) $\Delta E_{\mbox{\tiny mag}}/E_{\mbox{\tiny mag},0}$ (top right), angular momentum $|\vect{L}|$ (bottom left), and average shock radius $\bar{R}_{\mbox{\tiny Sh}}=(3\vShock/(4\pi))^{1/3}$.  Models $\model{10}{0.0}{00}$, $\model{12}{0.0}{00}$, and $\model{13}{0.0}{00}$ are represented by solid, dashed, and dotted lines, respectively.  The initial magnetic energy $E_{\mbox{\tiny mag},0}$ in these models is $2.3\times10^{-12}$~B, $2.3\times10^{-8}$~B, and $2.3\times10^{-6}$~B, respectively.  The dash-dotted lines in the top panels are proportional to $\exp{(t/\tau)}$, where the growth times are $\tau=85$~ms and $\tau=66$~ms in the top left and top right panels, respectively.  \label{fig:overviewNonRotating}}
\end{figure*}

An overview of the simulations with non-rotating initial conditions is given in Figure \ref{fig:overviewNonRotating}, in which we plot the time-evolution of selected globally integrated quantities for models $\model{10}{0.0}{00}$ (solid lines), $\model{12}{0.0}{00}$ (dashed lines), and $\model{13}{0.0}{00}$ (dotted lines): kinetic energy (upper left), relative magnetic energy change (upper right), angular momentum (lower left), and average shock radius (lower right).  
All quantities are obtained from integration over the shock volume $\vShock$, bounded by the surface of the PNS $\sPNS$ and the surface of the shock $\sShock$.  
The magnetic energy change is scaled with the initial magnetic energy for easy comparison across the models ($(E_{\mbox{\tiny mag}}-E_{\mbox{\tiny mag,0}})/E_{\mbox{\tiny mag,0}}$ is plotted).  

The kinetic energy of the settling flow beneath the shock is initially about $2\times10^{-3}$~B.  It begins to grow rapidly during the initial ramp-up phase of the SASI, which starts around 400~ms in all models.  
In particular, for the weak-field model, the post-shock kinetic energy grows exponentially with a nearly constant growth rate over the time period extending from $t\approx510$~ms to $t\approx720$~ms.  
The growth time during this epoch is about $\tau\approx85$~ms (cf. dash-dotted line in the top left panel in Figure \ref{fig:overviewNonRotating}).  
The kinetic energy in the models with a stronger initial magnetic field grows somewhat slower initially ($t\lesssim660$~ms), and then at a rate similar to that of the weak-field model.  
The growth slows down considerably for $t\gtrsim800$~ms, but the kinetic energy continues to grow throughout the nonlinear phase and reaches similar levels in all three models, with variability on a shorter timescale superimposed.  
When averaged over the time interval extending from 900~ms to 1100~ms we find\footnote{The temporal average of a variable $X$, over the interval $T=t_{2}-t_{1}$, is denoted $\langle X\rangle_{t_{1}}^{t_{2}}=\f{1}{T}\int_{t_{1}}^{t_{2}}X\,dt$.} $\timeAverage{E_{\mbox{\tiny kin}}}{0.9}{1.1}=$ 0.051~B, 0.048~B, and 0.044~B for models $\model{10}{0.0}{00}$, $\model{12}{0.0}{00}$, and $\model{13}{0.0}{00}$, respectively.  
While these time-averaged post-shock kinetic energies are slightly smaller in the models with a stronger initial magnetic field, we have not found convincing evidence that this slight decrease in global kinetic energy is a result of a stronger magnetic field.  
We find, however, that strong magnetic fields affect flows on small spatial scales (cf. Section \ref{sec:spectralAnalysis}).  

In terms of spherical harmonics, the SASI is characterized by exponentially growing power in low-order modes \citep{blondinMezzacappa_2006}.  
As a result of this, the shock surface deviates exponentially from its initially spherical shape.  
The obliquity of the shock front relative to the pre-shock accretion flow causes the non-radial post-shock kinetic energy to grow (also exponentially) at the expense of thermal energy \citep{blondin_etal_2003}.  
We have decomposed the post-shock kinetic energy into radial and non-radial components; $E_{\mbox{\tiny kin},\parallel}=\f{1}{2}\int_{\vShock}\rho u_{r}^{2}\,dV$ and $E_{\mbox{\tiny kin},\perp}=\f{1}{2}\int_{\vShock}\rho(u_{\vartheta}^{2}+u_{\varphi}^{2})\,dV$, respectively.  
The non-radial component grows much faster ($\tau\lesssim50$~ms) than the radial component, and the growth seen in Figure \ref{fig:overviewNonRotating} is due to a combination of the two.  
(The kinetic energy associated with the three velocity components become similar in the saturated state, and $E_{\mbox{\tiny kin},\perp}\approx2E_{\mbox{\tiny kin},\parallel}$.)  
Saturation of the post-shock kinetic energy may be due to the development of turbulence via secondary instabilities \citep[e.g.,][and Section \ref{sec:spectralAnalysis} in this paper]{guilet_etal_2010}.  
The exact details that determine the growth rate of the post-shock kinetic energy are tied directly to the physical origin of the SASI, which is not the focus of this paper.  
We focus primarily on magnetic field amplification in the flows that result from SASI activity.  

The magnetic energy grows at the expense of the turbulent kinetic energy below the shock (cf. Section \ref{sec:spectralAnalysis}).  
After an initial spurt, all the models shown in Figure \ref{fig:overviewNonRotating} experience an early period of exponential magnetic energy growth with essentially the same growth rate (cf. the temporal window from $t=650$~ms to $t=780$~ms).  
Such evolution is expected in a kinematic growth regime, in which the magnetic field's back-reaction on the fluid is negligible.  
The magnetic energy in the weak-field model ($\model{10}{0.0}{00}$, $B_{0}=1\times10^{10}$~G) grows exponentially at a nearly constant rate, with growth time $\tau\approx66$~ms, until the end of the simulation ($t=1100$~ms), and receives a total boost of about five orders of magnitude.  
The magnetic energy growth time is significantly shorter ($\sim25\%$) than the total kinetic energy growth time during the overlapping period of exponential growth.  
In the model with $B_{0}=1\times10^{12}$~G ($\model{12}{0.0}{00}$) we also find that $E_{\mbox{\tiny mag}}$ grows steadily until the end of the run ($t=1126$~ms).  
The magnetic energy in this model initially grows at the same rate as the weak-field model, but its growth rate clearly tapers off at later times ($t\gtrsim900$~ms).  
The strong-field model ($\model{13}{0.0}{00}$, $B_{0}=1\times10^{13}$~G) also exhibits exponential magnetic energy growth ($\tau\approx66$~ms) early on.  
Then, around $t\approx780$~ms, its growth rate drops almost discontinuously, and $E_{\mbox{\tiny mag}}$ grows by only about $50\%$ for the remainder of the simulation (until $t=1100$~ms).  
Model $\model{13}{0.0}{00}$ receives a total boost in magnetic energy of about a factor of 300.  
The abrupt change in the magnetic energy growth rate observed in the strong-field model occurs when magnetic fields become dynamically important in localized regions below the shock (cf. Section \ref{sec:radialProfiles}).  

At the end of the simulations the magnetic energy in models $\model{10}{0.0}{00}$, $\model{12}{0.0}{00}$, and $\model{13}{0.0}{00}$ has reached about $1.8\times10^{-7}$~B, $2.3\times10^{-4}$~B, and $8.9\times10^{-4}$~B, respectively.  
The magnetic energy in the weak-field model is many ($\sim$ five) orders of magnitude below the post-shock kinetic energy at this point.  
In the strong-field model it saturates below $10^{-3}$~B, which is also significantly less than the total kinetic energy in the post-shock flow.  
Also, the boost in magnetic energy in the strong-field model is almost an order of magnitude lower than the difference in average post-shock kinetic energy between the strong-field and the weak-field models, which is about $7\times10^{-3}$~B.  
At the end of the simulations the magnetic energy in the strong-field model is less than a factor of four larger than in model $\model{12}{0.0}{00}$, while it initially was a factor $10^{2}$ larger.  
We point out that the magnetic energies listed above are values recorded when the simulations were stopped after roughly one explosion time ($t\sim1$~s).  
(SASI-induced magnetic field amplification ceases once the explosion is initiated.)
However, the listed values should \emph{not} be interpreted as upper limits on the magnetic energy for the different initial magnetic fields.  
The magnetic energy growth rate is (for reasons we detail in this paper) underestimated by the numerical simulations, and we suspect that all models---independent of the initial magnetic field---will reach a saturated state, similar to the strong-field model, well within an explosion time.  
The issue of magnetic energy growth, saturation, and its effect on the post-shock flow will be discussed in later sections.  

The induced magnetic fields do not notably affect the global characteristics of the shock evolution.  
The plots of total angular momentum $|\vect{L}|$ in $\vShock$ and the average shock radius $\bar{R}_{\mbox{\tiny Sh}}$ show that these quantities reach similar values in all the non-rotating models.  
The angular momentum reaches a few $\times10^{47}$~g cm$^{2}$ s$^{-1}$, consistent with the comparable models of \citet{blondinMezzacappa_2007} and \citet{fernandez_2010}.  
Moreover, during the nonlinear evolution, after the period of exponential growth of the angular momentum in $\vShock$, we find $|\vect{L}|\lesssim f\dot{M}\bar{R}_{\mbox{\tiny Sh}}^{2}$, with $f\approx 0.25$ \citep[cf.][]{fernandez_2010}.  
The average shock radius exhibits significant variability, and briefly exceeds 500~km in some of the models.  
In particular, we find $\timeAverage{\bar{R}_{\mbox{\tiny Sh}}}{0.9}{1.1}=$ 484~km, 466~km, and 438~km, for models $\model{10}{0.0}{00}$, $\model{12}{0.0}{00}$, and $\model{13}{0.0}{00}$, respectively.  

The larger-amplitude oscillations in kinetic energy and the smaller average shock radius exhibited by the strong-field model ($\model{13}{0.0}{00}$) may be attributed to this model's somewhat different nonlinear evolution, which is due to the stochastic nature of the nonlinear SASI rather than to the stronger initial field.
Model $\model{13}{0.0}{00}$ first evolves into a typical spiral mode pattern (cf. Figure \ref{fig:machNumberAndVorticity}), but later develops a flow pattern reminiscent of the sloshing mode, with two oppositely directed high-speed streams, emanating from opposite sides of the shock, terminating on opposite sides of the PNS, or colliding head-on deep in the shocked cavity.  
The appearance of this flow pattern coincides with the turnover in the angular momentum seen in the lower left panel ($t\approx950$~ms).  
Note that there is little or no response in the magnetic energy evolution due to these rearrangements in the large scale flow.  
This is consistent with the magnetic field being amplified by small-scale rather than large-scale flows.  

Comparing Figure \ref{fig:overviewNonRotating} with Figure 12 of  \citetalias{endeve_etal_2010}, we note that spatial resolution ($\Delta l=1.17$~km in this paper versus $\Delta l=1.56$~km in \citetalias{endeve_etal_2010}) affects global magnetic quantities much more than global fluid quantities. 
In particular, models 3DB10Rm and 3DB12Rm in \citetalias{endeve_etal_2010} correspond to models $\model{10}{0.0}{00}$ and $\model{12}{0.0}{00}$ respectively.  
The increased spatial resolution has no significant impact on the post-shock kinetic energy, total angular momentum, average shock radius, or exponential growth time for the kinetic energy ($\tau\approx85$~ms).  
The magnetic energy in model 3DB12Rm is boosted by a factor of $2\times10^{3}$, while model 3DB10Rm received a boost of less than a factor of $10^{3}$.  
We also measured an exponential growth time for the magnetic energy of $\tau\approx71$~ms over several hundred milliseconds in \citetalias{endeve_etal_2010}.  
These results are somewhat different than the results presented in Figure \ref{fig:overviewNonRotating}, which show that the increased resolution results in a larger boost in the magnetic energy and shorter exponential growth time ($\tau \approx 66$~ms), and that the magnetic energy growth at late times depends on the initial magnetic field strength.  
The sensitivity to spatial resolution was also pointed out in our previous study, and it will be further discussed later in this paper.  

\subsection{Evolution of Spherically Averaged Radial Profiles and Saturation of Magnetic Energy}
\label{sec:radialProfiles}

\begin{figure*}
  \epsscale{1.0}
  \plottwo{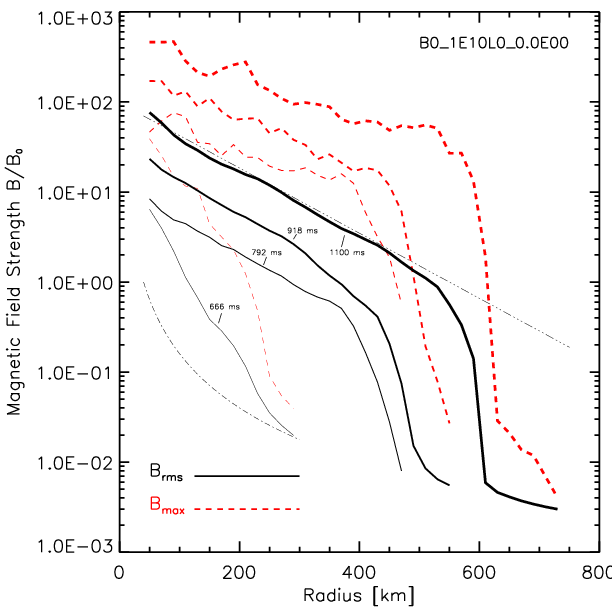}
                {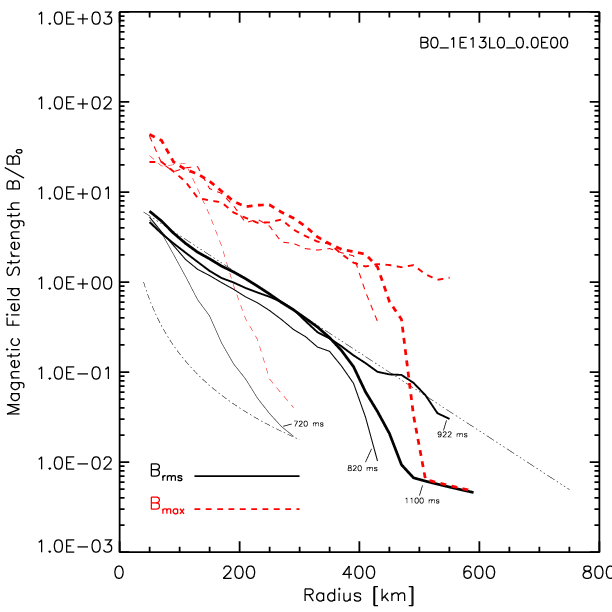}
  \plottwo{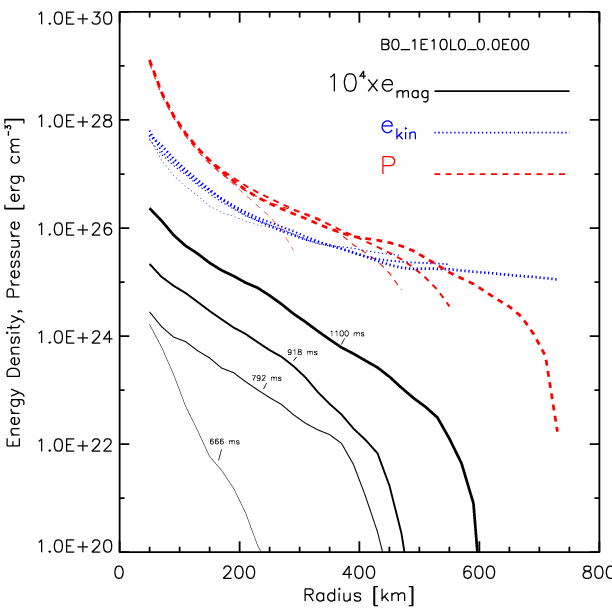}
                {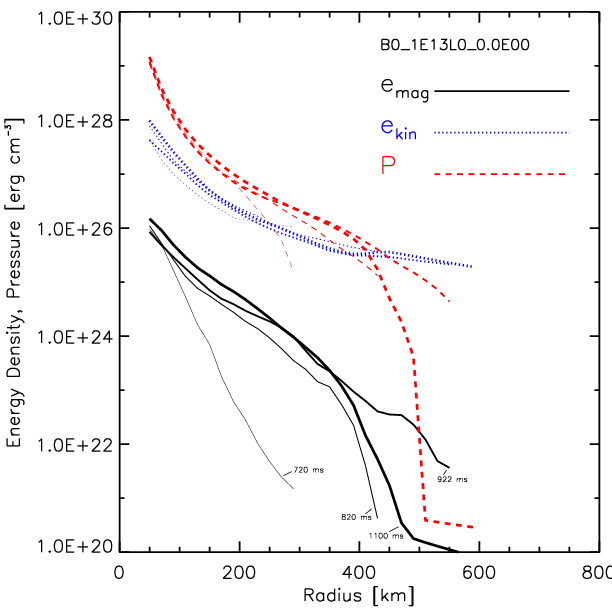}
  \caption{Spherically averaged radial profiles at selected times for the non-rotating weak-field (left panels) and the strong-field (right panel) models.  Upper panels: rms magnetic field strength (solid black) and maximum magnetic field strength (dashed red).  For reference, the initial magnetic field profile is plotted in each panel (dash dot).  We also plot reference lines proportional to $\exp{(-r/L_{B})}$ (dash-dotted) with $L_{B}=120$~km (weak-field case) and $L_{B}=100$~km (strong-field case).  The magnetic field strengths have been normalized to the initial value $B_{0}$ at $r=\rPNS$.  Lower panels: magnetic energy density (solid black), kinetic energy density (dotted blue), and thermal pressure (dashed red).  For model $\model{10}{0.0}{00}$ profiles are plotted for 666~ms, 792~ms, 918~ms, and 1100~ms (marked by diamonds in the upper right panel of Figure \ref{fig:overviewNonRotating}), while for model $\model{13}{0.0}{00}$ we plot radial profiles for times 720~ms, 820~ms, 922~ms, and 1100~ms (marked by squares in the upper right panel of Figure \ref{fig:overviewNonRotating}).  Thicker lines indicate a more advanced time state.  Note that the magnetic energy densities in the weak-field model (lower left panel) has been multiplied by a factor of $10^{4}$.  \label{fig:sphericalProfilesNonRotating}}
\end{figure*}

The magnetic energy in the strong-field model reaches saturation relatively early in the nonlinear evolution ($t\approx780$~ms).  
To help elucidate the physical conditions under which the magnetic energy growth in our simulations is quenched, we plot, in Figure \ref{fig:sphericalProfilesNonRotating}, spherically averaged radial profiles from the evolution of the weak-field model ($\model{10}{0.0}{00}$, left panels) and the strong-field model ($\model{13}{0.0}{00}$, right panels).  
In the upper panels we plot the rms magnetic field strength $B_{\mbox{\tiny rms}}$ (solid lines) and the maximum magnetic field strength $B_{\mbox{\tiny max}}$ (dashed red lines) versus radial distance from the center of the PNS.  
Values are computed in spherical shells centered on the origin of the computational domain $r=\sqrt{x^{2}+y^{2}+z^{2}}=0$, with thickness $\delta L=20$~km and volume $\delta V_{i}$, which includes all zones with radial coordinate $r\in[r_{i}-\delta L/2,r_{i}+\delta L/2)$, with $r_{i}=50$~km, $70$~km,$\ldots$,$(L-\delta L)/2$.  
The rms magnetic field is computed from the shell-volume-averaged\footnote{$\volumeAverage{X}{V}=\f{1}{V}\int_{V}X\,dV$ denotes the volume average of $X$ over the volume $V$.} magnetic energy density $B_{\mbox{\tiny rms}}=\sqrt{2\mu_{0}\volumeAverage{\edens{mag}}{\delta V_{i}}}$, while $B_{\mbox{\tiny max}}$ is simply the maximum magnetic field over all zones in each shell.  
In the lower panels we plot the shell-volume-averaged magnetic energy density $\volumeAverage{\edens{mag}}{\delta V_{i}}$ (solid lines), kinetic energy density $\volumeAverage{\edens{kin}}{\delta V_{i}}$ (dotted blue lines), and fluid pressure $\volumeAverage{P}{\delta V_{i}}$ (dashed red lines) versus radial distance.  
For each model we plot four profiles (time states) of each variable.  
The time states, which are also indicated with diamonds ($\model{10}{0.0}{00}$) and squares ($\model{13}{0.0}{00}$) in the upper right panel of Figure \ref{fig:overviewNonRotating}, are chosen to emphasize the temporal magnetic field evolution in each of the models and to contrast the two models.  

The spherically averaged radial profiles further illustrate the differences in magnetic field evolution of the weak-field and strong-field models.  
The magnetic field in model $\model{10}{0.0}{00}$ intensifies steadily throughout the nonlinear evolution, at all radii below the shock, and $B_{\mbox{\tiny rms}}$ evolves self-similarly during the later stages.  
Toward the end of the weak-field run, the rms magnetic field has received a boost of about two orders of magnitude near $\rPNS$, and stays above $B_{0}$ for $r\lesssim500$~km.  
The maximum magnetic field is typically an order of magnitude above the rms magnetic field, which is an indication of strong spatial intermittency in the magnetic field (see also Figures \ref{fig:magneticFieldStrength} and \ref{fig:pdfBFieldAndVorticityDotBField} below).  
In the lower left panel of Figure \ref{fig:sphericalProfilesNonRotating} we see that the magnetic energy density is many orders of magnitude smaller than the kinetic energy density and pressure for the shown times, consistent with kinematic magnetic field growth.  
Not even in localized regions does the magnetic energy density become comparable to the kinetic energy density or the pressure.  
At the end of the simulation, there are only a few zones in which the ratio of magnetic-to-kinetic energy exceeds $10^{-2}$.  

The strong-field model's magnetic energy evolution is not governed by kinematic growth (except for during the initial boost received at early times), but rather by dynamic interactions with the fluid in a saturated state (and numerical dissipation; see Sections \ref{sec:spectralAnalysis} and \ref{sec:magneticEnergyGrowthRates}).  
The magnetic energy in this model falls off the exponential growth curve around $t=780$~ms.  
The thin solid curve in the upper right panel in Figure \ref{fig:sphericalProfilesNonRotating} ($t=720$~ms) represents the transition from the initial magnetic field profile (dash-dot curve) to the saturated state.  
Although the post-shock flow is governed by vigorous turbulence at later times, there are only minor changes to the rms and maximum magnetic field profiles.  
The relative boost of $B_{\mbox{\tiny rms}}$ and $B_{\mbox{\tiny max}}$ in model $\model{13}{0.0}{00}$ is about an order of magnitude less than what is observed in model $\model{10}{0.0}{00}$.  
At the end of the strong-field model, the rms magnetic field exceeds $10^{13}$~G for $r\lesssim225$~km, while the maximum magnetic field exceeds $10^{14}$~G out to $r\approx200$~km.  

At the end of both runs the rms magnetic field follows an exponential decrease with radius, $B_{\mbox{\tiny rms}}$ is proportional to $\exp{(-r/L_{B})}$, where the characteristic length scale $L_{B}$ is about 120~km and 100~km for model $\model{10}{0.0}{00}$ and model $\model{13}{0.0}{00}$, respectively (cf. dash-dotted lines).  
From the upper panels in Figure \ref{fig:sphericalProfilesNonRotating} it is apparent that the exponential decrease in $B_{\mbox{\tiny rms}}$ with radius holds reasonably well in both models throughout the nonlinear evolution.  
Moreover, the averaged kinetic energy density below the shock has increased significantly compared to the initial condition (Figure \ref{fig:initialCondition}) and roughly follows a power-law in radius $r^{\alpha}$, with the exponent $\alpha$ varying between $-2.7$ and $-2.3$.  
The decrease in kinetic energy density with radius is mostly due to the decrease in mass density: the mass density falls off as $r^{-3}$ inside $r=150$~km, and almost as $r^{-2}$ outside $r=150$~km.  
During the runs, the spherically averaged pressure remains relatively quiescent inside $r=150$~km, where it falls off as $r^{-4}$.  

As noted above, the decrease in kinetic energy density (the source of magnetic energy) follows a power law with radius, while the magnetic energy density (and $B_{\mbox{\tiny rms}}$) decreases exponentially with radius.  
On the other hand, we find that the enstrophy $\f{1}{2}\volumeAverage{\omega^{2}}{\delta V_{i}}$ also decreases exponentially with radius, with a length scale comparable to (but somewhat shorter than) that of $\volumeAverage{\edens{mag}}{V_{i}}$.  
Moreover, by comparing the lower panels of Figure 6 in \citetalias{endeve_etal_2010} (showing $|\vect{B}|$) with Figure 11 in \citetalias{endeve_etal_2010} (showing $|\boldsymbol\omega|$) we see that the spatial distribution of magnetic field and vorticity is very similar.  
These observations support a similarity between vorticity and magnetic field \citep{batchelor_1950}.  
(We have not investigated the physical reasons for the particular spatial distribution of vorticity and magnetic field in further detail, but we plan to do so in a future study.)  

The relative boost in $B_{\mbox{\tiny rms}}$ decreases monotonically with the initial field strength in our models.  
The results from model $\model{12}{0.0}{00}$ (not shown) confirm this trend.  
This is because the models with stronger initial fields reach saturation during the simulation.  
(The growth rate is the same in all models during the kinematic regime.)  
Saturation occurs when the magnetic energy becomes comparable (locally) to the kinetic energy.  
In particular, we find that the kinematic growth regime ends when $B_{\mbox{\tiny max}}^{2}/(2\mu_{0})\lesssim\volumeAverage{\edens{kin}}{\delta V_{i}}$.  

For model $\model{12}{0.0}{00}$ we find $B_{\mbox{\tiny max}}^{2}/(2\mu_{0})\approx 0.1\times\volumeAverage{\edens{kin}}{\delta V_{i}}$ and $B_{\mbox{\tiny max}}^{2}/(2\mu_{0})\approx 0.3\times\volumeAverage{\edens{kin}}{\delta V_{i}}$ for $t=966$~ms and $t=1126$~ms, respectively, which represent time states after the magnetic energy growth has fallen off the exponential curve with growth time $\tau=66$~ms (Figure \ref{fig:overviewNonRotating}).  
In model $\model{13}{0.0}{00}$, the ratio $B_{\mbox{\tiny max}}^{2}/(2\mu_{0}\volumeAverage{\edens{kin}}{\delta V_{i}})$ stays above $0.3$ for the three most advanced time states shown in Figure \ref{fig:sphericalProfilesNonRotating}, and hovers around unity at the end of the simulation.  
In both models the ratio stays remarkably constant with distance from the PNS, varying by less than a factor of two, out beyond $r=300$~km (although exact details vary by model).  
Thus, turbulence-induced magnetic fields may impact dynamics in localized regions throughout the shock volume.  

\subsection{Full Spatial Distributions, Intermittency, and Saturation of Magnetic Energy}
\label{sec:spatialDistributions}

\begin{figure*}
  \epsscale{1.0}
  \plottwo{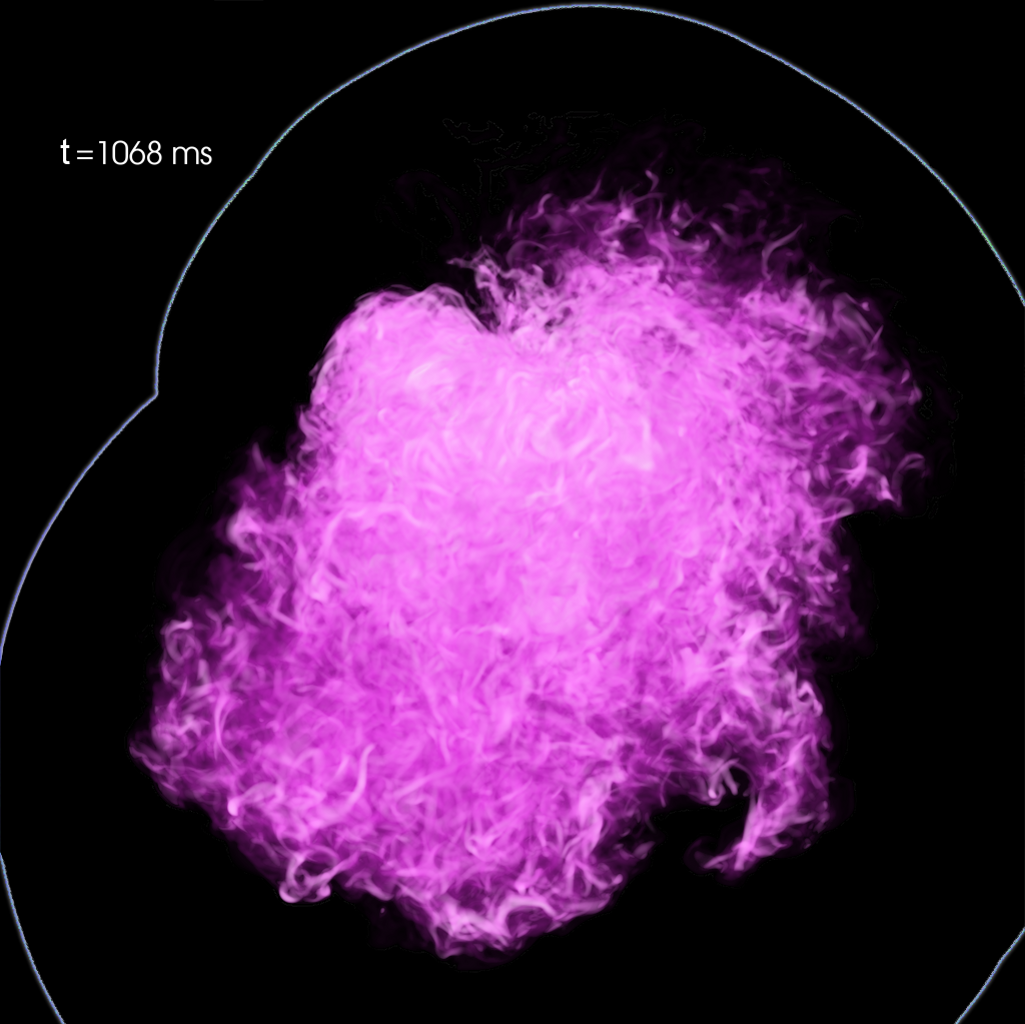}
                {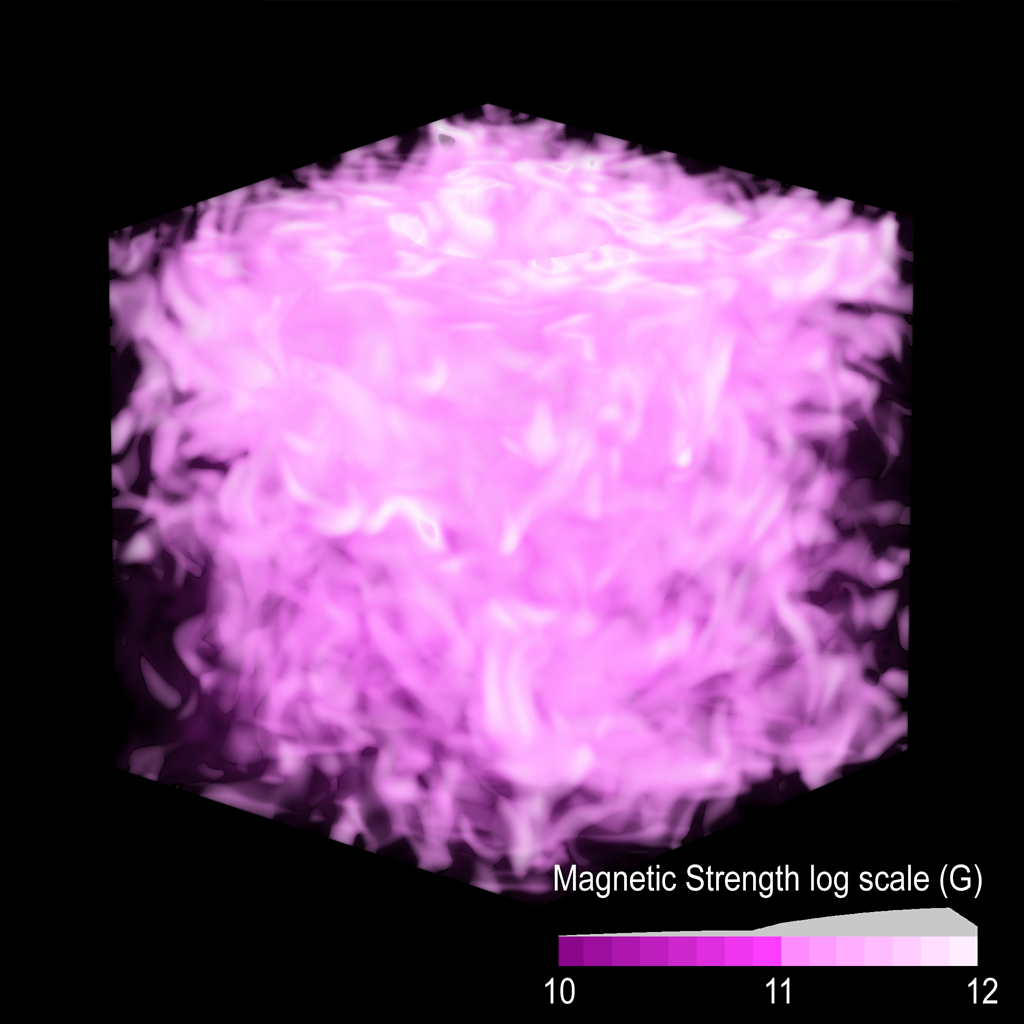}
  \caption{Magnetic field magnitude $|\vect{B}|$ near the end ($t=1068$~ms) of the weak-field simulation ($\model{10}{0.0}{00}$).  The left panel shows a global view of the magnetic fields below the shock (traced out with a white contour).  The right panel shows the magnetic field in a $200$~km$^{3}$ volume near the PNS.  \label{fig:magneticFieldStrength}}
\end{figure*}

Figure \ref{fig:magneticFieldStrength} shows volume renderings of the magnetic field magnitude in model $\model{10}{0.0}{00}$ near the end of the simulation ($t=1068$~ms).  
The left panel shows a global view of the amplified magnetic fields below the shock (white contour), and the right panel shows a zoomed in view in a $200$~km$^{3}$ volume near the PNS.  
These renderings illustrate the complicated, highly intermittent magnetic fields that develop from SASI-induced turbulent flows.  
The magnetic energy is concentrated in thin, folded, intense flux ropes.  
Notice in the left panel that amplified magnetic fields do not extend all the way to the shock, but are confined to a smaller volume, which is characterized by highly turbulent flows (see also the distribution of vorticity for model $\model{13}{0.0}{00}$ in the volume renderings in Figure \ref{fig:machNumberAndVorticity}, and the associated movies).  
The spatial distribution of the magnetic field (in particular the intermittency) in the strong-field model is similar to that of the weak-field model.  

\begin{figure*}
  \epsscale{1.0}
  \plottwo{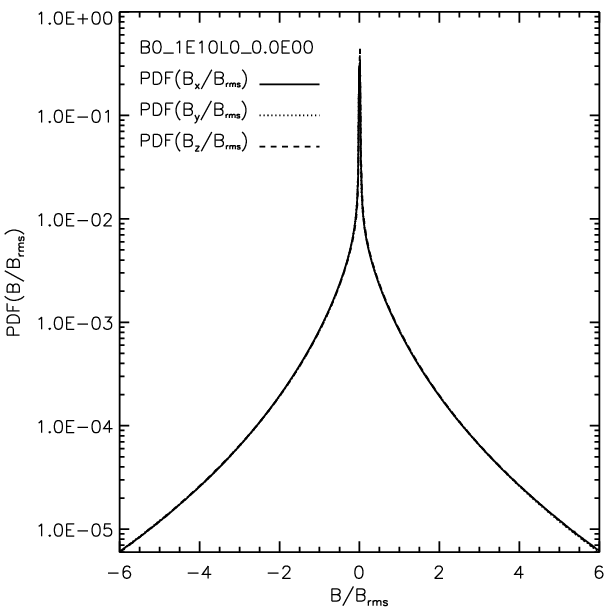}
                {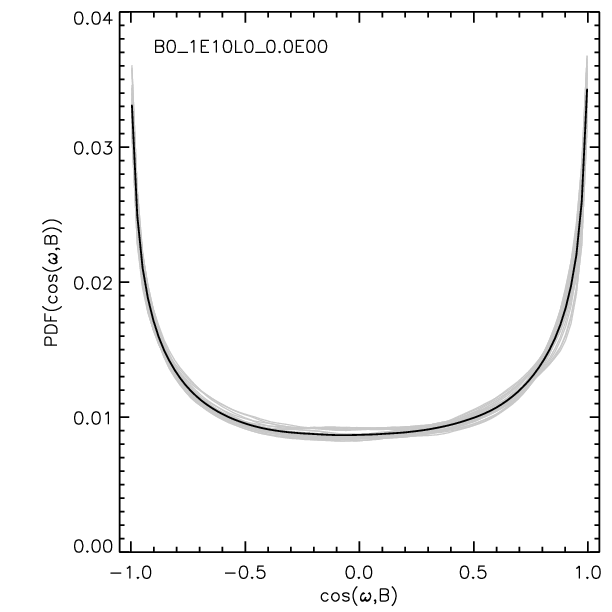}
  \caption{Normalized PDFs of the magnetic field components (left panel) and the cosine of the angle between vorticity and magnetic field (right panel) in model \model{10}{0.0}{00}.  The PDFs are constructed by averaging over the time period from $804$~ms to $918$~ms.  The PDFs of the $x$-, $y$-, and $z$-components of the magnetic field (solid, dotted, and dashed lines, respectively) are practically indistinguishable.  \label{fig:pdfBFieldAndVorticityDotBField}}
\end{figure*}

Normalized PDFs of individual components of the magnetic field from model \model{10}{0.0}{00} are plotted in the left panel of Figure \ref{fig:pdfBFieldAndVorticityDotBField}.  
The shape of the distributions is strongly peaked with extended exponential tails (similar to the vorticity distributions in Figure \ref{fig:pdfVelocityAndVorticity}), and the PDFs of the different magnetic field components are practically indistinguishable.  
The intermittency is high, $\mbox{Kurt}(B_{x}/B_{\mbox{\tiny rms}})\approx\mbox{Kurt}(B_{y}/B_{\mbox{\tiny rms}})\approx\mbox{Kurt}(B_{z}/B_{\mbox{\tiny rms}})\approx32.5$ (somewhat larger than, but comparable to, the intermittency of the vorticity field), which is consistent with the visual impression given in Figure \ref{fig:magneticFieldStrength}.  
We note that the PDFs are highly symmetric, which implies small overall polarity of the field; the magnitude of the mean values $\left[B_{x}/B_{\mbox{\tiny rms}}\right]_{\mbox{\tiny PDF}}$, $\left[B_{y}/B_{\mbox{\tiny rms}}\right]_{\mbox{\tiny PDF}}$, and $\left[B_{z}/B_{\mbox{\tiny rms}}\right]_{\mbox{\tiny PDF}}$ are all less than $0.01$.  
The PDF of the cosine of the angle between $\boldsymbol\omega$ and $\vect{B}$ is plotted in the right panel of Figure \ref{fig:pdfBFieldAndVorticityDotBField}, which shows that the vorticity and magnetic field tend to be aligned or anti-aligned, and gives further support to the similarity between the magnetic and vorticity fields.  
Similar distributions were also reported by \citet{brandenburg_etal_1996}.  

The ratio $B_{\mbox{\tiny max}}^{2}/(2\mu_{0}\volumeAverage{\edens{kin}}{\vShell})$ (used in Section \ref{sec:radialProfiles} to characterize saturation of the magnetic energy) is still only an approximate measure of the relative strength of the magnetic field since the kinetic energy density is averaged over the shell.  
The highly intermittent magnetic fields created by turbulence have been ``expelled" from the fluid \citep{thompsonDuncan_1993}.  
As evolution proceeds, an increasing percentage of the total magnetic energy resides in regions where the ratio of magnetic-to-kinetic energy $\beta_{\mbox{\tiny kin}}^{-1}(=v_{\mbox{\tiny A}}^{2}/|\vect{u}|^{2})$ exceeds $10^{-2}$, $10^{-1}$, and $1$: $55\%$, $10\%$, and $0.5\%$, respectively, for model $\model{12}{0.0}{00}$ at $t=966$~ms, while for $t=1126$ the respective percentages have increased to $72\%$, $20\%$, and $1.5\%$.  
(The percentages quoted for $t=1126$~ms are very similar to those quoted in \citetalias{endeve_etal_2010} for model 3DB12Ah, which was computed with the same resolution and initial condition as model $\model{12}{0.0}{00}$, but with a different initial perturbation.)  
The fraction of the total magnetic energy concentrated in regions with high $\beta_{\mbox{\tiny kin}}^{-1}$ stays relatively constant during the saturated state of the strong-field model ($t\ge780$~ms):  we find $\sim90\%$ ($\beta_{\mbox{\tiny kin}}^{-1}\ge10^{-2}$), $\sim50\%$ ($\beta_{\mbox{\tiny kin}}^{-1}\ge10^{-1}$), and $\sim7\%$ ($\beta_{\mbox{\tiny kin}}^{-1}\ge1$).  

\begin{figure*}
  \epsscale{1.0}
  \plotone{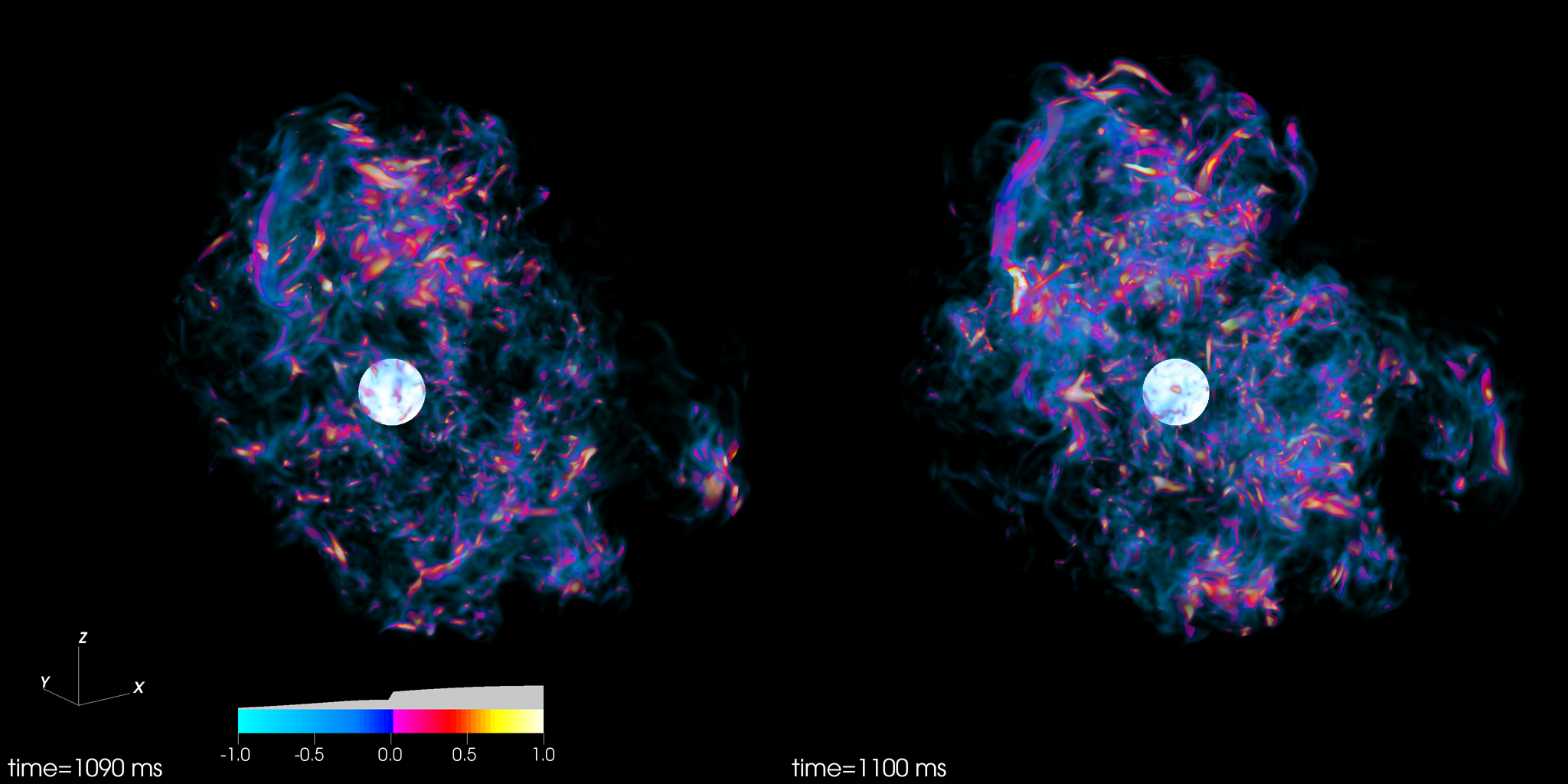}
  \caption{Select snapshots of the logarithm of the magnetic-to-kinetic energy ratio $\beta_{\mbox{\tiny kin}}^{-1}=v_{\mbox{\tiny A}}^{2}/|\vect{u}|^{2}$ late in the highly nonlinear magnetically saturated phase of the strong-field model ($\model{13}{0.0}{00}$).  The Alfv{\'e}n speed is $v_{\mbox{\tiny A}}=|\vect{B}|/\sqrt{\mu_{0}\rho}$.  The snapshots are separated by 10~ms, taken at $t=1090$~ms (left) and $t=1100$~ms (right).  \label{fig:inverseKineticBeta}}
\end{figure*}

Volume renderings in Figure \ref{fig:inverseKineticBeta} show the spatial distribution of $\beta_{\mbox{\tiny kin}}^{-1}$ late in the evolution of the strong-field model, illustrating the spatial \emph{and} temporal intermittency of turbulence-induced strong magnetic fields.  
The snapshots are temporally separated by $10$~ms, which is longer than the turbulent turnover time (cf. Section \ref{sec:spectralAnalysis}), but is significantly shorter than both the advection time and the Alfv{\'e}n crossing time (defined loosely as $\rShockAvg/|\vect{u}|$ and $\rShockAvg/v_{A}$ respectively).  
Concentrations of high $\beta_{\mbox{\tiny kin}}^{-1}$, which can briefly exceed unity in localized regions, are scattered throughout the shock volume.  
As noted above, about $50\%$ of the total magnetic energy resides in regions where $\beta_{\mbox{\tiny kin}}^{-1}\ge10^{-1}$, and these magnetic fields occupy less than $10\%$ of the total shock volume.  
(Movie 3 and Movie 4 in the online material show the time-evolution of $\beta_{\mbox{\tiny kin}}^{-1}$ from $t=1050$~ms to $t=1100$~ms and a full revolution of $\beta_{\mbox{\tiny kin}}^{-1}$ for $t=1100$~ms, respectively.)  

Alfv{\'e}n waves are likely excited by the SASI activity discussed in this paper.  
\citet{suzuki_etal_2008} performed simulations in spherical symmetry and investigated the role of Alfv{\'e}n waves on the core-collapse supernova explosion mechanism.  
They found that---for sufficiently strong magnetic fields ($\gtrsim2\times10^{15}$~G)---heating associated with Alfv{\'e}n wave energy dissipation may revive the stalled shock.  
For weaker magnetic fields no shock revival was observed.  
We have not attempted to identify Alfv{\'e}n waves, or energy dissipation due to Alfv{\'e}n waves, in our simulations (the highly dynamic nature of the SASI-driven flows makes this a nontrivial task).  
However, the magnetic fields attained in our strong-field model are significantly weaker than $10^{15}$~G and we do not expect Alfv{\'e}n wave heating due to the mechanism studied by \citet{suzuki_etal_2008} to result in significant energy deposition near the shock (or elsewhere) in our simulations.  
\citet{guilet_etal_2011} recently suggested a mechanism for magnetic field amplification in the vicinity of an Alfv{\'e}n surface (i.e., where $v_{\mbox{\tiny A}}=|\vect{u}|$).  
In their model, Alfv{\'e}n waves, excited for example by the SASI, may amplify near the Alfv{\'e}n surface and create a dynamic back-reaction.  
However, the turbulent nature of the hydromagnetic evolution in our simulations may result in unfavorable conditions for this mechanism to operate.  
In particular, regions where $v_{\mbox{\tiny A}}=|\vect{u}|$ appear and disappear in a highly intermittent manner (cf. Figure \ref{fig:inverseKineticBeta}, and associated movies).  

\subsection{Spectral Analysis}
\label{sec:spectralAnalysis}

Further important insight into the numerical simulations can be gained from a Fourier decomposition of the magnetic and kinetic energy.  
In particular, our analysis presented in \citetalias{endeve_etal_2010} lacked the ability to quantify the amount of turbulent kinetic energy available to amplify the magnetic field as well as the magnetic field's impact on the evolution of the small-scale flows.  
We seek to address these questions in this section.  

Following \citet{ryu_etal_2000} we compute Fourier amplitudes from components of the velocity and magnetic fields in the computational domain\footnote{The Fourier transforms are computed using the FFTW library documented at http://www.fftw.org.}
\begin{equation}
  \widehat{X}(\vect{k})
  =\f{1}{V_{\mbox{\tiny L}}}\int_{V_{\mbox{\tiny L}}}X(\vect{x})\times\exp\left( i \vect{k}\cdot\vect{x}\right)\,dV, 
  \label{eq:fourierTransform}
\end{equation}
where $X(\vect{x})$ represents $\sqrt{\rho}u_{j}$ or $B_{j}$, with $j\in\{x,y,z\}$.  
We then compute the kinetic and magnetic spectral energy density on a $k$-space shell
\begin{equation}
  \widehat{e}_{\mbox{\tiny kin}}(k)
  =\f{1}{2}\int_{k\mbox{-shell}}
  \sum_{j}|\widehat{\sqrt{\rho}u_{j}}|^{2}
  k^{2}\,d\Omega_{k}
\end{equation}
and
\begin{equation}
  \widehat{e}_{\mbox{\tiny mag}}(k)
  =\f{1}{2\mu_{0}}\int_{k\mbox{-shell}}
  \sum_{j}|\widehat{B_{j}}|^{2}
  k^{2}\,d\Omega_{k}, 
\end{equation}
respectively.  
The magnitude of the wave vector (wavenumber) is $k=|\vect{k}|=(k_{x}^{2}+k_{y}^{2}+k_{z}^{2})^{1/2}$ and $d\Omega_{k}$ is a solid angle element in Fourier space.  
Proper normalization of the Fourier transform ensures that integration of the spectral energy densities over $k$-space equals real-space integrals of the corresponding energy densities; i.e., 
\begin{equation}
  \int_{k_{\mbox{\tiny min}}}^{k_{\mbox{\tiny max}}} \widehat{e}_{\mbox{\tiny kin}}\,dk
  =\int_{V_{L}}e_{\mbox{\tiny kin}}\,dV
  \equiv \widehat{E}_{\mbox{\tiny kin}} \label{eq:parsevalKin}
\end{equation}
and
\begin{equation}
  \int_{k_{\mbox{\tiny min}}}^{k_{\mbox{\tiny max}}} \widehat{e}_{\mbox{\tiny mag}}\,dk
  =\int_{V_{L}}\edens{mag}\,dV
  \equiv \widehat{E}_{\mbox{\tiny mag}}, \label{eq:parsevalMag}
\end{equation}
where the integrals over $k$ extend from $\kMin=2\pi/L$ (defined by the spatial scale of the computational box $L$) to $\kMax=2\pi/\Delta l$ (limited by finite grid resolution).  
The real-space integrals extend over the volume of the computational domain $V_{L}$.  
Note that $\widehat{E}_{\mbox{\tiny mag}}$, the total magnetic energy in the computational domain, practically equals the magnetic energy below the shock wave $E_{\mbox{\tiny mag}}$, while a similar equality does not hold for the kinetic energy ($\widehat{E}_{\mbox{\tiny kin}}>E_{\mbox{\tiny kin}}$) because the kinetic energy of the supersonic accretion flow above the shock (included in the Fourier transform) is substantial, but contributes mostly to the spectrum at small $k$ (see Figure \ref{fig:compensatedKineticEnergySpectra}).  

\subsubsection{Varying the Initial Magnetic Field Strength}

\begin{figure*}
  \epsscale{1.0}
  \plottwo{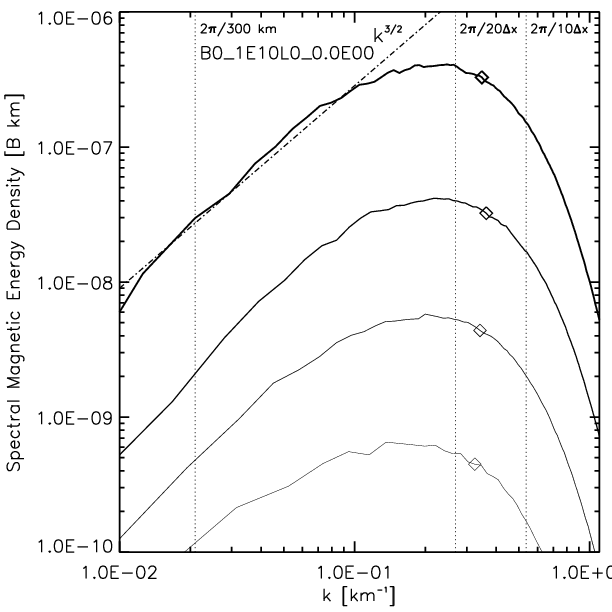}
                {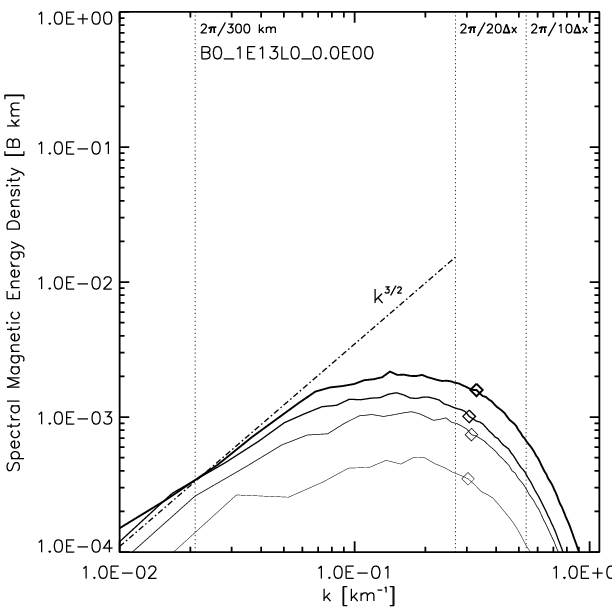}
  \caption{Temporal evolution of the spectral magnetic energy density $\edensk{mag}$ for the weak-field (left) and the strong-field (right) models, respectively.  The spectral distributions are plotted at 666~ms, 792~ms, 918~ms, and 1100~ms (model $\model{10}{0.0}{00}$) and 720~ms, 820~ms, 922~ms, and 1100~ms (model $\model{13}{0.0}{00}$), respectively (i.e., the same times for the respective models as those displayed in Figure \ref{fig:sphericalProfilesNonRotating}).  The dotted vertical reference lines denote spatial scales (from left to right) of 300~km, $20\times\Delta l$, and $10\times\Delta l$, where $\Delta l=1.17$~km is the size of a computational cell.  The dash-dot line in each panel is proportional to $k^{3/2}$.  The mean magnetic wavenumber $\kMeanMag$ (Eq. (\ref{eq:meanMagneticWaveNumber})) is indicated by diamonds.  \label{fig:spectralMagneticEnergyDensityNonRotating}}
\end{figure*}

The evolution of magnetic energy in Fourier space is shown in Figure \ref{fig:spectralMagneticEnergyDensityNonRotating}, in which we plot magnetic energy spectra during the nonlinear SASI for the weak-field model (left panel) and the strong-field model (right panel), for the same times used for the spherically averaged radial profiles displayed in Figure \ref{fig:sphericalProfilesNonRotating}.  
Initially (not shown), the magnetic energy spectrum decreases monotonically with increasing $k$.  
Then, most of the magnetic energy resides on relatively large scales ($k\lesssim0.1$), and $\edensk{mag}$ is roughly proportional to $k^{-2}$ for larger $k$-values.  
The energy spectra in Figure \ref{fig:spectralMagneticEnergyDensityNonRotating} exhibit features typical of MHD turbulence simulations \citep[see][for a recent comprehensive review]{brandenburgSubramanian_2005}.  
The spectral magnetic energy density increases with wavenumber, roughly as $\edensk{mag}\propto k^{3/2}$, for small $k$-values \citep[cf.][]{kazantsev_1968,brandenburgSubramanian_2005}. It reaches a maximum around 0.2-0.3~km$^{-1}$, where the turnover is due to numerical diffusivity (Figure \ref{fig:energySpectraResolutionStudy}), beyond which it decreases rapidly with increasing $k$.  

The spectral magnetic energy density remains basically self-similar over the time-intervals displayed in Figure \ref{fig:spectralMagneticEnergyDensityNonRotating}.  
The weak-field model exhibits the exponential growth on all scales typical of a kinematic small-scale dynamo \citep{brandenburgSubramanian_2005}, the peak value increasing by almost three orders of magnitude (from $\sim6\times10^{-10}$~B~km to $\sim4\times10^{-7}$~B~km), as seen in Figure \ref{fig:spectralMagneticEnergyDensityNonRotating}. In contrast the peak value saturates in the strong-field model, increasing by only a factor of about four (to $\sim2\times10^{-3}$~B~km). However the spectral shape stays relatively unchanged in both cases, with the full width at half maximum being roughly constant over time ($\sim0.40$~km$^{-1}$ and $\sim0.38$~km$^{-1}$ for the weak-field model and the strong-field model, respectively).  

The spectral shape in Figure \ref{fig:spectralMagneticEnergyDensityNonRotating} appears unchanged across the two models, even though the strong-field model saturates and the weak-field model does not.  
There are, however, small differences.  
At the end of the simulations ($t=1100$~ms), the normalized spectra $\edensk{mag}/\max(\edensk{mag})$ of models $\model{10}{0.0}{00}$ and $\model{12}{0.0}{00}$ lie practically on top of each other for all $k$.  
The corresponing spectrum of the strong-field model follows those of the weaker-field models for large $k$ (although the peak is slightly shifted to the left), but has ``excess" power for $k\lesssim0.1$~km$^{-1}$: the integral $\f{1}{\max(\edensk{mag})}\int_{\kMin}^{0.1}\edensk{mag}\,dk$ is about $65\%$ larger in the strong-field model than in the weaker-field models.  
In the simulations of non-helical MHD turbulence by \citet{haugen_etal_2004}, the magnetic energy spectra grow self-similarly until saturation ($\edensk{mag}\sim\edensk{kin}$), which occurs first on smaller spatial scales, and on larger scales later (i.e., the smallest wavenumber where $\edensk{mag}\sim\edensk{kin}$ moves to even smaller $k$), and the magnetic energy spectrum appears to align itself with the kinetic energy spectrum with $\edensk{mag}\gtrsim\edensk{kin}$, almost up to the forcing scale.  
This may suggest that the shape of the saturated and unsaturated spectra should differ more than displayed in Figure \ref{fig:spectralMagneticEnergyDensityNonRotating}.  
Despite the differences (saturated or not), we have not been able to find a reasonable explanation for why the spectral shape in the two models remain so similar.  

From the spectral magnetic energy distribution we obtain the mean magnetic wavenumber
\begin{equation}
  \kMeanMag=\f{1}{\widehat{E}_{\mbox{\tiny mag}}}\int_{k_{\mbox{\tiny min}}}^{k_{\mbox{\tiny max}}}k\edensk{mag}\,dk, 
  \label{eq:meanMagneticWaveNumber}
\end{equation}
and the characteristic spatial scale of the magnetic field $\lambdaMeanMag=2\pi/\kMeanMag$.  
(The mean magnetic wavenumber is indicated by a diamond on each of the energy spectra in Figure \ref{fig:spectralMagneticEnergyDensityNonRotating}.)  
During the initial ramp-up to nonlinear SASI evolution we find that $\lambdaMeanMag$ decreases rapidly with time, from $\lambdaMeanMag\approx60$~km initially to $\lambdaMeanMag\approx20$~km around $t=650$~ms, and stays relatively constant thereafter in all the non-rotating models.  
Magnetic field amplification in our simulations is caused by turbulent stretching of flux tubes.  
In a kinematic dynamo the characteristic scale of the magnetic field decreases exponentially.  
If the kinematic approximation remains valid, the decrease is halted by resistive dissipation when the spatial dimension of the field (the flux tube thickness) approaches the resistive scale \citep{shekochihin_etal_2002}.  
The kinematic approximation remains valid throughout the evolution of the weak-field model.  

The temporal constancy of $\lambdaMeanMag$ for $t\ge650$~ms in the weak-field model is a strong indication that numerical diffusion plays an important role in our simulations.  
In our numerical scheme, we have adopted the HLL Riemann solver \citep{harten_etal_1983}, which approximates the MHD Riemann problem by only considering the left- and right-propagating fast magnetosonic waves.  
This approximation results in diffusive evolution of intermediate waves (e.g., slow magnetosonic, Alfv{\'e}n, and entropy waves), and is the main source of dissipation in our simulations.  
(No other form of dissipation, physical or numerical, has been explicitly included in our simulations.)  
The inherent diffusivity of schemes based on the HLL Riemann solver also affects the evolution of small-scale structures (e.g., turbulence induced magnetic fields.  
See Appendix \ref{app:numericalDissipation} for further details on the source and nature of numerical dissipation of the magnetic energy in our simulations.)  

Moreover, in the strongly nonlinear regime of the SASI at $t\gtrsim750$~ms we find that $\lambdaMeanMag$ is somewhat larger ($\sim10\%$) in models with a stronger initial magnetic field.  
Specifically, we find $\timeAverage{\lambdaMeanMag}{0.8}{1.1}\approx18$~km for the weak-field model and $\timeAverage{\lambdaMeanMag}{0.8}{1.1}\approx20$~km for the strong-field model.  
This trend is consistent with the magnetic field becoming strong enough to cause a back-reaction on the fluid through the magnetic tension force and thereby limit the extent to which magnetic flux tubes are stretched and bent by the chaotic flow induced by the SASI.  
\citet{shekochihin_etal_2001} observed a strong anti-correlation between the magnetic field strength and the curvature of magnetic flux tubes in their small-scale dynamo simulations; i.e., that the strongest magnetic fields are less curved.  
(This effect could potentially be much stronger in simulations similar to ours, but performed at significantly higher spectral resolution, where the magnetic diffusion scale would move to larger $k$-values.)  

We find that the magnetic curvature radius $\lambdaCurvature$ and the magnetic rms scale $\lambdaRMS$ \citepalias[cf. Eqs. (16) and (17), respectively, in][]{endeve_etal_2010} evolve similarly to $\lambdaMeanMag$.  
In particular, for the weak-field model we find $\timeAverage{\lambdaCurvature}{0.8}{1.1}\approx9.5$~km and $\timeAverage{\lambdaRMS}{0.8}{1.1}\approx3.7$~km.  
The corresponding values for the strong-field model are about $10\%$ larger.  
Note that $\lambdaCurvature$ and $\lambdaRMS$ combined characterize the structure of the magnetic field.  
They measure respectively how sharply magnetic flux tubes are bent and how thinly they are stretched.  
Such information is not contained in $\lambdaMeanMag$ alone.  

\begin{figure*}
  \epsscale{1.0}
  \plottwo{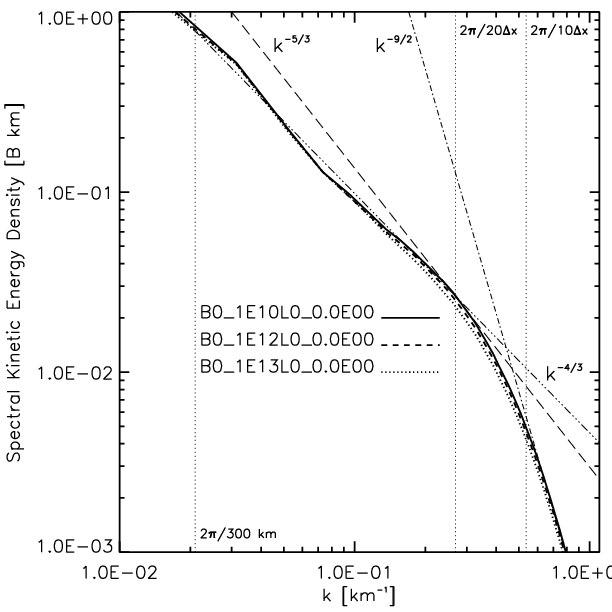}
                {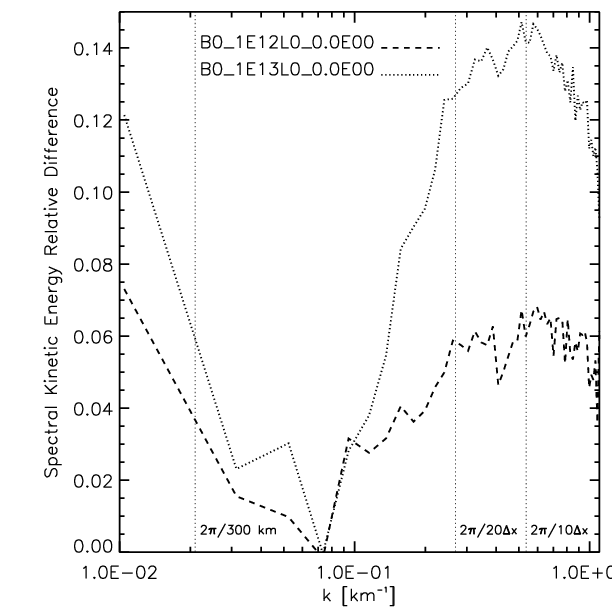}
  \caption{Left panel: (time-averaged) kinetic energy spectra $\edensk{kin}$ from non-rotating models $\model{10}{0.0}{00}$ (solid line), $\model{12}{0.0}{00}$ (dotted line), and $\model{13}{0.0}{00}$ (dotted line).  Right panel: difference in spectral kinetic energy relative to the weak-field reference model ($1-\edensk{kin}/\widehat{e}_{\mbox{\tiny kin}}^{\,\mbox{\tiny ref}}$), where $\widehat{e}_{\mbox{\tiny kin}}^{\,\mbox{\tiny ref}}$ is the spectral kinetic energy density of the weak-field model ($\model{10}{0.0}{00}$).  In both panels we include vertical reference lines to indicate spatial scales of 300~km, $20\times\Delta l$, and $10\times\Delta l$.  In the left panel we also include reference lines proportional to power-laws in $k$: $k^{-5/3}$ (long-dashed), $k^{-4/3}$ (dash-dotted), and $k^{-9/2}$ (dash-dot).  \label{fig:spectralKineticEnergyDensityNonRotating}}
\end{figure*}

Spectral kinetic energy distributions from the non-rotating models with different initial field strengths are shown in Figure \ref{fig:spectralKineticEnergyDensityNonRotating}.  
The stochastic nature of the SASI and the turbulent flows necessitates the use of temporally averaged spectra when cross-comparing the models; in particular,  
the kinetic energy spectra shown in Figure \ref{fig:spectralKineticEnergyDensityNonRotating} are averaged over the time period extending from 800~ms to 1100~ms (that is, we plot $\timeAverage{\edensk{kin}}{0.8}{1.1}$ versus $k$).  
The spectra show that the majority of the kinetic energy resides on relatively large spatial scales (small $k$; see the dotted vertical reference lines).  
The spectral kinetic energy density is roughly proportional to $k^{-4/3}$ for small $k$-values ($k\lesssim0.2$~km$^{-1}$), while for larger $k$-values ($k\gtrsim0.6$~km$^{-1}$) the flow is heavily influenced by numerical dissipation and the kinetic energy decreases rapidly with increasing wavenumber ($\edensk{kin}\propto k^{-9/2}$).  
Magnetic field amplification is driven by the turbulent flows, and it is the kinetic energy of the small-scale motions that is available to be tapped by the magnetic fields.  

When comparing the non-rotating models in the left panel of Figure \ref{fig:spectralKineticEnergyDensityNonRotating}, we see a decreasing trend in the spectral kinetic energy density for larger wavenumbers ($k\gtrsim0.2$) in models with a stronger initial magnetic field.  
(The decrease in kinetic energy on small scales is balanced by a corresponding increase in magnetic energy.)  
We emphasize this difference further in the right panel, where we plot the difference in $\edensk{kin}$ for the stronger field models relative to the weak-field model, $1-\edensk{kin}/\widehat{e}_{\mbox{\tiny kin}}^{\,\mbox{\tiny ref}}$.  
(The spectral kinetic energy density of the weak-field reference model is here denoted $\widehat{e}_{\mbox{\tiny kin}}^{\,\mbox{\tiny ref}}$.)  
The spectral kinetic energy density in model $\model{12}{0.0}{00}$ is reduced by up to $\sim6\%$, while in the strong-field model it is reduced by a maximum of $\sim15\%$ relative to the weak-field model.  
The largest difference is seen in the strongly diffusive regime around $k=0.6$~km$^{-1}$.  
These results demonstrate that even relatively weak initial magnetic fields can be amplified and impact the flow, although only on small spatial scales.  
For larger spatial scales ($k\lesssim0.06$) the differences are caused by differences in the pre-shock kinetic energy (due to differences in $\bar{R}_{\mbox{\tiny Sh}}$ and box size $L$).  

\begin{figure}
  \epsscale{1.0}
  \plotone{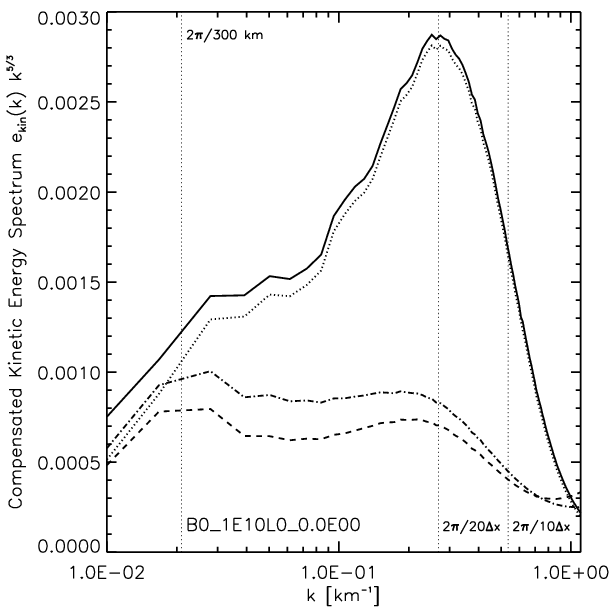}
  \caption{Compensated kinetic energy spectra from model \model{10}{0.0}{00}.  (The spectra are computed by averaging over the time period from $804$~ms to $918$~ms.)  The solid line corresponds to the kinetic energy spectrum shown as the solid line in left panel in Figure \ref{fig:spectralKineticEnergyDensityNonRotating}.  The kinetic energy spectrum obtained when setting the pre-shock flow velocity to zero $\edensk{kin}^{\,\mbox{\tiny I}}$ is represented by the dotted line.  Kinetic energy spectra with corrections for density stratification (also with the pre-shock flow excluded), $\edensk{kin}^{\,\mbox{\tiny II}}$ and $\edensk{kin}^{\,\mbox{\tiny III}}$, are represented by the dashed and dash-dot lines, respectively (see text for details).  Note the narrow inertial range in the stratification-corrected spectra ($\edensk{kin}^{\,\mbox{\tiny II}},\edensk{kin}^{\,\mbox{\tiny III}}\propto k^{-5/3}$) in $k\in[0.04,0.1]$~km$^{-1}$ (i.e., spatial scales from $\sim160$~km to $\sim60$~km).  \label{fig:compensatedKineticEnergySpectra}}
\end{figure}

The kinetic energy of the pre-shock flows, density stratification (resulting in the real-space power-law (in radius) in the spherically averaged kinetic energy density (Section \ref{sec:radialProfiles})), and compressibility (due to the presence of supersonic flows below the shock) contribute to the kinetic energy spectra in Figure \ref{fig:spectralKineticEnergyDensityNonRotating}.  
To further investigate details about the shape of the kinetic energy spectra due to these factors, in particular the $-4/3$ slope (as opposed to the $-5/3$ slope of Kolmogorov turbulence), we have computed (1) kinetic energy spectra with the pre-shock flow velocity set to zero $\edensk{kin}^{\,\mbox{\tiny I}}(k)$, (2) kinetic energy spectra with the pre-shock flow velocity set to zero \emph{and} corrected for the radial density stratification $\edensk{kin}^{\,\mbox{\tiny II}}(k)$ (i.e., we use $X=\sqrt{\volumeAverage{\rho}{V_{\mbox{\tiny L}}}}u_{j}$ with $j\in\{x,y,z\}$ in Eq. (\ref{eq:fourierTransform})), and (3) kinetic energy spectra with the pre-shock flow velocity set to zero \emph{and} corrected for radial density stratification, \emph{but} with local compressibility retained $\edensk{kin}^{\,\mbox{\tiny III}}(k)$ (i.e., we use $X=\sqrt{\bar{\rho}}u_{j}$, where $\bar{\rho}=\volumeAverage{\rho}{V_{\mbox{\tiny L}}}\times(\rho/\volumeAverage{\rho}{\delta V_{i}})$, in Eq. (\ref{eq:fourierTransform})).  
Results from these calculations are plotted in Figure \ref{fig:compensatedKineticEnergySpectra}, where we plot compensated kinetic energy spectra from the weak-field model (averaged over the time interval from $804$~ms to $918$~ms): $\edensk{kin}\times k^{5/3}$ (solid line), $\edensk{kin}^{\,\mbox{\tiny I}}\times k^{5/3}$ (dotted line), $\edensk{kin}^{\,\mbox{\tiny II}}\times k^{5/3}$ (dashed line), and $\edensk{kin}^{\,\mbox{\tiny III}}\times k^{5/3}$ (dash-dot line).  
(In Figure \ref{fig:compensatedKineticEnergySpectra}, $\edensk{kin}^{\,\mbox{\tiny II}}$ and $\edensk{kin}^{\,\mbox{\tiny III}}$ have been multiplied by a factor of three for convenient comparison with $\edensk{kin}$ and $\edensk{kin}^{\,\mbox{\tiny I}}$.)  

When comparing $\edensk{kin}$ and $\edensk{kin}^{\,\mbox{\tiny I}}$ in Figure \ref{fig:compensatedKineticEnergySpectra} we see that the supersonic pre-shock accretion flow contributes to the energy spectrum, mostly for small $k$-values, but the two spectra remain similar in shape.  
(By integrating the two spectra over all $k$ we find $\widehat{E}_{\mbox{\tiny kin}}\approx0.053$~B and $\widehat{E}_{\mbox{\tiny kin}}^{\,\mbox{\tiny I}}\approx0.040$~B.)  
On the other hand, the kinetic energy spectra change markedly when the density stratification is excluded from its computation.  
The $-4/3$ spectral slope seen in the left panel of Figure \ref{fig:spectralKineticEnergyDensityNonRotating} is due to density stratification from $\rPNS$ to $\rShock$.  
Effects due to compressibility are subdominant and do not change the shape of the spectrum in any significant way.  
(\citet{kritsuk_etal_2007} scaled the velocity with $\rho^{1/3}$ to recover Kolmogorov $-5/3$ scaling in spectra from simulations of supersonic isothermal turbulence.)  
Moreover, we observe a narrow inertial range in $k\in[0.04,0.1]$~km$^{-1}$ (i.e., spatial scales from $\sim160$~km to $\sim60$~km) where $\edensk{kin}^{\,\mbox{\tiny II}},\edensk{kin}^{\,\mbox{\tiny III}}\propto k^{-5/3}$.  
For larger $k$-values (around $k=0.2$~km$^{-1}$) we observe a bump in the $\edensk{kin}^{\,\mbox{\tiny II}}$ and $\edensk{kin}^{\,\mbox{\tiny III}}$ spectra \citep[e.g.,][and references therein]{ishihara_etal_2009}, which is less pronounced for $\edensk{kin}^{\,\mbox{\tiny III}}$.  
We note that the simulations by \citet{haugen_etal_2003,haugen_etal_2004} argue in favor of a $k^{-5/3}$ spectrum for non-helical MHD turbulence.  
(Given infinite spectral resolution, the $\edensk{kin}$ spectrum could possibly also follow $-5/3$ scaling for larger $k$, where the spectrum presumably would be less influenced by density stratification.)  

\begin{figure}
  \epsscale{1.0}
  \plotone{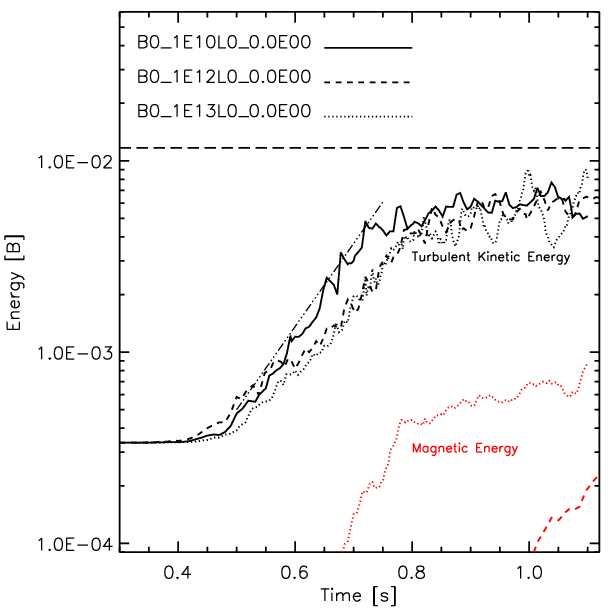}
  \caption{Turbulent kinetic energy (Eq. (\ref{eq:turbulentKineticEnergy}), black lines) and total magnetic energy (red lines) versus time in non-rotating models in which the initial magnetic field strength is varied: $\model{10}{0.0}{00}$ (solid), $\model{12}{0.0}{00}$ (dashed), and $\model{13}{0.0}{00}$ (dotted).  The dash-dotted line is proportional to $\exp{(t/\tau)}$, with $\tau=100$~ms.  The long-dashed horizontal line indicates the upper limit for turbulent kinetic energy, assuming Kolmogorov scaling applies for $k\ge\kTurb$.  ($10^{-2}~\mbox{B}=10^{49}~\mbox{erg}$.)  \label{fig:turbulentKineticEnergy}}
\end{figure}

The identification of Kolmogorov-like spectra in our simulations helps us associate post-shock flows with turbulence.  
In Figure \ref{fig:compensatedKineticEnergySpectra}, the peak in the $\edensk{kin}^{\,\mbox{\tiny II}}$ and $\edensk{kin}^{\,\mbox{\tiny III}}$ spectra for smaller $k$-values ($k\approx0.02$~km$^{-1}$; i.e., spatial scales around 300~km) is associated with the large scale SASI flows (cf. Figure \ref{fig:machNumberAndVorticity}), which drive post-shock turbulence.  
The peak is located around $k=0.05$~km$^{-1}$ (i.e., spatial scales around 125~km) in the earlier stages of SASI-development, when the spectrum also has a steeper-than-$-5/3$ slope for larger $k$.  
When the SASI develops nonlinearly, and the average shock radius begins to increase, the peak moves to smaller $k$-values, and the $\edensk{kin}^{\,\mbox{\tiny II}}$ and $\edensk{kin}^{\,\mbox{\tiny III}}$ spectra develop Kolmogorov slopes.  
Thus, the power in the large scale flows cascades to smaller-scale flows.  
In particular, integrating the $\edensk{kin}^{\,\mbox{\tiny I}}$ spectrum in Figure \ref{fig:compensatedKineticEnergySpectra} over $k$, from $k=0.04$ to $\kMax$, gives $0.022$~B.  
Thus, a large fraction (up to $\sim50\%$) of the kinetic energy below the shock may be associated with turbulence.  
These observations suggest that the SASI saturates due to the development of turbulence via secondary instabilities (e.g., the Kelvin-Helmholtz instability), which feed on the power in the low-order SASI modes \citep[e.g.,][]{guilet_etal_2010}.  
The turbulent energy is either dissipated via viscous heating, or converted into magnetic energy and dissipated via Joule heating.  
However, we find that significantly less than $50\%$ of the post-shock kinetic energy is accessed for magnetic field amplification (cf. Figure \ref{fig:turbulentKineticEnergy}).  

We use the unmodified kinetic energy spectrum $\edensk{kin}$ in our further analysis since it is related to the total kinetic energy via Eq. (\ref{eq:parsevalKin}), and therefore most useful for extracting quantitative information from our simulations.  

From the spectral kinetic energy density we obtain the turbulent kinetic energy in our simulations
\begin{equation}
  E_{\mbox{\tiny kin}}^{\mbox{\tiny tur}}
  =\int_{\kTurb}^{\kMax}\edensk{kin}\,dk.  
  \label{eq:turbulentKineticEnergy}
\end{equation}
For the purpose of studying magnetic field amplification, we have chosen to define turbulent flows to include flows residing on scales with $k\ge\kTurb=2\pi/\lambdaTurb$, where the turbulent spatial scale covers 25 grid cells $\lambdaTurb=25\times\Delta l\approx30$~km (for $\Delta l=1.17$~km).  
For reference, $\lambdaTurb$ is more than an order of magnitude smaller than the average shock radius, which again is comparable to the forcing scale of the turbulent flows (i.e., the scale of the supersonic downdrafts from the shock triple-point), but comparable to $\rPNS$ ($\sim25\%$ smaller).  
This particular choice for $\kTurb$ is motivated by several factors, including (1) most of the magnetic field amplification occurs on spatial scales with $k>\kTurb$ (Figures \ref{fig:spectralMagneticEnergyDensityNonRotating} and \ref{fig:energySpectraResolutionStudy}), (2) any dynamic effect of the magnetic field is seen on scales with $k\gtrsim\kTurb$ (Figure \ref{fig:spectralKineticEnergyDensityNonRotating}), and (3) the flow Taylor microscale, $\lambdaTaylor=\sqrt{\volumeAverage{u^{2}}{\vShock}/\volumeAverage{|\curl{\vect{u}}|^{2}}{\vShock}}$, which measures the average size of turbulent eddies \citep[e.g.,][]{ryu_etal_2000}, is comparable to $\lambdaTurb$ (about a factor of two smaller).  

In Figure \ref{fig:turbulentKineticEnergy} we plot the time evolution of the turbulent kinetic energy in the non-rotating models.  
The turbulent kinetic energy evolves similarly to the total kinetic energy below the shock (cf. top left panel in Figure \ref{fig:overviewNonRotating}).  It grows exponentially during the ramp-up of the SASI and reaches a saturation level, where the intermittent time variability is superimposed on a barely noticeable overall growth.  
During the exponential growth phase, the growth rate is somewhat lower and the saturation level about an order of magnitude below the total kinetic energy beneath the shock.  
The exponential growth time in the weak-field model is $\tau\approx100$~ms.  
The time-averaged saturation levels for the turbulent kinetic energy in the respective models are found to be $\timeAverage{\EKinTurb}{0.8}{1.1}=5.90\times10^{-3}$~B ($\model{10}{0.0}{00}$), $5.48\times10^{-3}$~B ($\model{12}{0.0}{00}$), and $5.11\times10^{-3}$~B ($\model{13}{0.0}{00}$), which implies $\sim7\%$ ($4.2\times10^{-4}$~B) and $\sim13\%$ ($7.9\times10^{-4}$~B) reductions with respect to the weak-field model for model $\model{12}{0.0}{00}$ and model $\model{13}{0.0}{00}$, respectively.  
These reductions in turbulent energy are comparable to the increase in magnetic energy in the respective models.  
Thus the magnetic energy grows at the expense of the turbulent kinetic energy.  

The saturation level for the turbulent kinetic energy in our models is only about a factor of two below what is obtained by (hypothetically) assuming Kolmogorov scaling ($\edensk{kin}\propto k^{-5/3}$) for $k\ge\kTurb$ (indicated by the long-dashed line in Figure \ref{fig:turbulentKineticEnergy}).  Thus, $\EKinTurb\approx 10^{-2}$~B may serve as an upper limit for turbulent kinetic energy in our models, and therefore also as a reasonable upper limit on the magnetic energy attainable in these simulations.  
(The turbulent kinetic energy may depend on the accretion rate ahead of the shock, which is held fixed in our simulations.  
Thus, the upper limit on $\EKinTurb$ is only approximate.)  
We find that the magnetic energy in model $\model{13}{0.0}{00}$ saturates at about $10\%$ of the turbulent kinetic energy.  
One must keep in mind that the magnetic energy is also heavily influenced by numerical dissipation during the saturated phase.  
It is entirely possible that the magnetic energy can grow beyond the levels seen in our simulations, but probably not much above the long-dashed horizontal line in Figure \ref{fig:turbulentKineticEnergy}.  
Small-scale dynamo simulations commonly show that the magnetic energy spectrum lies slightly above the kinetic energy spectrum on the smallest scales during saturation \citep[e.g.,][]{brandenburgSubramanian_2005}.  
The magnetic energy does not exceed the kinetic energy in any part of the spectrum in our simulations.  
This may be due to finite resolution and numerical dissipation on the smallest scales.  

\subsubsection{Varying the Spatial Resolution}

In \citetalias{endeve_etal_2010} we found that magnetic field amplification from SASI-induced turbulent flows is very sensitive to the spatial resolution adopted in the numerical simulations.  
(In general, increased spatial resolution results in stronger magnetic fields and improves the conditions for a dynamical influence of magnetic fields.)  
With the energy spectra we continue to study the effect of resolution in this section.  

\begin{figure*}
  \epsscale{1.0}
  \plottwo{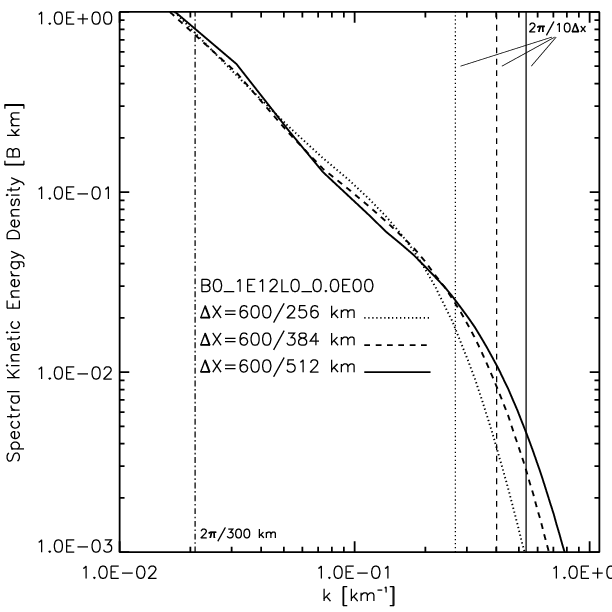}
                {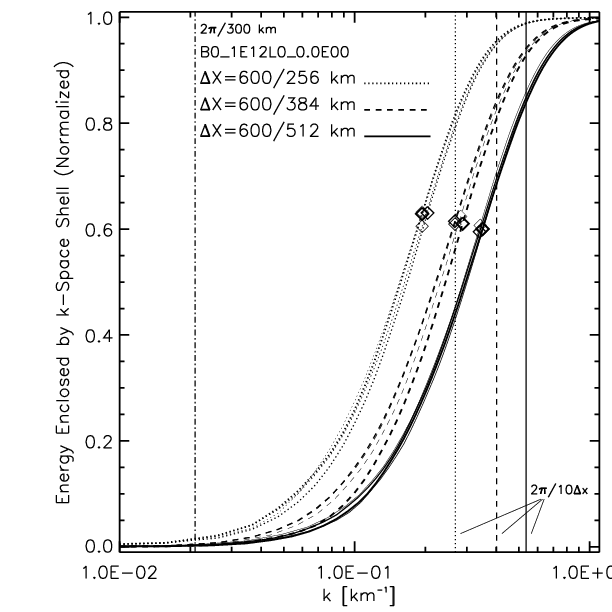}
  \caption{Energy spectra from simulations in which the spatial resolution has been varied.  Kinetic energy spectra are plotted in the left panel, and the fractional magnetic energy enclosed by $k$-space shell with radius $k$, $\fMagEncl$ (cf. Eq. (\ref{eq:eMagEnclosed})),  is plotted in the right panel.  Results are plotted for the non-rotating model with $B_{0}=1\times10^{12}$ ($\model{12}{0.0}{00}$).  The spatial resolution in these runs has been set to $\Delta l=2.34$~km (dotted lines), $1.56$~km (dashed lines), and $1.17$~km (solid lines).  The kinetic energy spectra are averaged over the time period extending from $800$~ms to $1100$~ms.  For each model, magnetic energy spectra are plotted at $t=800$~ms, $900$~ms, $1000$~ms, and $1100$~ms (thicker lines represent more advanced time states).  The mean magnetic wavenumber (cf. Eq. (\ref{eq:meanMagneticWaveNumber})) is denoted with a diamond on each spectrum in the right panel.  In both panels we include vertical reference lines indicating the spatial scale of 300~km (dash-dot line), and $10\times\Delta l$ (with line styles matching each of the models).  \label{fig:energySpectraResolutionStudy}}
\end{figure*}

Energy spectra from simulations of the non-rotating model with $B_{0}=1\times10^{12}$~G for various spatial resolutions are plotted in Figure \ref{fig:energySpectraResolutionStudy}.  
In the left panel we plot the spectral kinetic energy density.  
In the right panel we plot the fractional magnetic energy enclosed by the $k$-space shell with radius $k$
\begin{equation}
  \fMagEncl=
  \f{1}{\widehat{E}_{\mbox{\tiny mag}}}
  \int_{\kMin}^{k}\edensk{mag}(k')\,dk', 
  \label{eq:eMagEnclosed}
\end{equation}
where $\fMagEncl$ is normalized to the total magnetic energy $\widehat{E}_{\mbox{\tiny mag}}$ so that $f_{\mbox{\tiny mag}}(\kMax)=1$.  
Results from three simulations are presented, and the grid size has been varied by a factor of two: $\Delta l=2.34$~km (low resolution; dotted lines), $\Delta l=1.56$~km (medium resolution; dashed lines), and $\Delta l=1.17$~km (high resolution; solid lines).  
Kinetic energy spectra are averaged over a time period extending from $800$~ms to $1100$~ms, while magnetic energy spectra are plotted for $t=800$~ms, $900$~ms, $1000$~ms, and $1100$~ms.  

The kinetic energy spectra are very similar and follow each other closely for small wavenumbers ($k\lesssim0.2$~km$^{-1}$).  
Numerical dissipation influences the kinetic energy for larger $k$-values, and $\edensk{kin}$ begins to fall off more rapidly with increasing $k$.  
The fall-off starts at smaller $k$-values for the lower resolution runs: $\edensk{kin}$ falls below $10^{-3}$~B km$^{-1}$ around $k=0.5$~km$^{-1}$ in the low resolution run, and around $k=0.8$~km$^{-1}$ in the high resolution run.  

Since most of the kinetic energy below the shock resides on large scales, and includes the flows associated with the supersonic stream ahead of the shock triple-point (Figure \ref{fig:machNumberAndVorticity}), the total kinetic energy below the shock, $\eShock{kin}$, is insensitive to the spatial resolution.  
During the highly nonlinear operation of the SASI spiral mode we find $\timeAverage{\eShock{kin}}{0.8}{1.1}=0.045$~B, $0.043$~B, and $0.044$~B, for the low, medium, and high resolution model, respectively.  
Our definition of $\kTurb$, used in Eq. (\ref{eq:turbulentKineticEnergy}), is not optimal when comparing simulations computed with different spatial resolutions, since it results in smaller $\kTurb$ and more turbulent kinetic energy in models with larger $\Delta l$ ($\timeAverage{\EKinTurb}{0.8}{1.1}=8.6\times10^{-3}$~B, $6.3\times10^{-3}$~B, and $5.5\times10^{-3}$~B, for low, medium, and high resolution, respectively).  
For the purpose of quantifying the increase in kinetic energy on small scales due to higher resolution, we fix $\kTurb=0.2$~km$^{-1}$ and find that $\timeAverage{\EKinTurb}{0.8}{1.1}$ increases (linearly) by a factor of two when $\Delta l$ is decreased by a corresponding factor of two ($\timeAverage{\EKinTurb}{0.8}{1.1}=2.7\times10^{-3}$~B, $4.5\times10^{-3}$~B, and $5.5\times10^{-3}$~B for low, medium, and high resolution, respectively).  

The self-similar evolution of the magnetic energy spectra is clearly displayed in the right panel of Figure \ref{fig:energySpectraResolutionStudy}.  
The different models are well separated, while each model's temporally separated spectra fall practically on top of each other.  
The spectra are shifted to larger $k$-values when the resolution is increased.  
The shift to the right in the spectrum is a direct consequence of a corresponding shift of the spatial scale where numerical diffusion dominates.  
The characteristic scale of the magnetic field is roughly constant with time over the time period displayed, but decreases linearly with increasing spatial resolution.  
In particular, we find $\timeAverage{\lambdaMeanMag}{0.8}{1.1}\approx32$~km, $23$~km, and $18$~km, for the low, medium, and high resolution runs, respectively.  
Similarly, the magnetic rms scale $\lambdaRMS$ (the average flux tube thickness) decreases by almost a factor of two when $\Delta l$ is reduced by a factor of two (from $7.0$~km to $3.8$~km).  

The shift to smaller spatial scales---in particular the decrease in the flux tube thickness---afforded by higher spatial resolution results in stronger magnetic fields and an increase in the magnetic energy.  
The integrated magnetic energy below the shock in the low resolution model reaches saturation for $t<800$~ms, and does not grow much beyond $3\times10^{-6}$~B.  
(Saturation of magnetic energy in this model is solely due to numerical dissipation.  
The influence of magnetic fields on small scale flows emphasized in the right panel of Figure \ref{fig:spectralKineticEnergyDensityNonRotating} is not observed in lower-resolution models.)  
In the high resolution model the magnetic energy grows throughout the run, and reaches $2\times10^{-4}$~B near the end (dashed red line in Figure \ref{fig:turbulentKineticEnergy}).  
It is interesting to note that during the time span from 800~ms to 1100~ms, between $68\%$ and $77\%$ of the total magnetic energy resides on scales smaller than $\lambdaTurb$, nearly independent of spatial resolution.  
(There is a weak decrease in the percentage with increasing resolution).  
The corresponding percentage for spatial scales smaller than $10\times\Delta l$ varies between $16\%$ and $21\%$.  

We expect the magnetic energy spectra will continue to move to higher wavenumbers when the resolution is increased beyond that of our simulations.  
The shift to smaller spatial scales (smaller flux tube cross-section) is accompanied by stronger magnetic fields, and we expect the flux tube cross-section to decrease until the magnetic fields become strong enough to cause a back-reaction on the fluid through the Lorentz force.  
(\citet{haugen_etal_2003} presented converged magnetic energy spectra in their simulations of non-helical MHD turbulence.  
In their converged spectra, most of the magnetic energy resides at a wavenumber $\sim5$ times the minimum wavenumber in the computational domain.)  

\subsection{Magnetic Energy Growth Rates}
\label{sec:magneticEnergyGrowthRates}

In this section we focus on the relative importance of mechanisms that control the exponential growth rate of magnetic energy when the magnetic field is weak and the kinematic approximation remains valid.  
We also consider the impact of finite numerical resolution on the growth rate in our simulations.  

An eddy turnover time $\tauEddy=\lambdaMeanMag/\uRMSTurb$ is commonly invoked as the characteristic exponential growth time of magnetic fields in a turbulent small-scale dynamo \citep[e.g.,][]{kulsrudAnderson_1992}. Here $\lambdaMeanMag$ is the characteristic spatial scale of the magnetic field defined below Eq. (\ref{eq:meanMagneticWaveNumber}). The turbulent rms velocity is
\begin{equation}
  \uRMSTurb=\left(\f{2\EKinTurb}{\mShock}\right)^{1/2}, 
  \label{eq:rmsVelocity}
\end{equation}
where $\mShock$ is the mass in $\vShock$.  
The use of $\mShock$ in Eq. (\ref{eq:rmsVelocity}), instead of only the mass of the flow included in $\EKinTurb$, results in an underestimate of $\uRMSTurb$.  
On the other hand, $\EKinTurb$ (and therefore $\uRMSTurb$) is also sensitive to the definition of $\kTurb$, which may be larger than the value we use in Eq. (\ref{eq:turbulentKineticEnergy}) and result in smaller $\EKinTurb$.  
Nevertheless, Eq. (\ref{eq:rmsVelocity}) provides a reasonable order-of-magnitude estimate of $\uRMSTurb$.  
We find that $\uRMSTurb$ grows rapidly during the initial ramp up of the SASI and then levels off at later times.  
In the non-rotating models we find $\timeAverage{\uRMSTurb}{0.8}{1.1}\approx4000$~km s$^{-1}$.  
A turbulent rms velocity of several $\times10^{3}$~km s$^{-1}$ is consistent with an inspection of the subsonic flows below the shock: the average velocity among the zones with $|\vect{u}|/c_{S}\le 1$ is about $7000$~km s$^{-1}$.  
For the non-rotating models $\tauEddy$ is about $5$~ms during the highly nonlinear stage of strong SASI activity.  
(Another commonly used expression for the eddy turnover time, $\volumeAverage{|\curl{\vect{u}}|^{2}}{\vShock}^{-1/2}$, gives a similar result.)  

We now investigate the individual magnetic energy growth rates relevant to our simulations.  
Assuming a non-ideal electric field $-(\vect{u}\times\vect{B})+\eta\vect{J}$ with scalar resistivity $\eta$, the evolution equation for the magnetic energy density is easily obtained by dotting the Maxwell-Faraday (induction) equation with $\vect{B}/\mu_{0}$:
\begin{equation}
  \pderiv{\edens{mag}}{t}
  +\divergence{\vect{P}}
  =-\vect{u}\cdot\left(\vect{J}\times\vect{B}\right)
  -\f{1}{\mu_{0}}\vect{B}\cdot\curl{\left(\eta\vect{J}\right)}, 
  \label{eq:magneticEnergyEquation}
\end{equation}
where $\vect{P}=[\vect{u}(\vect{B}\cdot\vect{B})-\vect{B}(\vect{B}\cdot\vect{u})]/\mu_{0}$ and $\vect{J}=\left(\curl{\vect{B}}\right)/\mu_{0}$.  (See also Eq. (10) as well as the discussion in Section 3.3 in \citetalias{endeve_etal_2010}.)  The first and second terms on the right-hand-side of Eq. (\ref{eq:magneticEnergyEquation}) represent work done against the Lorentz force ($W_{\mbox{\tiny L}}$) and magnetic energy decay due to resistive (Joule) dissipation ($-Q_{\mbox{\tiny J}}$), respectively.  Kinetic energy of the flow is converted into magnetic energy if $W_{\mbox{\tiny L}}>0$.  

It is apparent from Eq. (\ref{eq:magneticEnergyEquation}) that the total magnetic energy growth rate $\tau_{\mbox{\tiny tot}}^{-1}=\langle\edens{mag}\rangle^{-1}\langle\partial\edens{mag}/\partial t\rangle$ equals the sum  $\rateLorentzWork+\ratePoynting+\rateDissipation$ of individual rates due to work done against the Lorentz force, accretion of magnetic energy (Poynting flux) through $\sPNS$, and resistive energy dissipation.  
(The angle brackets in the total rate imply an integral over a volume bounded by the surface $\sPNS$ and a spherical surface enclosing the accretion shock.)  
The Poynting flux through the spherical surface enclosing the accretion shock vanishes because $\vect{u}\parallel\vect{B}$ ahead of the shock .  
The Poynting flux through $\sPNS$ and resistive dissipation generally result in decay of the magnetic energy in the computational domain.  
The decay must be overcome by the Lorentz work term in order for the magnetic energy to increase.  

In \citetalias{endeve_etal_2010} we found flux tube stretching by turbulent flows driven by the spiral SASI mode to be the dominant mechanism for magnetic field amplification (see also Figure \ref{fig:magneticEnergyGrowthRatesB0_1E10L0_0_0E00} below).  
The magnetic energy growth rate due to work done against the Lorentz force is
\begin{eqnarray}
  \rateLorentzWork
  &=&
  \f{1}{E_{\mbox{\tiny mag}}}\int_{V}\vect{u}\cdot\left(\vect{J}\times\vect{B}\right)\,dV \nonumber \\
  &\approx&
  2\uRMSTurb/\lambdaMeanMag = 2 \tauEddy^{-1}, 
  \label{eq:growthRateLorentzWork}
\end{eqnarray}
where the turbulent rms velocity $\uRMSTurb$ and the characteristic scale of the magnetic field $\lambdaMeanMag$ have been used.  
(The factor two in the second part of Eq. (\ref{eq:growthRateLorentzWork}) stems from the factor of one half in the definition of magnetic energy, but is probably not important for this rough estimate.)  
The corresponding growth time is then approximately
\begin{equation}
  \timeLorentzWork
  \approx
  2.5~\mbox{ms}
  \left(\f{\lambdaMeanMag}{20~\mbox{km}}\right)
  \left(\f{\uRMSTurb}{4000~\mbox{km s}^{-1}}\right)^{-1}.  
  \label{eq:growthTimeLorentzWork}
\end{equation}

\begin{figure}
  \epsscale{1.0}
  \plotone{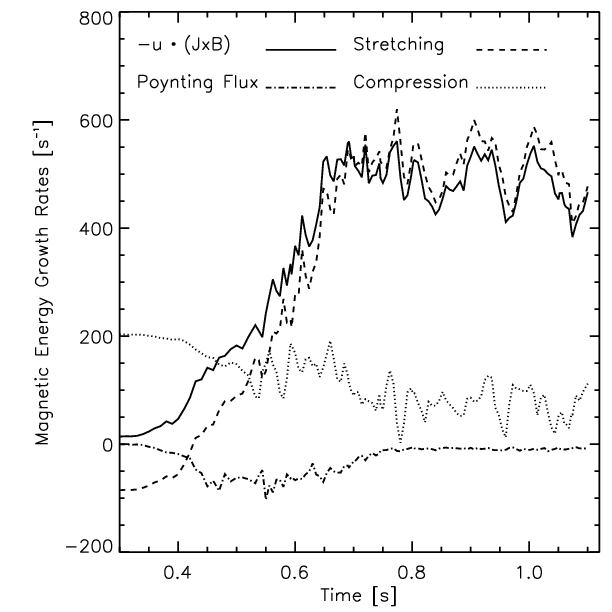}
  \caption{Magnetic energy growth rates versus time for the non-rotating weak-field model ($\model{10}{0.0}{00}$).  
  The growth rates are based on terms appearing in Eq. (\ref{eq:magneticEnergyEquation}) and are due to work done against the Lorentz force (solid line) and Poynting flux losses due to accretion through the spherical surface with $r=\rPNS$ (dash-dot line).  
  We also plot the growth rates due to compression (dotted line) and stretching (dashed line) (Eqs. (13) and (11) in \citetalias{endeve_etal_2010}, respectively).  \label{fig:magneticEnergyGrowthRatesB0_1E10L0_0_0E00}}
\end{figure}

The volume occupied by the PNS is excluded from our simulations, and the magnetic energy in $\vShock$ is also affected by accretion of magnetized matter through $\sPNS$.  
The decay rate due to this process is
\begin{equation}
  \ratePoynting
  =
  \f{1}{E_{\mbox{\tiny mag}}}\oint_{\sPNS}\vect{P}\cdot d\vect{S}
  \approx
  \f{3\dot{M}}{2\pi\rho_{0}L_{B}^{3}}, 
  \label{eq:growthRatePoynting}
\end{equation}
where in the rightmost estimate we have adopted the exponential decrease of magnetic field with radius over a characteristic length scale $L_{B}$ (cf. Figure \ref{fig:sphericalProfilesNonRotating}) to relate the magnetic energy in $\vShock$ to the field strength at the surface of the PNS: $E_{\mbox{\tiny mag}}\approx\f{B_{0}^{2}}{2\mu_{0}}\f{4\pi}{3}L_{B}^{3}$.  
The decay time due to accretion through the inner boundary is then approximately
\begin{eqnarray}
  \timePoynting
  &\approx&
  90~\mbox{ms}
  \left(\f{\rho_{0}}{3\times10^{10}~\mbox{g cm}^{-3}}\right)\times \nonumber \\
  &&\times\left(\f{L_{B}}{100~\mbox{km}}\right)^{3}
  \left(\f{\dot{M}}{0.36~M_{\odot}\mbox{ s}^{-1}}\right)^{-1}
  \label{eq:growthTimePoynting}
\end{eqnarray}
The average mass density around $r=\rPNS$, denoted $\rho_{0}$, stays fairly constant throughout the simulations.  

In Figure \ref{fig:magneticEnergyGrowthRatesB0_1E10L0_0_0E00} we plot the growth rates $\rateLorentzWork$ (solid line) and $\ratePoynting$ (dash-dot line) versus time for model $\model{10}{0.0}{00}$.  
This model exhibits exponential magnetic energy growth throughout with a growth time of about 66~ms.  
The growth rates plotted in Figure \ref{fig:magneticEnergyGrowthRatesB0_1E10L0_0_0E00} are computed from numerical approximations (second-order finite differences) to the integral expressions, and not the approximations provided by the rightmost expressions in Eqs. $(\ref{eq:growthTimeLorentzWork})$ and $(\ref{eq:growthTimePoynting})$.  
We also include the growth rates due to stretching $\rateStretching$ and compression $\rateCompression$ (Eqs. (11) and (13) in \citetalias{endeve_etal_2010}, respectively), and the plot shows that stretching dominates over compression.  
The rates remain quasi-steady for $t\gtrsim750$~ms, and in particular we find $\timeAverage{\rateLorentzWork}{0.9}{1.1}\approx480$~s$^{-1}$ and $\timeAverage{\ratePoynting}{0.9}{1.1}\approx9$~s$^{-1}$.  (We also find $\timeAverage{\rateStretching}{0.9}{1.1}\approx515$~s$^{-1}$, and $\timeAverage{\rateCompression}{0.9}{1.1}\approx76$~s$^{-1}$.)  
We note that there is good agreement between the numerically computed growth rates and the growth rates predicted by the estimates provided by the rightmost expressions in Eqs. $(\ref{eq:growthTimeLorentzWork})$ and $(\ref{eq:growthTimePoynting})$.  
Furthermore, the relative importance of these rates in determining the total magnetic energy growth rate becomes clear: since $\timeLorentzWork\ll\timePoynting$, accretion of magnetic energy through $\sPNS$ has virtually no effect on the growth of magnetic energy in $\vShock$.  

The discrepancy between the millisecond growth time predicted by Eq. (\ref{eq:growthRateLorentzWork}) and the numerically measured growth time ($\tau\approx66$~ms; Figure \ref{fig:overviewNonRotating}) suggests that numerical dissipation plays an important role in controlling the growth time for magnetic energy in our simulations.  
This is further supported by the results presented in Section \ref{sec:spectralAnalysis}, which show that the magnetic energy develops on spatial scales that are strongly affected by numerical dissipation (see also Appendix \ref{app:numericalDissipation}).  
If not suppressing field growth entirely, numerical dissipation tends to increase the magnetic energy growth time.  
The characteristic decay rate due to resistive dissipation of magnetic energy is
\begin{equation}
  \rateDissipation
  =
  \f{1}{E_{\mbox{\tiny mag}}}\int_{V}\f{1}{\mu_{0}}\vect{B}\cdot\curl{\left(\eta\vect{J}\right)}\,dV
  \approx
  \f{2\eta}{\lambdaDissipation^{2}}, 
  \label{eq:growthRateDissipation}
\end{equation}
where we have introduced the dissipation scale $\lambdaDissipation$.  
The decay time due to resistive dissipation is then
\begin{equation}
  \timeDissipation
  \approx
  R_{\mbox{\tiny m}}\left(\f{\lambdaDissipation}{\lambdaMeanMag}\right)^{2}\tau_{\vect{J}\times\vect{B}}, 
  \label{eq:growthTimeDissipation}
\end{equation}
where the magnetic Reynolds number is defined as $\reynoldsMag=\uRMSTurb\lambdaMeanMag/\eta$.  
The magnetic Reynolds number in the supernova environment is expected to be extremely large; on the order $10^{17}$ in the PNS \citep[][]{thompsonDuncan_1993}.  
As far as the magnetic energy growth rate is concerned, resistive effects are only relevant on very small scales, and the growth is most likely curbed by dynamical interactions with the fluid through magnetic tension forces before the magnetic field develops to resistive scales due to turbulent stretching of flux ropes \citep{thompsonDuncan_1993}.  
\citet{thompsonDuncan_1993} list (in their Table 1) the resistivity in the PNS convection zone ($\eta=1\times10^{-4}$~cm$^{2}$~s$^{-1}$).  
Adopting this value, the resistive decay time for a magnetic field varying on a spatial scale of, for example, $1$~m (i.e., much smaller than any scale resolved by our simulations) becomes very long ($\timeDissipation=5\times10^{7}$~s) compared to the explosion time ($\sim1$~s).  

Resistive effects are, however, important to consider in numerical MHD simulations of astrophysical systems.  
We do not explicitly include resistivity in our simulations, but the numerical scheme incorporates an effective numerical resistivity in the induction equation in order to stabilize the solution when discontinuities or underresolved gradients appear in the flow (see Appendix \ref{app:numericalDissipation} for further details).  
An approximation of the total growth rate can be obtained by combining Eqs. (\ref{eq:growthRateLorentzWork}) and (\ref{eq:growthRateDissipation}):
\begin{equation}
  \rateTotal
  \approx
  \rateLorentzWork
  \left[1-\f{1}{\reynoldsMag}\left(\f{\lambdaMeanMag}{\lambdaDissipation}\right)^{2}\right].  
  \label{eq:totalGrowthRate}
\end{equation}
In our simulations we have $\timeTotal\gg\timeLorentzWork$ and $\lambdaDissipation\lesssim\lambdaMeanMag$.  
Defined this way, the magnetic Reynolds number in our simulations is therefore somewhat larger than, but still close to, unity ($\reynoldsMag\gtrsim 1$).  
This conclusion is consistent with the observations from the energy spectrum plots above, which show that a sizable fraction of the magnetic energy resides on spatial scales where numerical diffusion is significant.  
Our simulations are therefore likely to grossly underestimate the magnetic energy growth rates that can be expected under more realistic physical conditions (i.e., where $\reynoldsMag\gg1$).  

We point out here that the saturation of magnetic energy observed in model $\model{13}{0.0}{00}$ does \emph{not} mean that $\rateLorentzWork\approx0$~s$^{-1}$ for this model.  
We find that the amplified magnetic fields in model $\model{13}{0.0}{00}$ result in about a $10\%$ reduction in $\rateLorentzWork$ relative to the weak-field model, which, because of numerical dissipation, results in a significant reduction in the total growth rate $\rateTotal$ (we expect $\rateDissipation$ to be the same in both models).  

In Appendix \ref{app:numericalDissipation} we measure numerically the magnetic energy decay rate due to resistive dissipation in one of our simulations.  
We find $\rateDissipation\approx380$~s$^{-1}$, which is comparable to, but still somewhat smaller than $\rateLorentzWork$.  
The decay rates $\rateDissipation$, along with $\lambdaDissipation(\approx\lambdaMeanMag)$ and $\eta_{\mbox{\tiny num}}$ (Appendix \ref{app:numericalDissipation}), and $\rateLorentzWork$ do not vary significantly with time during the highly nonlinear stage of the SASI.  
Thus, the numerically measured growth time ($\rateTotal\approx66$~ms; Figure \ref{fig:overviewNonRotating}) is mostly the result of two large and competing processes: growth due to $\rateLorentzWork$ and decay due to $\rateDissipation$.  
For increasing spatial resolution (i.e., increasing $\reynoldsMag$) we expect $\rateTotal\to\rateLorentzWork$ (cf. Eq. (\ref{eq:totalGrowthRate})).  

Indeed, we have carried out simulations with different spatial resolutions and measured the magnetic energy growth rate when an epoch of exponential growth can be identified. 
We find that the growth rate increases with increasing resolution: for the lowest resolution model ($\Delta l=2.34$~km) the exponential growth time is about 150~ms, while in a simulation with $\Delta l=0.78$~km the exponential growth time decreases to about 50~ms\footnote{To conserve computational resources, this model was not run to completion but until the computational domain consisted of $1280^{3}$ zones ($t\approx880$~ms).  At this time the SASI is still ramping up and the magnetic energy growing rapidly.}.  
In fact, a divergent increase in the magnetic energy growth rate with increasing magnetic Reynolds number (i.e., resolution) has been reported in direct numerical simulations of MHD turbulence \citep[e.g.,][]{haugen_etal_2004} and recently in simulations of turbulent star formation using adaptive mesh refinement \citep[e.g.,][]{syr_etal_2010,federrath_etal_2011}.  
These authors show results from simulations in which the resolution has been doubled several times, and they find that the magnetic energy growth rate increases as a power law with increasing magnetic Reynolds number.  

The sensitivity of the magnetic field evolution to numerical resolution does raise concerns about what aspects of our simulations are relevant to core-collapse supernovae.  
Dissipation due to finite grid resolution tends to suppress magnetic energy growth.  
At face value, our simulations (falsely) result a negative assessment on the efficiency of SASI-induced magnetic field amplification.  
However, as the resolution is increased the growth rate increases and the resulting magnetic fields become stronger.  
Further analysis suggests that the simulations grossly underestimate the growth rates and fields that may obtain in the supernova environment.  
The SASI-induced turbulent magnetic field amplification mechanism is a robust result from our simulations.  
Only the growth rate, saturation amplitude, and the dynamical impact of the amplified magnetic fields remain uncertain.  
An important consequence of the implied millisecond growth time is that any weak seed magnetic fields may be amplified to saturation levels ($|\vect{B}|\approx\sqrt{\mu_{0}\rho}|\vect{u}|$) in a core-collapse supernova if the SASI operates and drives vigorous turbulent flows below the shock.  
The kinetic energy available to amplify the magnetic energy (some fraction of $\EKinTurb$) is not sufficient for magnetic fields generated in this way to play a principal role in the explosion dynamics.  
We cannot, however, completely rule out the possibility that SASI-generated magnetic fields play a secondary role in the dynamics leading to core-collapse supernovae.  

\begin{figure*}
  \epsscale{1.0}
  \plottwo{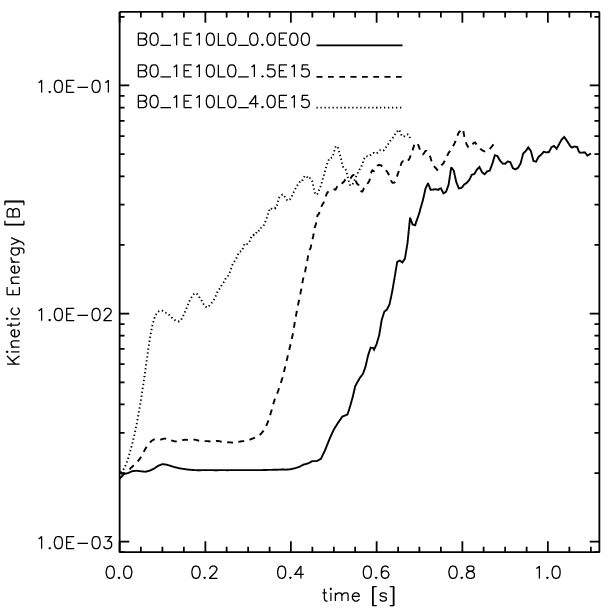}
                {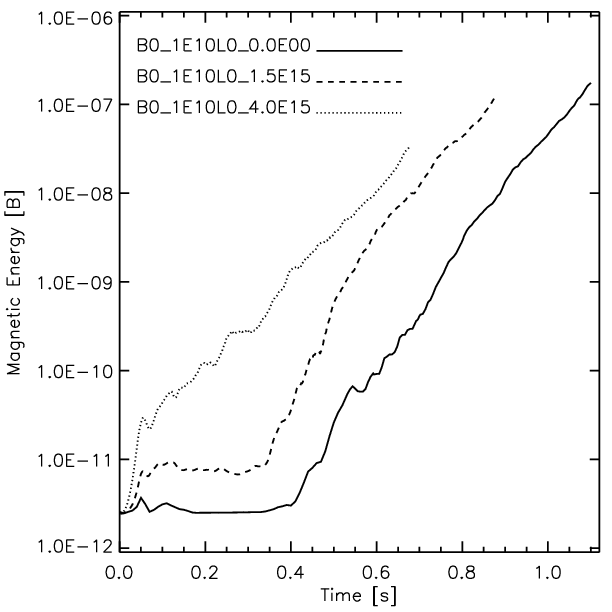}
  \caption{Kinetic energy (left) and magnetic energy (right) versus time from simulations with varying degree of initial rotation.  The specific angular momentum in the pre-shock flow has been set to $l=0.0$~cm$^{2}$~s$^{-1}$ (solid; $\model{10}{0.0}{00}$), $1.5\times10^{15}$~cm$^{2}$~s$^{-1}$ (dashed; $\model{10}{1.5}{15}$), and $4.0\times10^{15}$~cm$^{2}$~s$^{-1}$ (dotted; $\model{10}{4.0}{15}$).  The magnetic field strength at $r=\rPNS$ is initially $B_{0}=1\times10^{10}$~G in all the models.  \label{fig:kineticAndMagneticEnergy_rotating}}
\end{figure*}

\subsection{Simulations with Initial Rotation}
\label{sec:rotatingModels}

Results from rotating models are shown in Figure \ref{fig:kineticAndMagneticEnergy_rotating}, in which we plot kinetic energy (left panel) and magnetic energy (right panel) versus time.  
Rotating models with $B_{0}=1\times10^{10}$~G, and $l_{0}=1.5\times10^{15}$~cm$^{2}$~s$^{-1}$ ($\model{10}{1.5}{15}$; dashed lines) and $l_{0}=4.0\times10^{15}$~cm$^{2}$~s$^{-1}$ ($\model{10}{4.0}{15}$; dotted lines) are compared with the corresponding non-rotating model ($\model{10}{0.0}{00}$; solid lines).  

The most notable difference between these models is the earlier onset of the SASI observed in the rotating models.  
The post-shock flow is set into rotation about the $z$-axis as the pre-shock material with angular momentum advects downstream.  
The kinetic energy in the post-shock flow in model $\model{10}{1.5}{15}$ increases initially by $\sim50\%$, and settles momentarily into a quiescent state, which lasts for about 200~ms.  
Then, for $t\gtrsim300$~ms, the nonlinear phase of the SASI sets in, and the kinetic energy begins to grow exponentially with a growth time $\tau\approx55$~ms, which is notably faster than in the non-rotating model.  
Model $\model{10}{4.0}{15}$ receives a stronger initial perturbation due to more angular momentum ahead of the shock, and the kinetic energy in this model grows rapidly by a factor of $\sim5$ before settling into a short, quasi-steady state with $E_{\mbox{\tiny kin}}\sim10^{49}$~erg.  
The kinetic energy begins to grow again for $t\gtrsim200$~ms.  
The kinetic energy in all models eventually reaches similar levels in the strongly nonlinear phase of the SASI; when averaged over the last $200$~ms of each run we find $\timeAverage{E_{\mbox{\tiny kin}}}{0.68}{0.88}=0.052$~B and $\timeAverage{E_{\mbox{\tiny kin}}}{0.48}{0.68}=0.050$~Bs for models $\model{10}{1.5}{15}$ and $\model{10}{4.0}{15}$ respectively.  
(We reported $\timeAverage{E_{\mbox{\tiny kin}}}{0.9}{1.1}=0.051$~B for model $\model{10}{0.0}{00}$ in Section \ref{sec:timeGlobal}.)

The earlier onset of the nonlinear SASI in the rotating models is consistent with \citet{blondinMezzacappa_2007} and \citet{yamasakiFoglizzo_2008}.  
However, model $\model{10}{4.0}{15}$ is perturbed relatively hard when the rotating pre-shock material advects downstream, and the model does not settle into a quiescent state, as is observed in $\model{10}{1.5}{15}$.  
We think it is very likely that the early SASI development in model $\model{10}{4.0}{15}$ is partially a result of our method of initiating the rotating models.  
Nevertheless, the purpose of these simulations is to study the effect of rotation on turbulent magnetic field amplification during the non-linear phase, and our rotating models are suitable for this purpose.  

The evolution of the magnetic energy below the shock during nonlinear SASI-operation (right panel in Figure \ref{fig:kineticAndMagneticEnergy_rotating}) in the rotating models is not significantly different from model $\model{10}{0.0}{00}$.  
All models exhibit exponential magnetic energy growth during the late stages.  
The magnetic energy in model $\model{10}{1.5}{15}$ grows exponentially with a growth time $\tau\approx44$~ms during the early stages (from $t\approx340$~ms to $t\approx550$~ms), and grows at a rate similar to model $\model{10}{0.0}{00}$ later on ($t\gtrsim600$~ms).  
The magnetic energy in model $\model{10}{4.0}{15}$ grows exponentially at a somewhat slower rate than the other models ($\tau\approx85$~ms).  
However, all models have reached similar levels at the end of the respective runs.  
In particular, we find $E_{\mbox{\tiny mag}}\approx1.2\times10^{-7}$~B ($t=878$~ms) for $\model{10}{1.5}{15}$ and $E_{\mbox{\tiny mag}}\approx3.4\times10^{-8}$~B ($t=678$~ms) for $\model{10}{4.0}{15}$.  

\section{MAGNETIZATION OF PROTONEUTRON STARS}
\label{sec:pnsMagnetization}

In \citetalias{endeve_etal_2010} we pointed out that the underlying PNS may be significantly magnetized due to SASI-induced magnetic field amplification.  
In this section we estimate in a similar manner the degree of PNS magnetization predicted by the current set of simulations.  

Adopting Eq. (\ref{eq:magneticEnergyEquation}), the magnetic energy in the volume occupied by the PNS $\vPNS$ at some time $t>t_{0}$ is
\begin{eqnarray}
  E_{\mbox{\tiny mag}}(t)
  &=&E_{\mbox{\tiny mag}}(t_{0}) \nonumber \\
  & &
  +\int_{t_{0}}^{t}\,dt'
  \left(
    \int_{V_{\mbox{\tiny PNS}}}\left(W_{\mbox{\tiny L}}-Q_{\mbox{\tiny J}}\right)\,dV 
  \right. \nonumber \\
  & &
  \left.
    \hspace{1.5cm}
   -\int_{\partial V_{\mbox{\tiny PNS}}}\vect{P}\cdot d\vect{S}
  \right). 
  \label{eq:pnsMagneticEnergy}
\end{eqnarray}
Here $\vect{P}$ is the Poynting flux through the surface of the PNS, 
and $W_{\mbox{\tiny L}}$ and $Q_{\mbox{\tiny J}}$ are obtained from $\vect{u}$ and $\vect{B}$, which must be computed with an appropriate physical model of the PNS.  
The dissipative term $Q_{\mbox{\tiny J}}$ also involves the resistivity $\eta$.  
Resistive dissipation is not likely to suppress field amplification in the PNS \citep{thompsonDuncan_1993}, but may be important to the long-term evolution of the neutron star magnetic field (strength and topology).  

Evaluation of the volume integral on the right-hand-side of Eq. (\ref{eq:pnsMagneticEnergy}) involves numerical simulations of the hydro-magnetic evolution inside the PNS during the explosion phase of core-collapse supernovae and subsequent PNS cooling, and includes neutrino radiation-magnetohydrodynamic simulations of dense nuclear matter.  
Such calculations are well beyond the scope of this study.  
Earlier works have suggested numerous mechanisms for field amplification in the PNS, including winding by differential rotation \citep[e.g.,][]{wheeler_etal_2002}; the magneto-rotational instability \citep{akiyama_etal_2003}; and convective dynamo action, driven by entropy gradients, lepton gradients, or both \citep[e.g.,][]{thompsonDuncan_1993,bonanno_etal_2003,bonanno_etal_2005}.  
All these mechanisms operate inside or on the surface of the PNS and rely on rotation.  
We exclude the PNS from our simulations and do not address field amplification mechanisms in its interior.  
Our simulations, however, focus on field amplification by the SASI exterior to the PNS, which is often ignored in models addressing the origin of pulsar magnetism.  

From our simulations we compute the increase in magnetic energy in the volume occupied by the PNS due to the Poynting flux through the surface bounding it,
\begin{equation}
  E_{\mbox{\tiny mag},\vect{P}}(t)
  =-\int_{t_{0}}^{t}\,dt'\int_{\partial V_{\mbox{\tiny PNS}}}\vect{P}\cdot d\vect{S}.  
  \label{eq:pnsMagneticEnergyPoyntingFlux}
\end{equation}
We then estimate the PNS magnetic field due to SASI activity $\langle B_{\mbox{\tiny PNS},\vect{P}}\rangle=(2\mu_{0}E_{\mbox{\tiny mag},\vect{P}}/V_{\mbox{\tiny PNS}})^{1/2}$ \citepalias[cf. Eq. (18) in][]{endeve_etal_2010}.  Results from these estimates for the rotating and non-rotating models with varying initial magnetic field strengths are listed in Table \ref{tab:pnsMagnetization}.  

\begin{table}
  \begin{center}
    \caption{PNS Magnetic field estimates.  \label{tab:pnsMagnetization}}
    \begin{tabular}{cccc}
    Model & $t_{\mbox{\tiny end}}$ (ms) & $E_{\mbox{\tiny mag},\vect{P}}$ (Erg) & $\langle B_{\mbox{\tiny PNS},\vect{P}}\rangle$ (G) \\
    \tableline
    \tableline
    \model{10}{0.0}{00} & 1100 & $1.14\times10^{44}$ & $3.3\times10^{12}$  \\
    \model{10}{1.5}{15} & 878 & $6.70\times10^{43}$ & $2.5\times10^{12}$  \\
    \model{10}{4.0}{15} & 678 & $2.52\times10^{43}$ & $1.5\times10^{12}$ \\
    \tableline
    \model{12}{0.0}{00} & 1126 & $3.16\times10^{47}$ & $1.7\times10^{14}$  \\
    \model{12}{1.5}{15} & 1000 & $1.15\times10^{48}$ & $3.3\times10^{14}$  \\
    \model{12}{4.0}{15} &   644  & $1.74\times10^{47}$ & $1.3\times10^{14}$ \\
    \tableline
    \model{13}{0.0}{00} & 1100 & $4.49\times10^{48}$ & $6.5\times10^{14}$ \\
    \tableline
    \tableline
    \end{tabular}
    \tablecomments{Magnetic energy accumulated on the proto-neutron star in computed models.  The inferred magnetic field strength resulting from SASI-induced magnetic field amplification is also listed.}
  \end{center}
\end{table}

Our results show that the magnetic energy generated by SASI activity may result in significant magnetization of the PNS.  
The magnetic energies generated in some of the models meet the energy requirements to power the total flare energy released per SGR \emph{and} the persistent X-ray emission \citep{thompsonDuncan_2001}.  
The models with the weakest initial magnetic field predict field strengths in the range of ordinary pulsars (a few $\times10^{12}$~G), while the models with stronger initial magnetic fields predict fields in the magnetar range (exceeding $10^{14}$~G).  
The magnetic field in the strong-field model ($\model{13}{0.0}{00}$) saturates dynamically, and this model may represent an upper limit to the fields attainable from this process.  
On the other hand, the weak-field models do not reach saturation.  
The magnetic energy in these models continues to grow at an underestimated rate, and the maximum attainable field strength/energy is also limited by finite grid resolution.  
The PNS field strengths predicted by these models are therefore artificially low.  
It then seems likely, given infinite grid resolution, that PNS magnetic fields can exceed $10^{14}$~G due to the SASI alone, independent of the initial magnetic field strength.  
Moreover, as finite grid resolution severely limits the exponential growth rate of magnetic energy, the duration of SASI operation may be less critical.  
The amount of initial rotation in the models does not seem to affect the degree of PNS magnetization.  

The field strengths listed in Table \ref{tab:pnsMagnetization} should also be corrected for additional magnetic field amplification as the PNS cools and contracts.  
From conservation of magnetic flux through the PNS surface, contraction from a 40~km radius to a radius of about 15~km boosts the surface field by a factor of $\sim7$.  

We point out that the PNS magnetic fields resulting from the turbulent flows driven by the SASI are likely small-scale and disordered.  
A connection to the dipolar magnetic field structure inferred for neutron stars is currently missing, and, of course, the SASI alone cannot fully explain the origin of pulsar magnetism.  
However, pulsar magnetic fields are thought to consist of a global dipole field superimposed with higher order multipole (small-scale) components, and pulsar magnetism is most likely a result of the combined action of multiple amplification mechanisms.  
While the inferences we can make are limited by resolution (which affects the magnetic growth rate) and the absence of important physics (which determines the time to explosion), our simulations suggest that the SASI could in principle make a nontrivial contribution.  

\section{SUMMARY, DISCUSSION, AND CONCLUSIONS}
\label{sec:discussionConclusions}

We present results from three-dimensional MHD simulations of the SASI.  
The simulations are initiated from a configuration that resembles the early stalled shock phase in a core-collapse supernova, albeit with simplified physics that excludes critical components of a supernova model (e.g., neutrino transport, self-gravity, and the PNS itself).
On the other hand, our simulations are computed with a spatial resolution that is currently inaccessible to state-of-the-art supernova models in three spatial dimensions, and they may therefore provide valuable insight into MHD developments in core-collapse supernovae.  
In particular we study the evolution and amplification of magnetic fields in SASI-driven flows in order to assess the effects of the amplified magnetic fields on supernova dynamics, and the possible role of the SASI in magnetizing the PNS.  

This paper is a continuation and extension of the study initiated in \citetalias{endeve_etal_2010}.  
The simulations reported here were performed with higher spatial resolution (up to $1280^{3}$ grid cells) and cover a broader parameter range than the 3D simulations presented in \citetalias{endeve_etal_2010}: we have varied the strength of the initial magnetic field and the degree of rotation in the flow ahead of the shock (including no rotation).  
We have also varied the spatial resolution in some of the simulations, and extended the analysis from \citetalias{endeve_etal_2010} to include a Fourier decomposition of the kinetic energy and magnetic energy in the simulations.  
Our main findings are
\begin{itemize}
  \item[1.] The SASI-driven turbulence that develops is essentially non-helical, and shares similarities with convectively driven MHD turbulence \citep[e.g.,][]{brandenburg_etal_1996}.  
  (See also ``box turbulence" simulations by \cite{haugen_etal_2004}.)  
  When corrected for density stratification, the kinetic energy spectra associated with the post-shock flow develop Kolmogorov-like $-5/3$ scaling (i.e., $\edensk{kin}^{\,\mbox{\tiny II}},\edensk{kin}^{\,\mbox{\tiny III}}\propto k^{-5/3}$; Section \ref{sec:spectralAnalysis}) in a narrow wavenumber range.  
  Moreover, inspection of the time evolution of the kinetic energy spectra reveals that the power in low-order SASI modes (i.e., large scale flows) cascades to higher-order modes (i.e., smaller-scale flows), and that a significant fraction (up to $\sim50\%$) of the post-shock kinetic energy can be associated with turbulence (although a smaller fraction is involved in magnetic field amplification).  
  This further suggests that the non-linear SASI saturates due to the development of turbulence via secondary instabilities \citep[e.g.,][]{guilet_etal_2010}.  
  
  \item[2.] The magnetic energy grows exponentially with time in turbulent flows driven by the SASI, as long as the kinematic regime obtains.  
  Our simulations develop flows characteristic of the SASI spiral mode \citep[e.g.,][]{blondinMezzacappa_2007}.  
  These flows drive vigorous turbulence below the shock ($\uRMSTurb\sim4000$~km~s$^{-1}$), which amplifies magnetic fields by stretching.  
  The resulting magnetic field is highly intermittent and consists of thin, intense magnetic flux ropes.  

  \item[3.] Simulations initiated with weak or moderate rotation evolve similarly to non-rotating models as far as the magnetic field amplification mechanism is concerned.  
  However, models with initial rotation develop the nonlinear spiral SASI flows earlier, and exponential magnetic energy growth sets in sooner.  
  The earlier onset of the SASI in models with initial rotation is consistent with the results of \citet{blondinMezzacappa_2007} and \citet{yamasakiFoglizzo_2008}.  

  \item[4.] The magnetic energy grows at the expense of the kinetic energy available in the turbulent flows driven by the SASI.
  Our simulations show that strong magnetic fields emerge on small (turbulent) spatial scales, and reduce the turbulent kinetic energy on those scales.  
  For our reference spatial resolution, magnetic fields impact flows on scales with wavenumber $k>\kDynamic\approx0.1-0.2$~km$^{-1}$ ($\lambdaDynamic=2\pi/\kDynamic\lesssim30-60$~km) and peak around $k=0.6$~km$^{-1}$ ($\sim10$~km) (Figure \ref{fig:spectralKineticEnergyDensityNonRotating}).  
  That is, magnetic fields do not affect the portion of the kinetic energy spectrum with $k\lesssim\kDynamic$.  
  The turbulent kinetic energy (that is, the kinetic energy on spatial scales below some specified cutoff) in models with larger magnetic fields is reduced compared to models initiated with weaker magnetic fields, indicating a dynamical impact of the amplified magnetic field.  

  \item[5.] The magnetic field evolution in our simulations remains very sensitive to the spatial resolution.  
  Key parameters extracted from simulations performed with increasing spatial resolution do not converge in the range covered in this study.  
  Both the final magnetic energy attained and the rate at which the magnetic energy grows increase with increasing grid resolution.  
  In particular, estimates using data extracted from our simulations suggest that the magnetic energy may grow exponentially on a millisecond timescale under physically realistic conditions, with very large magnetic Reynolds numbers, as opposed to the $\sim50$-$60$~ms timescale measured directly in our runs.  

  \item[6.] The magnetic energy saturates when the magnetic energy density becomes comparable to the kinetic energy density (i.e., $|\vect{B}|/\sqrt{\mu_{0}\rho}\gtrsim|\vect{u}|$) in localized regions of the flow.  
  Only our ``strong-field'' model (with the largest initial magnetic field) reaches this saturated state.  
  The subsequent magnetic field evolution remains highly dynamic: strong fields are advected through the flow, are temporarily weakened, and then reemerge in a seemingly stochastic manner.

  \item[7.] The magnetic fields amplified by the SASI are not likely to play an important role in the explosion dynamics (but see further discussion below).  
  The presence of amplified magnetic fields does not result in noticeable effects on the global shock dynamics in our simulations, and this can be understood as a matter of simple energetics.  
  Magnetic energy grows at the expense of kinetic energy, and the kinetic energy content in the post-shock flow during vigorous SASI activity ($\sim5\times10^{-2}$~B) is not enough for magnetic fields to become energetically significant to the explosion ($\sim1$~B).  
  This was also pointed out in \citetalias{endeve_etal_2010}.  
  Moreover, the turbulent kinetic energy---which powers SASI-driven field amplification---accessible for magnetic field amplification only amounts to about $10\%$ of the total kinetic energy below the shock.  
  We further point out that our estimate for turbulent kinetic energy is \emph{not} critically sensitive to the numerical resolution (Section \ref{sec:spectralAnalysis} and Figure \ref{fig:turbulentKineticEnergy}).  
  A rapidly rotating (millisecond period) PNS would provide an energy reservoir large enough to power magnetically-driven explosions \citep[e.g.,][]{burrows_etal_2007}, but it is not likely that rotation would be this strong in most supernova progenitors \citep[e.g.,][]{heger_etal_2005}.  
  These observations suggest a rather passive role of magnetic fields in the overall dynamics of at least most supernovae.  

  \item[8.] Our simulations suggest that SASI-induced magnetic field amplification may play an important role in determining the strength of the magnetic field in proto-neutron stars and young pulsars.  
  Upon integrating the Poynting flux through the surface encompassing the PNS, we estimate that the magnetic energy accumulated on the PNS may account for magnetic field strengths exceeding $10^{14}$~G.  
  This is stronger than the canonical dipole field inferred for typical pulsars, and in this connection two points must be emphasized. 
  First, SASI-driven amplification is expected to cease when the explosion takes off, so that different delay times to explosion (which may for example be a function of progenitor mass) may result in different degrees of PNS magnetization. 
  Second, the SASI-amplified portion of the field accumulated by the PNS will at least initially be disordered and not of the large-scale, dipolar character of the fields inferred from pulsar spindown.  
\end{itemize}

Despite the pessimism of point 7 above regarding the relevance of SASI-amplified magnetic fields to the explosion mechanism, we caution that the sensitivity of magnetic field amplification and evolution to numerical resolution prevents us from completely dismissing magnetic fields as unimportant to supernova dynamics in weakly rotating progenitors.  

Certainly, our simulations cannot accurately describe the dynamical interaction between the magnetic field and the fluid on small scales.
An initially weak magnetic field is amplified exponentially in turbulent flows when the flux tubes are stretched and their cross sectional area decreases.  
In a realistic post-shock supernova environment, where $\reynoldsMag$ is extremely large, field amplification is likely quenched by dynamic back-reaction on the fluid before the flux tube thickness reaches the resistive scale \citep{thompsonDuncan_1993}.  
The resistive decay time then remains much longer than the dynamical timescale of hydro-magnetic interactions.  
But in numerical simulations the flux tube cross section inevitably approaches the grid scale, and numerical dissipation sets in and prevents further strengthening of the magnetic field.  
This occurs in all our simulations.  
(The strong-field model ($\model{13}{0.0}{00}$) develops dynamically relevant magnetic fields, but is also strongly affected by numerical dissipation.)  

Our simulations suggest that magnetic fields become dynamically relevant on spatial scales smaller than $\lambdaDynamic\sim30$~km (Figure \ref{fig:spectralKineticEnergyDensityNonRotating}).  
The global shock dynamics remains unaffected by the presence of magnetic fields (e.g., Figure \ref{fig:overviewNonRotating}).  
However, we cannot rule out the possibility that flows on scales larger than $\lambdaDynamic$ could ultimately be affected by hydro-magnetic interactions emerging from small-scale turbulent flows.  
Simulations of non-helical MHD turbulence \citep[e.g.,][]{haugen_etal_2004} show that the magnetic energy grows exponentially on the turnover time, on all spatial scales during the kinematic regime.  
(We also observe exponential growth on all scales in our runs during this regime.)  
The kinematic regime ends when the magnetic energy becomes comparable to the kinetic energy.  
This occurs on a scale by scale basis.  
Magnetic energy growth slows down considerably after this equipartition, which occurs first on the smallest spatial scales, and the magnetic energy spectrum settles somewhat above the kinetic energy spectrum.  
(We also observe that magnetic energy growth is quenched when $\edens{mag}\sim\edens{kin}$, but the magnetic energy spectrum stays below the kinetic energy spectrum for all $k$ in our simulations.)  
At later times in MHD turbulence simulations, the largest spatial scale at which $\edensk{mag}\gtrsim\edensk{kin}$ (i.e., $\lambdaDynamic$) increases, and may approach the driving scale of the turbulent forcing.  
For helical MHD turbulence, which may be more relevant when a rapidly rotating PNS is included in the model, $\lambdaDynamic$ can even grow beyond the forcing scale \citep[e.g.,][]{meneguzzi_etal_1981,brandenburg_2001}.  
However, the timescale for this process is relatively slow, and increases with $\reynoldsMag$ \citep{brandenburg_2001}.  
Nevertheless, it would be desirable to determine the largest scale at which the magnetic energy equilibrates with the kinetic energy in SASI-driven flows.  
The lack of sufficient spectral resolution in our simulations prevents us from determining whether magnetic fields can become strong on large enough scales to alter post-shock flows in a significant way.  

The SASI may play an important role in improving the conditions for successful neutrino-driven explosions \citep[e.g.,][]{bruenn_etal_2006,buras_etal_2006,mezzacappa_etal_2007,marekJanka_2009,suwa_etal_2010,muller_etal_2012}.  
If amplified magnetic fields can alter the evolution of the SASI and change the conditions (making them more, or less, favorable) for energy deposition by neutrinos, then magnetic fields may play a secondary but relevant role in the dynamics of a broader range of core-collapse supernovae (i.e. not just those arising from rapidly rotating progenitor stars).  
This point was also argued by \citet{obergaulingerJanka_2011}, who studied magnetic field amplification in non-rotating collapsed stellar cores using axisymmetric simulations that included the PNS and neutrino transport.  
They found that the SASI and convection contribute to magnetic field amplification, and observed the most pronounced shock expansion in the model where the magnetic field was strong enough to alter the post-shock flow topology.  
(This model was initiated with a strong pre-collapse magnetic field.)  

However, axial symmetry severely constrains magnetic field evolution driven by the SASI \citepalias[see][]{endeve_etal_2010} and, most likely, also convectively driven field amplification.   
Simulations similar to those of \citet{obergaulingerJanka_2011} in full 3D, where the SASI spiral mode can develop and drive turbulent field amplification, are therefore highly desired.  
Such simulations will improve on our simulations in (at least) two important ways:
\begin{itemize}
  \item[1.] A significant amount of magnetic energy (comparable to that in $\vShock$) is lost through the boundary at $r=\rPNS$ in our models, and not accounted for in the subsequent dynamics.  
  3D simulations with the PNS included do not suffer from this artificial limitation, and will allow us to better assess the role of SASI-induced magnetic fields.  
  \item[2.] Simulations that include neutrino transport develop neutrino-driven convection, both in the PNS and in the shocked mantle.  This convective activity will impact the evolution of magnetic fields, and possibly also the SASI.  We will then be able to study magnetic field evolution in a much more physically realistic supernova environment.  
Moreover, with neutrino transport included, we will be able to directly address the role of magnetic fields on neutrino-powered explosions.  
\end{itemize}

The constraint on numerical resolution in order to properly describe turbulent flows may still be computationally prohibitive, especially when additional (necessary) physics components are added to the models.  
This may be partially circumvented with the use of adaptive mesh refinement techniques and improved numerical algorithms.  
Local (or semi-global) simulations \citep[e.g.,][]{obergaulinger_etal_2009}, adopting physical conditions and forcing functions relevant to the supernova environment (i.e., derived from global multi-physics simulations), may also be necessary to study turbulent magnetic field evolution and its impact on supernova dynamics in more detail.  
More investigations, using both local and global simulations, are needed to better understand the role of magnetic fields in core-collapse supernovae.  

In summary, we conclude from our simulations that magnetic fields in core-collapse supernovae may be amplified exponentially by turbulence on a millisecond timescale; i.e., much shorter than the time between core bounce/shock formation and initiation of the explosion.  
Details of the impact on explosion dynamics by SASI-amplified magnetic fields remain unclear, but on energetic grounds alone the role of these magnetic fields is likely sub-dominant.  
The simulations further suggest that small-scale neutron star magnetic fields in the $10^{14}-10^{15}$~G range may be formed, which may be sufficient to power some of the energetic activity that define AXPs and SGRs.  

\acknowledgments

This research was supported by the Office of Advanced Scientific Computing Research and the Office of Nuclear Physics, U.S. Department of Energy.  
This research used resources of the Oak Ridge Leadership Computing Facility at the Oak Ridge National Laboratory provided through the INCITE program.  
We are grateful for support from members of the National Center for Computational Sciences during the execution and analysis of the simulations, especially Bronson Messer.  
We also thank an anonymous referee for comments that helped us improve the manuscript.

\appendix

\section{DISSIPATION OF MAGNETIC ENERGY IN THE NUMERICAL SIMULATIONS}
\label{app:numericalDissipation}

Here we briefly describe the dominant source of magnetic energy dissipation in our numerical simulations.  
We evolve the MHD equations with a second-order, semi-discrete, central-upwind, finite volume scheme for hyperbolic conservation laws, combined with the constrained transport (CT) method of \citet{evansHawley_1988} for divergence-free magnetic field evolution \citep[see][and the references therein; in particular, the MC-HLL-UCT scheme in \citet{londrilloDelZanna_2004}]{kurganov_etal_2001,londrilloDelZanna_2004}.  
Furthermore, we adopt the HLL Riemann solver \citep{harten_etal_1983} to compute the fluxes and electric fields needed to evolve the system of MHD equations.  
The HLL Riemann solver considers only the fastest left- and right-propagating characteristic waves of the underlying hyperbolic system (fast magnetosonic waves for MHD).  
This approximation results in diffusive evolution of intermediate waves (e.g., slow magnetosonic, Alfv{\'e}n, and entropy waves) and is the main source of dissipation in our simulations.  

The discretization of the computational domain results in cubic computational cells with sides $\Delta x=\Delta y=\Delta z=\Delta l=L/N$.  
We adopt standard finite volume index notation to associate variables with the cells in the Cartesian grid: the coordinates of the geometric center of a cell with index triplet ($i,j,k$) are denoted ($x_{i},y_{j},z_{k}$).  
Finite volume variables centered on the geometric center are also assigned the index triplet.  
Superscripts $n$ and $n+1$ denote time states, and the time step $\Delta t$ increments time from $t^{n}$ to $t^{n+1}$.  
For example, the volume-averaged (in angle brackets) $x$-component of the velocity in the cell at time $t^{n}$ is denoted $\langle u_{x} \rangle_{i,j,k}^{n}$.  
Magnetic field components are centered on the faces of computational cells in the CT method.  
For example, the $x$-component of the area-averaged magnetic field, centered on the outer face of cell ($i,j,k$) with coordinates ($x_{i+\f{1}{2}},y_{j},z_{k}$) at time $t^{n}$, is denoted $\langle B_{x} \rangle_{i+\f{1}{2},j,k}^{n}$.  
(For uniform grid spacing we have $x_{i+\f{1}{2}}=x_{i}+\Delta x/2$.)  

An integration of the magnetic induction equation over the cell face with normal parallel to the $x$-coordinate direction and time interval $\Delta t$ results in (after application of Stoke's theorem and replacing time-integrals of electric field components with the rectangle rule) the time-explicit finite volume update formula for the area-averaged $x$-component of the magnetic field
\begin{equation}
  \langle B_{x} \rangle_{i+\f{1}{2},j,k}^{n+1}
  =\langle B_{x} \rangle_{i+\f{1}{2},j,k}^{n}
  +\f{\Delta t}{\Delta z}\left(\langle E_{y} \rangle_{i+\f{1}{2},j,k+\f{1}{2}}^{n}-\langle E_{y} \rangle_{i+\f{1}{2},j,k-\f{1}{2}}^{n}\right)
  -\f{\Delta t}{\Delta y}\left(\langle E_{z} \rangle_{i+\f{1}{2},j+\f{1}{2},k}^{n}-\langle E_{z} \rangle_{i+\f{1}{2},j-\f{1}{2},k}^{n}\right), 
  \label{eq:finiteVolumeBx}
\end{equation}
where the face-averaged magnetic field is
\begin{equation}
  \langle B_{x} \rangle_{i+\f{1}{2},j,k}^{n}
  =\f{1}{\Delta y\Delta z}\int_{y_{j-\f{1}{2}}}^{y_{j+\f{1}{2}}}\int_{z_{k-\f{1}{2}}}^{z_{k+\f{1}{2}}}B_{x}(x_{i+\f{1}{2}},y,z,t^{n})\,dy\,dz, 
\end{equation}
and the line-averaged $z$-component of the electric field (centered on the cell-edge) is
\begin{equation}
  \langle E_{z} \rangle_{i+\f{1}{2},j+\f{1}{2},k}^{n}
  =\f{1}{\Delta z}\int_{z_{k-\f{1}{2}}}^{z_{k+\f{1}{2}}}E_{z}(x_{i+\f{1}{2}},y_{j+\f{1}{2}},z,t^{n})\,dz.  
\end{equation}
Update formulae for the other magnetic field components are obtained in an analogous manner.  
The update given by Eq. (\ref{eq:finiteVolumeBx}) is exactly the forward Euler method and results in first-order temporal accuracy.  
We obtain second-order temporal accuracy with a Runge-Kutta method \citep[e.g.,][]{shu_1997}.  

The key to stable and accurate magnetic field evolution with the CT method is the specification of the edge-centered electric field components.  
The $z$-component of the edge centered electric field with spatial coordinates $(x_{i+\f{1}{2}},y_{j+\f{1}{2}},z_{k})$ is computed with an HLL-type formula \citep[cf.][]{londrilloDelZanna_2004}
\begin{eqnarray}
  \langle E_{z} \rangle_{p}^{n}
  &=&
  \f{\alpha_{x}^{+}\alpha_{y}^{+}\left[ E_{z}^{\mbox{\tiny SW}} \right]_{p}^{n}
     +\alpha_{x}^{+}\alpha_{y}^{-}\left[ E_{z}^{\mbox{\tiny NW}} \right]_{p}^{n}
     +\alpha_{x}^{-}\alpha_{y}^{+}\left[ E_{z}^{\mbox{\tiny SE}} \right]_{p}^{n}
     +\alpha_{x}^{-}\alpha_{y}^{-}\left[ E_{z}^{\mbox{\tiny NE}} \right]_{p}^{n}}
  {\left(\alpha_{x}^{+}+\alpha_{x}^{-}\right)\left(\alpha_{y}^{+}+\alpha_{y}^{-}\right)} \nonumber \\
  & &
  +\f{\alpha_{x}^{+}\alpha_{x}^{-}}{\left(\alpha_{x}^{+}+\alpha_{x}^{-}\right)}
  \left(\left[ B_{y}^{\mbox{\tiny E}} \right]_{p}^{n}-\left[ B_{y}^{\mbox{\tiny W}} \right]_{p}^{n}\right)
  -\f{\alpha_{y}^{+}\alpha_{y}^{-}}{\left(\alpha_{y}^{+}+\alpha_{y}^{-}\right)}
  \left(\left[ B_{x}^{\mbox{\tiny N}} \right]_{p}^{n}-\left[ B_{x}^{\mbox{\tiny S}} \right]_{p}^{n}\right), 
  \label{eq:HLLElectricField}
\end{eqnarray}
where $\alpha_{x}^{\pm}=\max(0,\pm\lambda_{x}^{\pm,\mbox{\tiny SW}},\pm\lambda_{x}^{\pm,\mbox{\tiny NW}},\pm\lambda_{x}^{\pm,\mbox{\tiny SE}},\pm\lambda_{x}^{\pm,\mbox{\tiny NE}})$, and $\lambda_{x}^{\pm}=u_{x}\pm c_{x}^{f}$ are the characteristic wave speeds associated with the fast magnetosonic modes.  
The maximum is taken over wave speeds computed in the four cells sharing the edge indexed ($i+\f{1}{2},j+\f{1}{2},k$), which are denoted with superscripts $\mbox{SW}$, cell ($i,j,k$); $\mbox{NW}$, cell ($i,j+1,k$); $\mbox{SE}$, cell ($i+1,j,k$); and $\mbox{NE}$, cell ($i+1,j+1,k$), respectively.  
(We have simplified the notation in Eq. (\ref{eq:HLLElectricField}) by replacing the subscript indices $i+\f{1}{2},j+\f{1}{2},k$ with $p$.)  
For example, for first-order spatial accuracy we simply set $\left[ E_{z}^{\mbox{\tiny SW}} \right]_{i+\f{1}{2},j+\f{1}{2},k}^{n}=\langle B_{x} \rangle_{i+\f{1}{2},j,k}^{n} \langle u_{y} \rangle_{i,j,k}^{n}-\langle u_{x} \rangle_{i,j,k}^{n}\langle B_{y} \rangle_{i,j+\f{1}{2},k}^{n}$.  
Magnetic field components are centered on cell faces, and cells indexed ($i+1,j,k$) and ($i+1,j+1,k$) share $\left[B_{y}^{\mbox{\tiny E}}\right]_{i+\f{1}{2},j+\f{1}{2},k}^{n}=\langle B_{y} \rangle_{i+1,j+\f{1}{2},k}^{n}$, which is assigned superscript $\mbox{E}$.  
Similarly, cells ($i,j,k$) and ($i,j+1,k$) share $\left[B_{y}^{\mbox{\tiny W}}\right]_{i+\f{1}{2},j+\f{1}{2},k}^{n}=\langle B_{y} \rangle_{i,j+\f{1}{2},k}^{n}$, which is assigned superscript $\mbox{W}$.  
To simplify the presentation, we only briefly discuss the first-order scheme.  
However, all our calculations are done with a second-order scheme.  
For second-order spatial accuracy we use monotonic linear interpolation (via the multivariable minmod limiter) to reconstruct variables to the appropriate edges \citep[e.g.,][and references therein]{kurganov_etal_2001}, and evaluate the electric field through Eq. (\ref{eq:HLLElectricField}).  

The HLL electric field given by Eq. (\ref{eq:HLLElectricField}) contains explicit dissipation due to the second and third terms on the right-hand-side.  
(We refer to the first term on the right-hand side as the ideal part of the electric field.)  
The dissipation terms mimic the effect of physical resistivity due to a non-ideal electric field $-\vect{u}\times\vect{B}+\eta \vect{J}$.  
This becomes evident by considering a subsonic flow with a weak magnetic field (appropriate for the turbulent post-shock flows in our simulations).  
Then, $\lambda_{x}^{\pm}\approx\pm c_{x}^{f}\approx\pm c_{S}$ and $\alpha_{x}^{\pm}\approx c_{S}$, where $c_{S}$ is the sound speed.  
(Similarly we have $\alpha_{y}^{\pm}\approx c_{S}$.)
With these approximations the electric field in Eq. (\ref{eq:HLLElectricField}) becomes
\begin{eqnarray}
  \langle E_{z} \rangle_{p}^{n}
  &=&
  \f{1}{4}
  \left(
    \left[ E_{z}^{\mbox{\tiny SW}} \right]_{p}^{n}
    +\left[ E_{z}^{\mbox{\tiny NW}} \right]_{p}^{n}
    +\left[ E_{z}^{\mbox{\tiny SE}} \right]_{p}^{n}
    +\left[ E_{z}^{\mbox{\tiny NE}} \right]_{p}^{n}
  \right) \nonumber \\
  & &+\f{\eta_{\mbox{\tiny num}}}{\mu_{0}}
  \left(
    \left[\langle B_{y} \rangle_{i+1,j+\f{1}{2},k}^{n}-\langle B_{y} \rangle_{i,j+\f{1}{2},k}^{n}\right]/\Delta x
    -\left[\langle B_{x} \rangle_{i+\f{1}{2},j+1,k}^{n}-\langle B_{x} \rangle_{i+\f{1}{2},j,k}^{n}\right]/\Delta y
  \right), 
  \label{eq:HLLElectricFieldApproximate}
\end{eqnarray}
where the ``numerical resistivity'' $\eta_{\mbox{\tiny num}}=\f{1}{2}\,c_{S}\,\Delta l\,\mu_{0}$ decreases linearly with $\Delta l$.  
In particular, we note that the terms inside the parenthesis of the second term on the right-hand side of Eq. (\ref{eq:HLLElectricFieldApproximate}) is a numerical approximation for $(\curl{\vect{B}})_{z}=\mu_{0}J_{z}$ at ($x_{i+\f{1}{2}},y_{j+\f{1}{2}},z_{k}$).  
(Thus, the analogy to non-ideal MHD with scalar resistivity used in the discussion in Section \ref{sec:magneticEnergyGrowthRates} is appropriate.)  
These terms prevent growth of grid scale oscillations and act to stabilize the evolution of the magnetic field.  
However, this numerical dissipation becomes non-negligible when the magnetic field develops a flux rope structure with the flux rope thickness comparable to a few grid cells.  

\begin{figure}
  \epsscale{1.0}
  \plotone{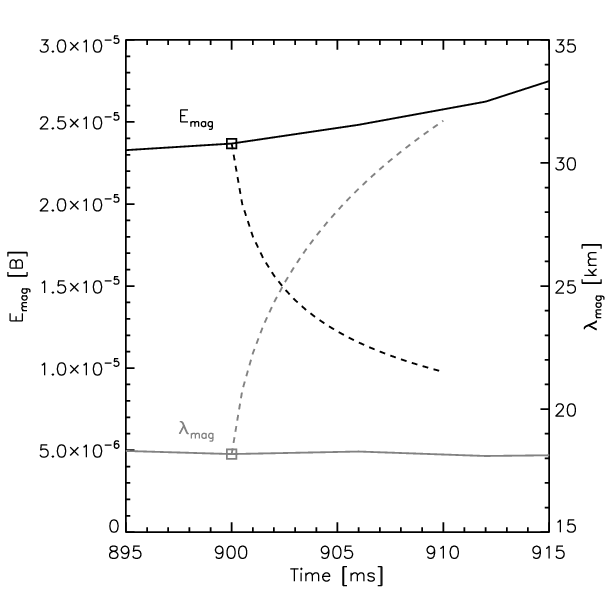}
  \caption{Evolution of the total magnetic energy (black lines) and the characteristic spatial scale of the magnetic field $\lambdaMeanMag$ (gray lines; cf. Eq. (\ref{eq:meanMagneticWaveNumber}) and proceeding text).  Solid lines represent data from model $\model{12}{0.0}{00}$, while dashed lines are from a simulation initiated from model $\model{12}{0.0}{00}$ at $t=900$~ms and evolved with the induction equation for ten milliseconds, with the fluid held fixed to the state at $t=900$~ms, and only the second and third terms on the right-hand side of Eq. (\ref{eq:HLLElectricField}) included.  \label{fig:magneticEnergyDissipation}}
\end{figure}

To investigate the amount of magnetic energy dissipation in our simulations we have initiated a simulation with data from one of our models (model $\model{12}{0.0}{00}$ at $t=900$~ms) and evolved the magnetic field with the induction equation with only the dissipative terms in Eq. (\ref{eq:HLLElectricField}) (the second and third term on the right-hand-side) retained.  
The fluid variables ($\rho$, $\vect{u}$, $\edens{int}$, etc.) are held fixed to their initial values ($t=900$~ms).  
The results are plotted in Figure \ref{fig:magneticEnergyDissipation}, where we plot the magnetic energy $E_{\mbox{\tiny mag}}$ (black lines) and the characteristic spatial scale of the magnetic field $\lambdaMeanMag$ (cf. Eq. (\ref{eq:meanMagneticWaveNumber}) and proceeding text; gray curves) versus time.  
We see that the magnetic energy decays rapidly with time in the absence of the ideal part of the electric field (black dashed curve).  
Initially, it decays on a millisecond time scale (the decay rate is about 380~s$^{-1}$ at $t=900$~ms).  
The characteristic spatial scale of the magnetic field increases as a result of diffusion (gray dashed curve), which results in a decrease in the decay rate.  
The magnetic energy increases with time, and $\lambdaMeanMag$ stays roughly constant in the full run (solid lines).  
These results confirm our claim in Section \ref{sec:spectralAnalysis}, that numerical diffusion plays an important role in our simulations, and support our claims in Section \ref{sec:magneticEnergyGrowthRates}, that our simulations grossly underestimate the magnetic energy growth rate due to turbulence-driven magnetic field amplification in CCSNe as a result of the SASI.  
Thus, we expect the magnetic energy to grow on millisecond time scales, if dissipative effects can be ignored in the supernova environment.  

The diffusive evolution of intermediate waves by the HLL Riemann solver motivates us to improve our MHD scheme by including more wave families in the Riemann solver in future applications.  
In particular, the HLLD Riemann solver \citep{miyoshiKusano_2005} includes Alfv{\'e}n and entropy modes and is an attractive option (perhaps in combination with an improved interpolation scheme).  
Indeed, \citet{mignone_etal_2009} have developed an HLLD-type scheme for relativistic MHD, which shows significantly improved resolution on small scales when compared with the corresponding HLL-type scheme.  

\end{document}